\documentclass[11pt,a4paper]{article}

\usepackage{afterpage,amsmath,amssymb,array,authblk,calc,color,float,framed,graphicx,mathtools,multirow,nccmath,rotating,setspace,subcaption,xhfill}
\usepackage[colorlinks=true,breaklinks=true,allcolors=blue]{hyperref}
\usepackage[margin=2cm]{geometry}
\usepackage[export]{adjustbox}

\usepackage{epstopdf}

\usepackage{caption}
\usepackage[backend=biber,style=numeric-comp,sorting=none,maxbibnames=99,giveninits=true]{biblatex}
\DeclareFieldFormat*{title}{#1}
\DeclareFieldFormat*{number}{(#1)}
\DeclareFieldFormat*{pages}{#1}
\renewbibmacro{in:}{}
\renewbibmacro{volume+number+eid}{%
	\printfield{volume}%
	\printfield{number}
	\setunit{\addcomma\space}%
	\printfield{eid}}
\renewbibmacro*{publisher+location+date}{%
	\printlist{publisher}%
	\iflistundef{location}
	{\setunit*{\addcomma\space}}
	{\setunit*{\addcomma\space}}%
	\printlist{location}%
	\setunit*{\addcomma\space}%
	\usebibmacro{date}%
	\newunit}

\definecolor{LightGray}{rgb}{0.93,0.93,0.93}
\definecolor{Gray}{rgb}{0.4,0.4,0.4}

\newcommand{\algbox}[1]{\vspace{0.0cm}\fcolorbox{LightGray}{LightGray}{\parbox{0.99\textwidth}{\color{black}\vspace{-0.2cm}#1 \vspace{-0.2cm}}}\vspace{0.2cm}}
\newtheorem{algorithm}{Algorithm}
\newcommand{\comment}[1]{\textcolor{Gray}{\% #1}}
\newcommand{\wrt}[1]{\hspace{0.15cm} {\rm d}{#1}}

\begin{document}
	\title{Random walk models of anisotropic diffusion on rectangular and hexagonal lattices}
	\author{Luke P. Filippini\thanks{Corresponding author. \\ \indent\phantom{.} \textit{E-mail addresses:} \href{mailto:luke.filippini@hdr.qut.edu.au}{luke.filippini@hdr.qut.edu.au} (L. P. Filippini), \href{mailto:adrianne.jenner@qut.edu.au}{adrianne.jenner@qut.edu.au} (A. L. Jenner), \\ \indent{}\phantom{..}\href{mailto:elliot.carr@qut.edu.au}{elliot.carr@qut.edu.au} (E. J. Carr)}, Adrianne L. Jenner and Elliot J. Carr}
	\affil{School of Mathematical Sciences, Queensland University of Technology, GPO Box 2434, Brisbane, QLD 4001, Australia.}
	
	\date{}
	
	\maketitle
	
	\section*{Abstract}
	The diffusive transport of particles in anisotropic media is a fundamental phenomenon in computational, medical and biological disciplines. While deterministic models (partial differential equations) of such processes are well established, their inability to capture inherent randomness, and the assumption of a large number of particles, hinders their applicability. To address these issues, we present several equivalent (discrete-space discrete-time) random walk models of diffusion described by a spatially-invariant tensor on a two-dimensional domain with no-flux boundary conditions. Our approach involves discretising the deterministic model in space and time to give a homogeneous Markov chain governing particle movement between (spatial) lattice sites over time. The spatial discretisation is carried out using a vertex-centred element-based finite volume method on rectangular and hexagonal lattices, and a forward Euler discretisation in time yields a nearest-neighbour random walk model with simple analytical expressions for the transition probabilities. For each lattice configuration, analysis of these expressions yields constraints on the time step duration, spatial steps and diffusion tensor to ensure the probabilities are between zero and one. We find that model implementation on a rectangular lattice can be achieved with a constraint on the diffusion tensor, whereas a hexagonal lattice overcomes this limitation (no restrictions on the diffusion tensor). Overall, the results demonstrate good visual and quantitative (mean-squared error) agreement between the deterministic model and random walk simulations for several test cases. All results are obtained using MATLAB code available on GitHub (\href{https://github.com/lukefilippini/Filippini2025}{https://github.com/lukefilippini/Filippini2025}).

	\section{Introduction}
\begin{figure}[t]
		\begin{subfigure}{0.45\textwidth}
			\centering
			\includegraphics[width=\textwidth]{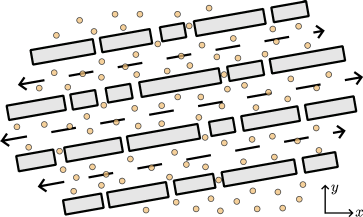}
			\caption{Particle diffusion in an anisotropic medium}
		\end{subfigure}
		\hfill
		\begin{subfigure}{0.45\textwidth}
			\centering
			\includegraphics[width=\textwidth]{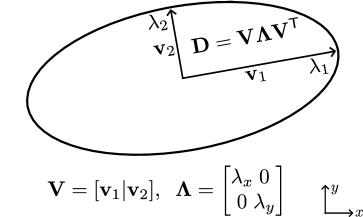}
			\caption{Ellipse representation of the diffusion tensor}
		\end{subfigure}
		\caption{\textbf{Representations of diffusion in anisotropic media.} (a) Particle diffusion in two-dimensional anisotropic media. The movement of particles (orange circles) occurs primarily along aligned structures (solid grey boxes) of the surrounding environment (indicated by dashed arrows). (b) Representation of the tensor describing diffusion in (a) using an ellipse. The eigenvalues ($\lambda_{1}$ and $\lambda_{2}$) and eigenvectors ($\mathbf{v}_{1}$ and $\mathbf{v}_{2}$) of the eigendecomposition $\mathbf{D}=\mathbf{V}\boldsymbol{\Lambda}\mathbf{V}^{\mathsf{T}}$ of the diffusion tensor $\mathbf{D}$ correspond to the length and magnitude of the ellipse axes (large arrows), respectively.}
		\label{fig:anisotropic_diffusion}
	\end{figure}

	Particle transport governed by diffusion is fundamental to computational, medical and biological physics. This process is influenced by the surrounding environment and often exhibits non-uniform behaviour due to spatial variations in the diffusion rate. In many applications, the presence of anisotropy in the medium implies that the rate of diffusion, typically quantified as a tensor, also depends on the direction in which a particle moves (see Figure \ref{fig:anisotropic_diffusion}). 
	Examples include diffusion filtering in image processing \cite{black_et_al_1998,yu_acton_2002}, cancer progression in the brain \cite{engwer_et_al_2015,painter_hillen_2013,suveges_et_al_2021}, and thermal conduction in plasma physics \cite{sharma_et_al_2007,van_es_et_al_2016}. The behaviour of such phenomena is frequently investigated using mathematical models to obtain insight into the relationship between diffusion coefficients and the movement of particles (e.g. cells, molecules).
	
	Mathematical models of diffusion processes are typically deterministic or stochastic. To elaborate, deterministic methods utilise partial differential equations to model the density of collective particles as a continuous function over space and time. These continuum solutions can facilitate analytical insight into the effect of diffusion coefficients on the transport process and can be obtained with minimal computational cost. Although, their validity requires a large number of particles, and the point estimates of particle density cannot capture inherent randomness and subtle details in the underlying phenomena \cite{gavagnin_yates_2018}. These limitations are overcome by stochastic random walk models, which consider particles individually and govern movement over discrete space and time using a set of probabilistic rules. However, this approach can be computationally expensive for a large number of particles \cite{klank_et_al_2018} and is relatively unsuitable for mathematical analysis \cite{gavagnin_yates_2018}. To account for the strengths and limitations of both approaches, it is beneficial to develop equivalent deterministic and random walk models such that the appropriate method can be used based on the application. 
	
	There are two classical approaches for obtaining equivalent deterministic and random walk models of particle diffusion. The first begins with a random walk model defined by phenomenological probabilities governing the diffusion process and utilises Taylor series expansions to derive equivalent partial differential equations. This approach yields analytical expressions for the diffusion coefficients in terms of the random walk parameters \cite{codling_et_al_2008,gavagnin_yates_2018,ibe_2013,redner_2001}. Alternatively, one can start with a deterministic model and discretise it in space and time to obtain a set of transition probabilities which define the equivalent random walk model. These probabilities govern particle movement between lattice sites of the spatial discretisation and are expressed in terms of the diffusion coefficients and discretisation parameters \cite{anderson_chaplain_1998,cai_et_al_2006,carr_2025,lotstedt_meinecke_2015,meinecke_et_al_2016,meinecke_lotstedt_2016}. In both cases, the relationships between transition probabilities, diffusion coefficients and other parameters (e.g. spatial steps, time step duration) are well understood for isotropic homogeneous media (spatially and directionally invariant diffusion rate) \cite{cai_et_al_2006,carr_2025,carr_simpson_2019,codling_et_al_2008,gavagnin_yates_2018,ibe_2013,redner_2001}. The latter approach has also been applied to layered heterogeneous media, where the diffusion rate varies spatially \cite{carr_2025}. However, extension of either method to anisotropic media presents key challenges.
	
	Deriving an equivalent partial differential equation from a random walk model of anisotropic diffusion is difficult \cite{gavagnin_yates_2018}. Currently, this approach is limited to lattice-free random walks that assume a spatially-invariant diffusion tensor \cite{ibe_2013}. For structured random walks, the methodology for defining probabilities phenomenologically is unclear due to directional and (possible) spatial variation of the diffusion rate. In comparison, discretising partial differential equations to obtain a set of transition probabilities is relatively straightforward. However, the discretisation must be configured appropriately to ensure that the obtained transition probabilities are between zero and one \cite{carr_2025,lotstedt_meinecke_2015,meinecke_et_al_2016,meinecke_lotstedt_2016,sharma_et_al_2007}. This can impose a constraint on the (fixed) diffusion tensor in addition to amenable conditions on the (adjustable) discretisation parameters \cite{edwards_zheng_2010}. Hence, it is important to configure the lattice and parameters of the discretisation appropriately to mitigate or eliminate a constraint on the diffusion tensor.
	
	In this paper, we derive equivalent (discrete-space and discrete-time) random walk models for the diffusion equation,
	\begin{gather}\label{eq:diffusion_eq}
		\frac{\partial u}{\partial t} + \boldsymbol{\nabla} \cdot \mathbf{q} = 0, \quad \mathbf{q}(\mathbf{x},t) = -\mathbf{D}\boldsymbol{\nabla}u, \quad \mathbf{x} \in \Omega,
	\end{gather}
	subject to,
	\begin{gather}
		u(\mathbf{x},0) = f(\mathbf{x}), \quad \mathbf{x} \in \Omega \cup \partial\Omega, \label{eq:IC} \\
		\mathbf{D}\boldsymbol{\nabla}u \cdot \hat{\mathbf{n}} = 0, \quad \mathbf{x}\in \partial\Omega \label{eq:no_flux},
	\end{gather}
	where $u(\mathbf{x},t)$ represents the density of particles located at $\mathbf{x} = (x,y)$ at time $t$, $f(\mathbf{x})$ describes the initial density of particles, $\partial\Omega$ corresponds to the boundary of the domain $\Omega$, and $\hat{\mathbf{n}}$ denotes the outward-facing unit normal from $\partial\Omega$. Here, we consider particle diffusion in a two-dimensional domain governed by a symmetric positive definite diffusion tensor,
	\begin{gather}\label{eq:D_tensor}
		\mathbf{D} = \begin{bmatrix} D_{xx} & D_{xy} \\ D_{xy} & D_{yy} \end{bmatrix},
	\end{gather}
	where the diffusivities $D_{xx} > 0$, $D_{yy} > 0$ and $D_{xy} \neq 0$ are spatially-invariant, and $\det(\mathbf{D}) = D_{xx}D_{yy} - D_{xy}^2 > 0$.
	
	Our approach involves discretising the diffusion equation (\ref{eq:diffusion_eq}) in space and time to give a homogeneous Markov chain which governs the movement of particles between lattice sites. The spatial discretisation is carried out using a vertex-centred element-based finite volume method (or control volume finite element method) \cite{marcondes_sepehrnoori_2010,moukalled_et_al_2016,oliver_et_al_2025,sheikholeslami_2019,voller_2009} on rectangular and hexagonal lattices. A forward Euler discretisation in time gives a nearest-neighbour random walk of non-interacting particles \cite{simpson_et_al_2009} with simple analytical expressions for the transition probabilities. For each lattice configuration, analysis of these expressions gives constraints on the spatial steps, time step duration and diffusion tensor to ensure that the probabilities are between zero and one. Implementation of the models on a rectangular lattice requires a condition on the diffusion tensor, whereas a hexagonal lattice overcomes this limitation (suitable for any diffusion tensor). Overall, results demonstrate good agreement between the deterministic model (\ref{eq:diffusion_eq})--(\ref{eq:D_tensor}) and random walk simulations for several test cases.
	
	The subsequent sections of this work are organised as follows. In section \ref{sec:deterministic}, we outline the spatial discretisation of the deterministic model (\ref{eq:diffusion_eq})--(\ref{eq:D_tensor}) on rectangular and hexagonal lattices. Then, the general approach for deriving an equivalent random walk model from the resulting system of differential equations is described in section \ref{sec:stochastic}. In section \ref{sec:transition_probs}, we present the full set of transition probabilities for the equivalent random walk models and analyse these expressions to determine constraints on the deterministic model parameters. Finally, we compare the deterministic model (\ref{eq:diffusion_eq})--(\ref{eq:D_tensor}) with random walk simulations in section \ref{sec:discussion}. The work is then summarised in section \ref{sec:conclusions} with suggestions for future work. 
	
	\section{Spatial discretisation of the deterministic model}\label{sec:deterministic}
\begin{figure}[p!]
		\centering
		\begin{subfigure}{\textwidth}
			\includegraphics[width=\textwidth]{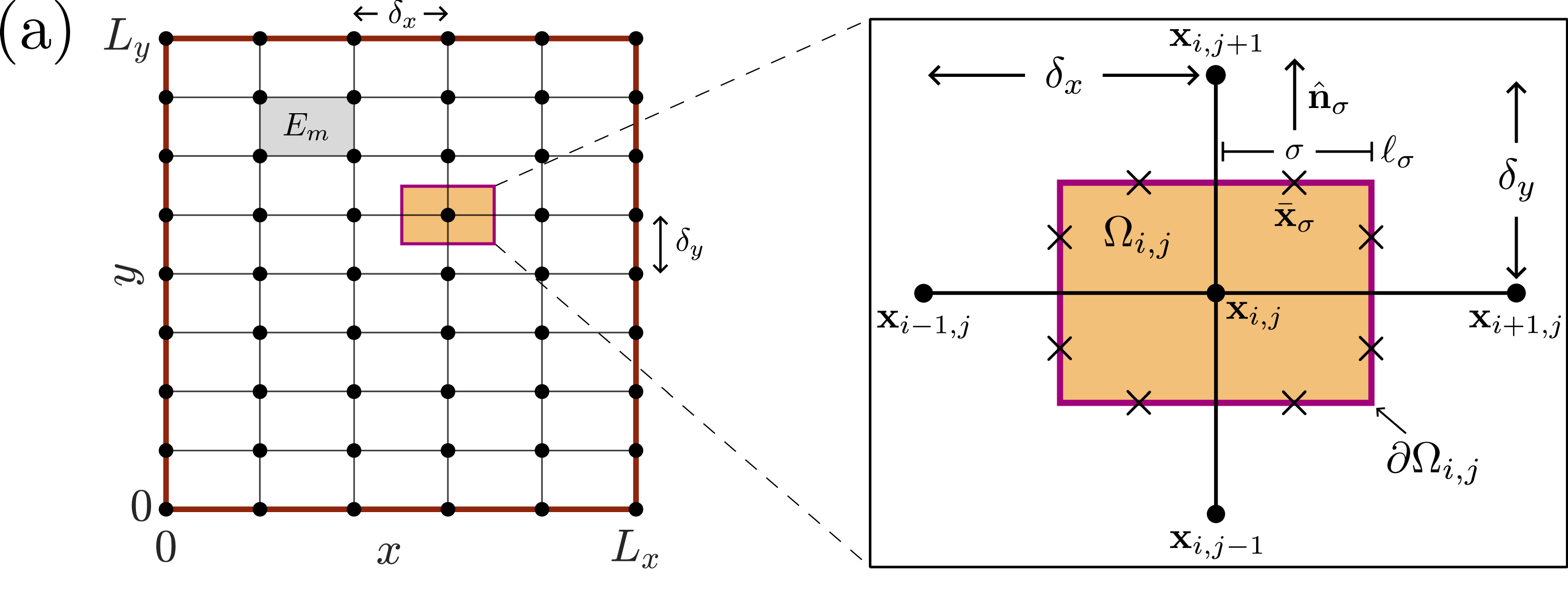}
			\label{fig:rectangular_lattice}
		\end{subfigure}
		\vfill
		\begin{subfigure}{\textwidth}
			\includegraphics[width=\textwidth]{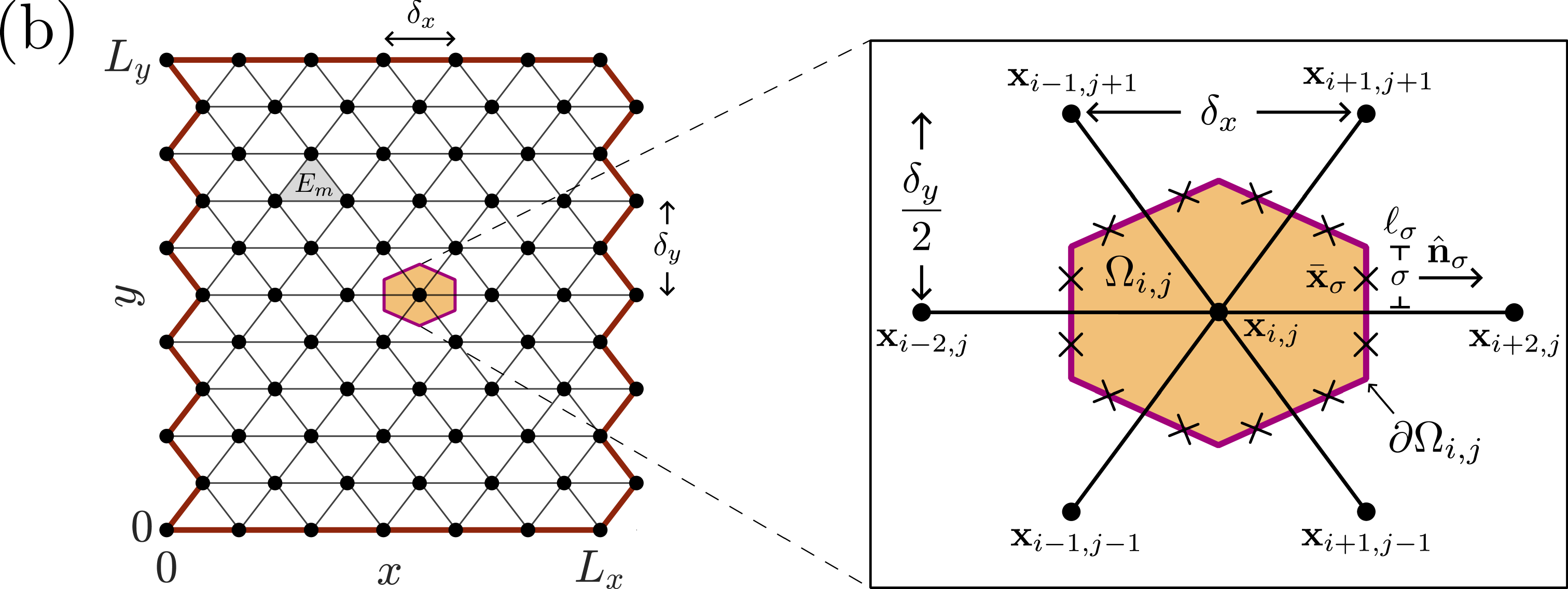}
			\label{fig:flat-top_lattice}
		\end{subfigure}
		\vfill
		\begin{subfigure}{\textwidth}
			\includegraphics[width=\textwidth]{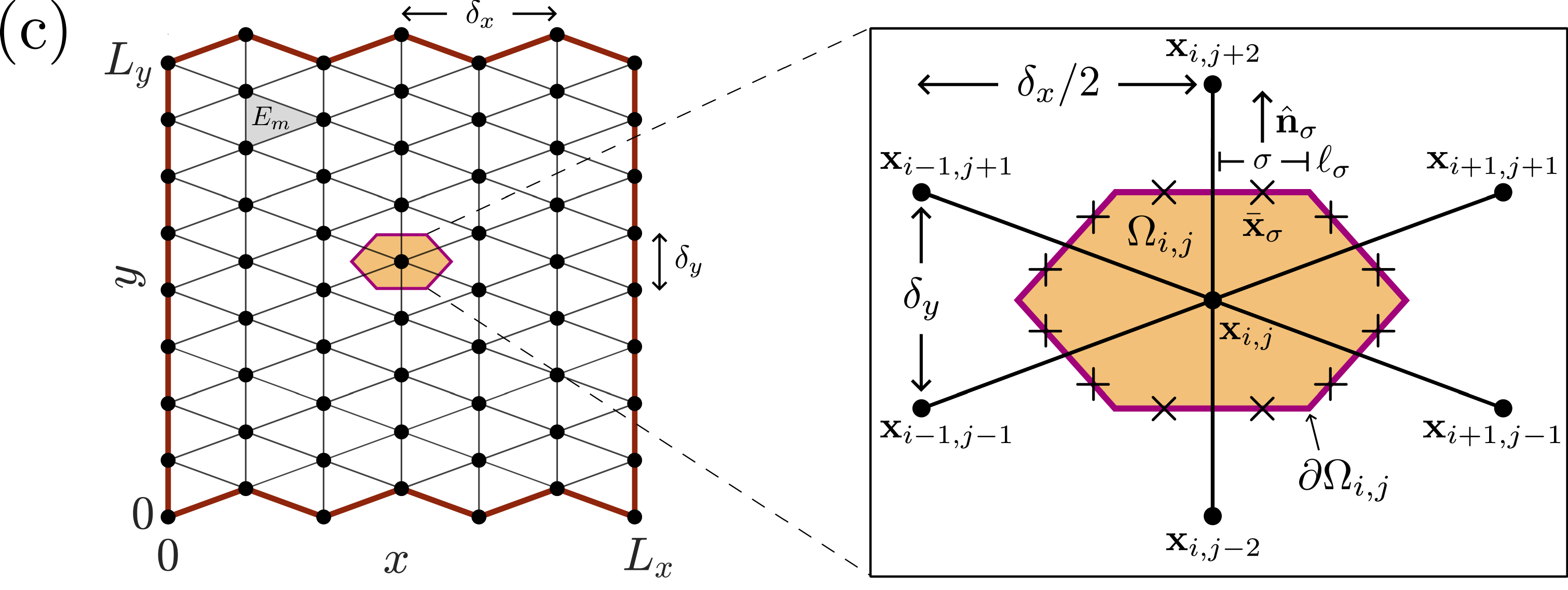}
			\label{fig:pointy-top_lattice}
		\end{subfigure}
		\caption{\textbf{Rectangular and hexagonal lattice configurations}. (a) Rectangular and (b)--(c) hexagonal lattices for the spatial discretisation of the deterministic model (\ref{eq:diffusion_eq})--(\ref{eq:D_tensor}). The problem domain $\Omega = [0,L_{x}] \times [0,L_{y}]$ is partitioned exactly by the (a) rectangular lattice (defined by structured rectangular elements), whereas the (b) flat-top and (c) pointy-top hexagonal lattices (defined by structured triangular elements) approximate $\Omega$. For each lattice configuration, $\delta_{x}$ and $\delta_{y}$ represent the spatial steps between lattice sites in the $x$ and $y$ directions, respectively. Around each lattice site $\mathbf{x}_{i,j}=(x_{i},y_{j})$ (black circles), we construct a control volume $\Omega_{i,j}$ (shaded orange) using a vertex-centred approach, implying the centroids of the elements $E_{m}$ (shaded grey), which have the lattice site $\mathbf{x}_{i,j}$ as a vertex, correspond to the vertices of the control volume boundary $\partial\Omega_{i,j}$ (shaded purple). An approximation for the flux across the control volume boundary is obtained by separating $\partial\Omega_{i,j}$ into line segments $\sigma \in \mathcal{E}_{i,j}$, where $\mathcal{E}_{i,j}$ denotes the set of line segments which constitute $\partial\Omega_{i,j}$, computing the outward (relative to $\sigma$) flux at each segment midpoint $\bar{\mathbf{x}}_{\sigma}$ (black crosses) in the direction of the unit normals $\hat{\mathbf{n}}_{\sigma}$ and multiplying by the segment lengths $\ell_{\sigma}$.}
		\label{fig:lattice_examples}
	\end{figure}

	The random walk models presented in this work are obtained from a spatial discretisation of the deterministic model (\ref{eq:diffusion_eq})--(\ref{eq:D_tensor}). In this section, the model is discretised in space using a vertex-centred element-based finite volume method (EbFVM) \cite{marcondes_sepehrnoori_2010,moukalled_et_al_2016,oliver_et_al_2025,sheikholeslami_2019,voller_2009} on three independent configurations of a lattice partitioning the problem domain $\Omega=[0,L_{x}]\times[0,L_{y}]$. We denote these as \textit{rectangular}, \textit{flat-top} and \textit{pointy-top} configurations \cite{red_blob_2013}, where the first corresponds to a lattice defined by structured rectangular elements, and the latter are hexagonal lattices defined by structured triangular elements. The rectangular lattice partitions $\Omega$ exactly, whereas the hexagonal lattices approximate the problem domain (examples are shown in Figure \ref{fig:lattice_examples}). 
	Firstly, we approximate the particle density $u(\mathbf{x},t)$ at each lattice site $\mathbf{x}_{i,j} = (x_{i},y_{j})$, denoted as $u_{i,j} \approx u(\mathbf{x}_{i,j},t)$, where the $x$ and $y$ positions of each site depend on the lattice configuration. These are given by
	\begin{gather}\label{eq:lattice_sites}
		x_{i} = \begin{cases} (i-1)\delta_{x}, & \text{if rectangular}, \\ (i-1)\delta_{x}/2, & \text{if hexagonal},\end{cases} \quad y_{j} = \begin{cases} (j-1)\delta_{y}, & \text{if rectangular}, \\ (j-1)\delta_{y}/2, & \text{if hexagonal},\end{cases}
	\end{gather}
	for $i=1,\hdots,N_{x}$ and $j=1,\hdots,N_{y}$, where $N_{x}$ and $N_{y}$ represent the total number of lattice sites in the $x$ and $y$ directions, respectively. Additionally, the spatial steps $\delta_{x}$ and $\delta_{y}$ are defined as
	\begin{gather}\label{eq:spatial_steps}
		\delta_{x} = \begin{cases} L_{x}/(N_{x}-1), & \text{if rectangular}, \\ 2L_{x}/(N_{x}-2), & \text{if flat-top}, \\ 2L_{x}/(N_{x} - 1), & \text{if pointy-top},\end{cases} \quad \delta_{y} = \begin{cases} L_{y}/(N_{y}-1), & \text{if rectangular}, \\ 2L_{y}/(N_{y}-1), & \text{if flat-top}, \\ 2L_{y}/(N_{y}-2), & \text{if pointy-top}. \end{cases}
	\end{gather}	
	The lattice sites for the hexagonal configurations are only defined if the sum of the position indices is even (i.e. $\mathbf{x}_{i,j} = (x_{i},y_{j})$ is only defined if $i + j$ is even). As a consequence, $N_{x}$ and $N_{y}$ must be even and odd, respectively, for the flat-top configuration and vice versa for the pointy-top configuration. This implies that the number of lattice sites in each row of the flat-top configuration is given by $N_{x}/2$, whereas the number of lattice sites in each column depends on whether the column index $j$ is even or odd. Similar observations are made for the pointy-top configuration, which has $N_{y}/2$ lattices sites in each column. For the rectangular configuration, the number of lattice sites in each row and column are given by $N_{x}$ and $N_{y}$, respectively.
	
	Next, we construct a control volume $\Omega_{i,j}$ around each lattice site $\mathbf{x}_{i,j}$ and define a discrete control-volume form of the diffusion equation (\ref{eq:diffusion_eq}). Integrating (spatially) over each control volume and applying the divergence theorem yields the finite volume equations
	\begin{gather}\label{eq:FVEs}
		\frac{{\rm d}u_{i,j}}{{\rm d}t} = -\frac{1}{V_{i,j}}\oint_{\partial\Omega_{i,j}}\mathbf{q}(\mathbf{x},t)\cdot\hat{\mathbf{n}}_{i,j}\wrt{s},
	\end{gather}
	where $V_{i,j}$ denotes the control volume area, $\partial\Omega_{i,j}$ corresponds to the boundary of the control volume $\Omega_{i,j}$, and $\hat{\mathbf{n}}_{i,j}$ is the outward-facing unit normal with respect to $\partial\Omega_{i,j}$. Note that $\mathbf{x}_{i,j}$ is centred within $\Omega_{i,j}$, and the vertices of $\partial\Omega_{i,j}$ coincide with the centroids of the lattice elements which have $\mathbf{x}_{i,j}$ as a vertex (see Figure \ref{fig:lattice_examples}). The flux integral in (\ref{eq:FVEs}) is separated into individual components corresponding to the distinct line segments of each control volume boundary,
	\begin{gather}\label{eq:flux_sum}
		\oint_{\partial\Omega_{i,j}}\mathbf{q}(\mathbf{x},t)\cdot\hat{\mathbf{n}}_{i,j}\wrt{s} = \hspace{-0.1cm}\sum_{\hspace{0.1cm}\sigma \hspace{0.05cm} \in \hspace{0.05cm} \mathcal{E}_{i,j}}\int_{\sigma}\mathbf{q}(\mathbf{x},t)\cdot\hat{\mathbf{n}}_{\sigma}\wrt{s},
	\end{gather}
	where $\mathcal{E}_{i,j}$ denotes the set of line segments which constitute $\partial\Omega_{i,j}$, and $\hat{\mathbf{n}}_{\sigma}$ is the outward-facing unit normal relative to the line segment $\sigma$. Each integral in (\ref{eq:flux_sum}) is then approximated using a midpoint rule,
	\begin{gather}\label{eq:MPFA}
		\sum_{\hspace{0.1cm}\sigma \hspace{0.05cm} \in \hspace{0.05cm} \mathcal{E}_{i,j}}\int_{\sigma}\mathbf{q}(\mathbf{x},t)\cdot\hat{\mathbf{n}}_{\sigma}\wrt{s} \approx \hspace{-0.1cm}\sum_{\hspace{0.1cm}\sigma \hspace{0.05cm} \in \hspace{0.05cm} \mathcal{E}_{i,j}}(\mathbf{q}(\bar{\mathbf{x}}_{\sigma},t)\cdot\hat{\mathbf{n}}_{\sigma})\ell_{\sigma},
	\end{gather}
	where $\bar{\mathbf{x}}_{\sigma} = (\bar{x}_{\sigma},\bar{y}_{\sigma})$ and $\ell_{\sigma}$ represent the midpoint and length of the line segment $\sigma$, respectively (see Figure \ref{fig:lattice_examples}).
	
	Finally, we approximate the flux terms $\mathbf{q}(\bar{\mathbf{x}}_{\sigma},t) \cdot \hat{\mathbf{n}}_{\sigma}$ in (\ref{eq:MPFA}) using an element-based approach. As implied in Figure \ref{fig:lattice_examples}, each control volume $\Omega_{i,j}$ is divided into sub-control volumes contained within different lattice elements. For each element $E_{m}$ which has the lattice site $\mathbf{x}_{i,j}$ as a vertex, we calculate flux terms $\mathbf{q}(\bar{\mathbf{x}}_{\sigma},t)\cdot\hat{\mathbf{n}}_{\sigma}$ contained within the element under the assumption that the particle density $u(\mathbf{x},t)$ varies linearly or bilinearly in space (within $E_{m}$). This is commonly achieved using an interpolating function to approximate $u(\mathbf{x},t)$ within each rectangular element (bilinear),
	\begin{gather}\label{eq:rec_interpolant}
		g_{m}(\mathbf{x}) = \alpha_{m,1}x + \alpha_{m,2}y + \alpha_{m,3}xy + \alpha_{m,4},
	\end{gather}
	or triangular element (linear),
	\begin{gather}\label{eq:hex_interpolant}
		g_{m}(\mathbf{x}) = \alpha_{m,1}x + \alpha_{m,2}y + \alpha_{m,3},
	\end{gather}
	implying $g_{m}(\mathbf{x}) \approx u(\mathbf{x},t)$ for $\mathbf{x}=(x,y)\in E_{m}$. The coefficients in (\ref{eq:rec_interpolant}) and (\ref{eq:hex_interpolant}), which ensure equality of the interpolating function and particle density at the vertices of the element $E_{m}$, are well known \cite{oliver_et_al_2025,sheikholeslami_2019} and given in Appendix \ref{sec:appendix_interp}. This yields an approximation $\boldsymbol{\nabla}g_{m}(\mathbf{x}) \approx \boldsymbol{\nabla}u(\mathbf{x},t)$ at any position $\mathbf{x}\in E_{m}$. Evaluating $\boldsymbol{\nabla}g_{m}(\mathbf{x})$ at the line segment midpoints $\bar{\mathbf{x}}_{\sigma} \in E_{m}$ gives the following approximation for the flux terms $\mathbf{q}(\bar{\mathbf{x}}_{\sigma},t)\cdot\hat{\mathbf{n}}_{\sigma}$ (within $E_{m}$):
	\begin{gather}\label{eq:flux_approx}
		\mathbf{q}(\bar{\mathbf{x}}_{\sigma},t)\cdot\hat{\mathbf{n}}_{\sigma} \approx -\mathbf{D}\boldsymbol{\nabla}g_{m}(\bar{\mathbf{x}}_{\sigma})\cdot\hat{\mathbf{n}}_{\sigma},
	\end{gather}
	where $\boldsymbol{\nabla}g_{m}(\bar{\mathbf{x}}_{\sigma}) = [\alpha_{m,1} + \alpha_{m,3}\bar{y}_{\sigma},\alpha_{m,2} + \alpha_{m,3}\bar{x}_{\sigma}]^{\mathsf{T}}$ and $\boldsymbol{\nabla}g_{m}(\bar{\mathbf{x}}_{\sigma}) = [\alpha_{m,1},\alpha_{m,2}]^{\mathsf{T}}$ for rectangular and triangular elements, respectively. Substitution of the approximation (\ref{eq:flux_approx}) for each flux term into (\ref{eq:FVEs})--(\ref{eq:MPFA}) gives the complete spatial discretisation of the deterministic model (\ref{eq:diffusion_eq})--(\ref{eq:D_tensor}),
	\begin{gather}\label{eq:FVEs_final}
		\frac{{\rm d}u_{i,j}}{{\rm d}t} = \frac{1}{V_{i,j}}\sum_{\hspace{0.1cm}\sigma \hspace{0.05cm} \in \hspace{0.05cm} \mathcal{E}_{i,j}}(\mathbf{D}\boldsymbol{\nabla}g(\bar{\mathbf{x}}_{\sigma})\cdot\hat{\mathbf{n}}_{\sigma})\ell_{\sigma},
	\end{gather}
	where $g(\bar{\mathbf{x}}_{\sigma}) = g_{m}(\bar{\mathbf{x}}_{\sigma})$ if $\bar{\mathbf{x}}_{\sigma} \in E_{m}$. Note that flux terms $\mathbf{q}(\bar{\mathbf{x}}_{\sigma},t)\cdot\hat{\mathbf{n}}_{\sigma}$ evaluated on the boundary of the domain $\partial\Omega$ do not contribute to the FVEs in accordance with the no-flux condition (\ref{eq:no_flux}). For specific details regarding implementation of the spatial discretisation, the reader is referred to MATLAB code available on GitHub (\href{https://github.com/lukefilippini/Filippini2025}{https://github.com/lukefilippini/Filippini2025}).

	\section{Random walk model}\label{sec:stochastic}
	In this section, we construct equivalent random walk models from the spatial discretisation of the deterministic model (\ref{eq:diffusion_eq})--(\ref{eq:D_tensor}) on rectangular and hexagonal lattices (see section \ref{sec:deterministic}). For each lattice configuration, we obtain a system of differential equations,
	\begin{gather}\label{eq:density_ODEs}
		\frac{{\rm d}\mathbf{U}}{{\rm d}t} = \mathbf{A}\mathbf{U}, \quad \mathbf{U}(0) = \mathbf{U}_0,
	\end{gather}
	which describes the evolution of particle density at each lattice site over time, where $\mathbf{U} = (U_{1},\hdots,U_{N_{\ell}})^{\mathsf{T}}$ contains the particle density at each lattice site, $\mathbf{U}_{0} = (f(\mathbf{X}_{1}),\hdots,f(\mathbf{X}_{N_{\ell}}))^{\mathsf{T}}$ denotes the initial particle density at each lattice site and $\mathbf{A}$ is a $N_{\ell} \times N_{\ell}$ banded matrix containing the coefficients (omitted for brevity) of the particle densities appearing in the finite volume equations (\ref{eq:FVEs_final}). These coefficients are expressed in terms of the spatial steps $\delta_{x}$ and $\delta_{y}$ and components of the diffusion tensor $\mathbf{D}$. Here, $N_{\ell}$ represents the total number of lattice sites and is defined as $N_{\ell} = N_{x}N_{y}$ and $N_{\ell}=N_{x}N_{y}/2$ for the rectangular and hexagonal lattices, respectively. Moreover, $U_{k}$ denotes the particle density at the lattice site  $\mathbf{X}_{k}$ whose position in $\mathbf{U}$ is mapped from the site location in the specified lattice configuration. In other words, $U_{k} = u_{i,j}$  at $\mathbf{X}_{k} = \mathbf{x}_{i,j}$ where the mapping is defined as 
	\begin{gather}\label{eq:map}
		k = \begin{cases}(j-1)N_{x} + i, & \text{if rectangular}, \\ (j-1)N_{x}/2 + \lceil i/2 \rceil, & \text{if flat-top}, \\ (i-1)N_{y}/2 + \lceil j/2 \rceil, & \text{if pointy-top},\end{cases}
	\end{gather}
	for $i = 1,\hdots,N_{x}$ and $j=1,\hdots,N_{y}$, noting that $\lceil x \rceil$ is the ceiling function. The mapping (\ref{eq:map}) is carried out column-wise for the pointy-top hexagonal configuration as opposed to the row-wise approach employed for the rectangular and flat-top hexagonal configurations.
	
	To obtain transition probabilities governing particle movement between lattice sites, we convert the system of differential equations (\ref{eq:density_ODEs}) to one concerned with the number of particles within each control volume. Here, we utilise the relationship $U_{k} = N_{k}S_{p}/V_{k}$, where $N_{k}$ represents the number of particles within the control volume $\Omega_{k}$ (assumed to be at the lattice site $\mathbf{X}_{k}$) \cite{carr_2025}. Additionally, $V_{k}$ represents the area of $\Omega_{k}$, and $S_{p} = \sum_{k=1}^{N_{\ell}}f(\mathbf{X}_{k})V_{k}/N_{p}$ is a scaling constant where $N_{p}$ denotes the total number of particles in the problem domain $\Omega$. To clarify, $V_{k} = V_{i,j}$ for $\Omega_{k} = \Omega_{i,j}$ (associated with $\mathbf{X}_{k} = \mathbf{x}_{i,j}$) where the mapping is given by (\ref{eq:map}). Substituting this relationship into (\ref{eq:density_ODEs}) yields an equivalent system of differential equations for the number of particles at each lattice site,
	\begin{gather}\label{eq:particle_ODEs}
		\frac{{\rm d}\mathbf{N}}{{\rm d}t} = \mathbf{B}\mathbf{N}, \quad \mathbf{N}(0) = \mathbf{N}_{0},
	\end{gather}
	where $\mathbf{N} = (N_{1},\hdots,N_{N_{\ell}})^{\mathsf{T}}$, $\mathbf{N}_{0} = \text{round}(S_{p}(f(\mathbf{X}_{1})V_{1},\hdots,f(\mathbf{X}_{N_{\ell}})V_{N_{\ell}}))^{\mathsf{T}}$ and $\mathbf{B} = \mathbf{V}\mathbf{A}\mathbf{V}^{-1}$ noting $\mathbf{V} = {\rm diag}(V_{1},\hdots,V_{N_{\ell}})$. Here, the initial number of particles, $\mathbf{N}_0$, is rounded to ensure an integer number of particles at each lattice site. Equation (\ref{eq:particle_ODEs}) is then discretised using a one-step method to give
	\begin{gather}\label{eq:stoch_markov}
		\mathbf{N}_{n}^{\mathsf{T}} = \mathbf{N}_{n-1}^{\mathsf{T}}\mathbf{P}, \quad n = 1,\hdots,N_{t},
	\end{gather}
	where $\mathbf{N}_{n}^{\mathsf{T}}=(N_{1,n},\hdots,N_{N_{\ell},n})$ contains the number of particles at each lattice site at time $t_{n} = n\tau$. Here, $\tau = T/N_{t}$ represents the time step duration, where $T$ and $N_{t}$ represent the total duration and number of time steps, respectively. Moreover, $\mathbf{P}$ is a $N_{\ell} \times N_{\ell}$ mapping matrix whose form depends on the chosen time discretisation method. Under the assumption that $\mathbf{P}$ is a right stochastic matrix (the entries of $\mathbf{P}$ are non-negative and each row sums to one), equation (\ref{eq:stoch_markov}) can be interpreted as a homogeneous Markov chain. This implies that the matrix $\mathbf{P}$ defines the transition probabilities governing particle movement between lattice sites, where we let $p_{k,m}$ correspond to the probability (entry in row $k$ and column $m$ of $\mathbf{P}$) that a particle located at $\mathbf{X}_{k}$ at time $t=t_{n-1}$ moves to $\mathbf{X}_{m}$ at time $t = t_{n}$. Note that the indices $k$ and $m$ are obtained according to (\ref{eq:map}), and $p_{k,k}$ denotes the probability that a particle remains at $\mathbf{X}_{k}$.
	
	In this work, the matrix $\mathbf{P}$ is obtained from a forward Euler discretisation of the system of differential equations (\ref{eq:particle_ODEs}). Discretising from time $t = t_{n-1}$ to $t = t_{n}$, we obtain the stochastic matrix $\mathbf{P} = \mathbf{I} + \tau\mathbf{C}$, where $\mathbf{I}$ is the $N_{\ell} \times N_{\ell}$ identity matrix, and $\mathbf{C} = \mathbf{B}^{\mathsf{T}} = (\mathbf{V}\mathbf{A}\mathbf{V}^{-1})^{\mathsf{T}}$ is a banded coefficient matrix where each row sums to zero. This implies that each row of $\mathbf{P}$ sums to one, although constraints on the model parameters, including the time step duration $\tau$, spatial steps $\delta_{x}$ and $\delta_{y}$ and diffusion tensor $\mathbf{D}$, are required to ensure the entries of $\mathbf{P}$ are between zero and one (see sections \ref{sec:rec_probs}--\ref{sec:pointy-top_probs}). These constraints are recovered from analysing simple analytical expressions for the transition probabilities, and are equivalent to those for ensuring monotonicity of the deterministic model solution. To elaborate, $\mathbf{P} = \mathbf{V}^{-\mathsf{T}}(\mathbf{I} + \tau\mathbf{A})\mathbf{V}$ is a stochastic matrix if the entries of $\mathbf{I} + \tau\mathbf{A}$ are non-negative, which is the well-known monotonicity condition for the forward Euler discretisation $\mathbf{U}_{n} = (\mathbf{I} + \tau\mathbf{A})\mathbf{U}_{n-1}$ of (\ref{eq:density_ODEs}), where $\mathbf{U}_{n}$ corresponds to $\mathbf{U}$ at time $t_{n} = n\tau$ \cite{holmes_2007}. When these conditions are met, the forward Euler discretisation of (\ref{eq:particle_ODEs}) yields a random walk model where particles can only move to neighbouring lattice sites during a single time step.
	
	The random walk model governing particle movement on the rectangular and hexagonal lattices is outlined in Algorithm \ref{alg:random_walk}. At each time step, we generate a uniform random number $r\sim\mathcal{U}(0,1)$ for each particle, allowing it to move from the current site $\mathbf{X}_{k}$ to the new site $\mathbf{X}_{m}$ if $r \in (P_{k,m-1},P_{k,m})$, where $P_{k,m}$ are cumulative probabilities defined as $P_{k,m} = \sum_{n=1}^{m}p_{k,n}$. Prior to the random walk, we compute the initial number of particles as in (\ref{eq:particle_ODEs}) and subsequently update the scaling constant: $S_{p}^{*} = \sum_{k=1}^{N_{\ell}}f(\mathbf{X}_{k})V_{k}/N_{p}^{*}$, where $N_{p}^{*} = \sum_{k=1}^{N_{\ell}}N_{k,0}$  is the updated number of particles. Thus, the (stochastic) particle density at time $t=t_{n}$ is given by $U_{k,n} = N_{k,n}S_{p}^{*}/V_{k}$.
	
	\bigskip\noindent
	\algbox{
		\begin{algorithm}[Random walk model]\mbox{}\\
			\label{alg:random_walk}
			\emph{
			\begin{tabular}{@{}l}
				$U_{k,0} = f(\mathbf{X}_{k})$  for $k = 1,\hdots,N_{\ell}$ \comment{initial particle density at site $k$}\\
				$S_{p} = \sum_{k=1}^{N_{\ell}}U_{k,0}V_{k}/N_{p}$ \comment{scaling constant}\\
				$N_{k,0} = \text{round}(U_{k,0}V_{k}/S_{p})$ for $k = 1,\hdots,N_{\ell}$ \comment{initial number of particles at site $k$}\\
				$S_{p}^{*} = \sum_{k=1}^{N_{\ell}}U_{k,0}V_{k}/N_{p}^{*}$, where $N_{p}^{*} = \sum_{k=1}^{N_{\ell}}N_{k,0}$ \comment{updated scaling constant}\\
				$P_{k,0} = 0$ and $P_{k,m} = \sum_{n=1}^{m} p_{k,n}$ \comment{cumulative probabilities}\\
				\textbf{for} $n = 1,\hdots,N_{t}$ \comment{loop over time steps}\\
				\qquad $N_{k,n} = N_{k,n-1}$ for $k = 1,\hdots,N_{\ell}$ \comment{number of particles at site $k$ and time $t = t_{n-1}$}\\
				\qquad \textbf{for} $k = 1,\hdots,N_{\ell}$ \comment{loop over lattice sites}\\
				\qquad \qquad \textbf{for} $m = 1,\hdots,N_{k,n-1}$ \comment{loop over number of particles at site $k$}\\
				\qquad \qquad \qquad Sample $r \sim \mathcal{U}(0,1)$ \comment{random number in $[0,1]$}\\
				\qquad \qquad \qquad Find $m$ such that $r\in(P_{k,m-1},P_{k,m})$ \comment{move from site $k$ to site $m$ at $t = t_{n}$}\\
				\qquad \qquad \qquad $N_{k,n} = N_{k,n-1} - 1$, $N_{m,n} = N_{m,n-1} + 1$ \comment{update number of particles}\\ 
				\qquad \qquad \textbf{end}\\
				\qquad \textbf{end}\\
				\qquad $U_{k,n} = N_{k,n}S_{p}^{*} / V_{k}$ for $k = 1,\hdots,N_{\ell}$ \comment{particle density at site $k$ and time $t = t_{n}$}\\
				\textbf{end}\\
			\end{tabular}
			}
	\end{algorithm}}
	
	\section{Analysis of transition probabilities}\label{sec:transition_probs}
	In the following subsections, we (i) provide analytical expressions for the transition probabilities which define the random walk models and (ii) analyse constraints on the time step duration $\tau$, spatial steps $\delta_{x}$ and $\delta_{y}$ and diffusion tensor $\mathbf{D}$ to ensure the probabilities are between zero and one. For each lattice configuration, we let $p_{i,j}^{k,m}$ denote the probability that a particle moves from $\mathbf{x}_{i,j} = (x_{i},y_{j})$ at time $t_{n-1} = (n-1)\tau$ to $\mathbf{x}_{k,m} = (x_{k},y_{m})$ at time $t_{n} = t_{n-1} + \tau$. This notation directly relates each transition probability to the relevant lattice sites rather than to the corresponding row and column positions in the matrix $\mathbf{P}$ (see section \ref{sec:stochastic}).
	
	\subsection{Rectangular lattice}\label{sec:rec_probs}
	Firstly, we consider the stochastic matrix $\mathbf{P} = \mathbf{I} + \tau\mathbf{C}$ whose entries correspond to the probabilities of particle movement on a rectangular lattice (see Figure \ref{fig:lattice_examples}). These entries are presented in Figure \ref{fig:rec_probs} for interior, edge and corner lattice sites. In each case, the probabilities of particle movement occurring along the coordinate axes in a horizontal (to $\mathbf{x}_{i-1,j}$ or $\mathbf{x}_{i+1,j}$) or vertical (to $\mathbf{x}_{i,j-1}$ or $\mathbf{x}_{i,j+1}$) direction are respectively defined in terms of the following expressions:
	\begin{gather}\label{eq:rec_horiz_vert}
		\rho_{h} = \frac{\tau(3\delta_{y}^2D_{xx} - \delta_{x}^2D_{yy})}{\delta_{x}^2\delta_{y}^2}, \quad
		\rho_{v} = \frac{\tau(3\delta_{x}^2D_{yy} - \delta_{y}^2D_{xx})}{\delta_{x}^2\delta_{y}^2}.
	\end{gather}
	Additionally, the expressions
	\begin{gather}\label{eq:rec_diag_plus}
		\rho_{d^{+}} = \frac{\tau(\delta_{y}^2D_{xx} + 4\delta_{x}\delta_{y}D_{xy} + \delta_{x}^2D_{yy})}{\delta_{x}^2\delta_{y}^2},
	\end{gather}
	and 
	\begin{gather}\label{eq:rec_diag_minus}
		\rho_{d^{-}} = \frac{\tau(\delta_{y}^2D_{xx} - 4\delta_{x}\delta_{y}D_{xy} + \delta_{x}^2D_{yy})}{\delta_{x}^2\delta_{y}^2},
	\end{gather}
	are used to define the probabilities of particle movement to the northeast or southwest sites (to $\mathbf{x}_{i+1,j+1}$ or $\mathbf{x}_{i-1,j-1}$) or to the northwest or southeast sites (to $\mathbf{x}_{i-1,j+1}$ or $\mathbf{x}_{i+1,j-1}$), respectively. Finally, we utilise a linear combination of the expressions 
	\begin{gather}\label{eq:remain_expr}
		\rho_{xx} = \frac{\tau D_{xx}}{\delta_{x}^2}, \quad \rho_{xy} = \frac{\tau D_{xy}}{\delta_{x}\delta_{y}}, \quad \rho_{yy} = \frac{\tau D_{yy}}{\delta_{y}^2},
	\end{gather}	
	to define the probability of a particle remaining in its current position. For each case of lattice sites, we provide a schematic of the local lattice (central and neighbouring sites labelled with position indices) to ease visual association between each lattice site and the corresponding probability of a particle moving to, or remaining at, that location (see Figure \ref{fig:rec_probs}). Note that the corresponding row and column positions in the matrix $\mathbf{P}$ for the probability $p_{i,j}^{k,m}$ are given by $m_{i,j}$ and $m_{k,m}$, respectively, where $m_{i,j} = (j-1)N_{x} + i$ is the mapping function from (\ref{eq:map}).
	\afterpage{
	\begin{figure}[H]
		\begin{minipage}[c]{0.65\textwidth}
			\vspace{0pt}
			{\bfseries (a) Interior ($i=2,\hdots,N_{x}-1$ and $j=2,\hdots,N_{y}-1$)}:
			\begin{fleqn}[10pt]
				\begin{alignat*}{3}
					&p_{i,j}^{i-1,j+1} = \frac{\rho_{d^{-}}}{8}, & \quad &p_{i,j}^{i,j+1} = \frac{\rho_{v}}{4}, & \quad &p_{i,j}^{i+1,j+1} = \frac{\rho_{d^{+}}}{8}, \\ 
					&p_{i,j}^{i-1,j} = \frac{\rho_{h}}{4}, & &p_{i,j}^{i,j} = 1 - \rho_{r}, & &p_{i,j}^{i+1,j} = \frac{\rho_{h}}{4}, \\
					&p_{i,j}^{i-1,j-1} = \frac{\rho_{d^{+}}}{8}, & &p_{i,j}^{i,j-1} = \frac{\rho_{v}}{4}, & &p_{i,j}^{i+1,j-1} = \frac{\rho_{d^{-}}}{8},
				\end{alignat*}
			\end{fleqn}
			where $\rho_{r} = 3(\rho_{xx} + \rho_{yy})/2$.
		\end{minipage}
		\begin{minipage}[c]{0.325\textwidth}
			\centering
			\vspace{0pt}
			\includegraphics[width=\textwidth]{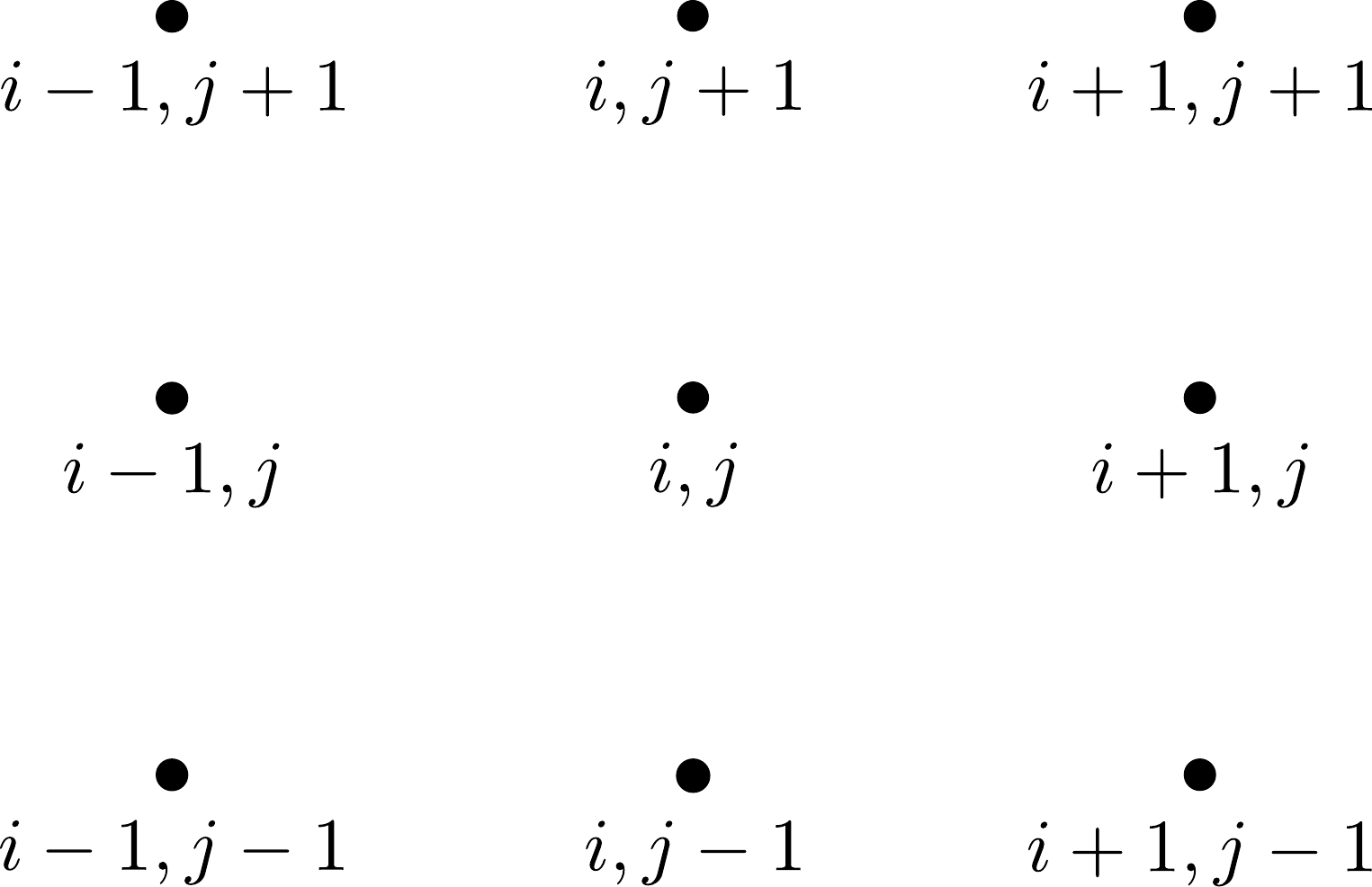}
		\end{minipage}
		\phantomcaption
	\end{figure}
	\vfill
	\vspace{-\intextsep}\noindent\dotfill
	\vfill
	\begin{figure}[H]\ContinuedFloat
		\begin{minipage}[c]{0.65\textwidth}
			\vspace{0pt}
			{\bfseries (b) Western edge ($i=1$ and $j=2,\hdots,N_{y}-1$)}:
			\begin{fleqn}[10pt]
				\begin{alignat*}{3}
					&p_{i,j}^{i,j+1} = \frac{\rho_{v}}{4}, & \quad &p_{i,j}^{i+1,j+1} = \frac{\rho_{d^{+}}}{4}, & \quad &p_{i,j}^{i,j} =  1 - \rho_{r}, \\
					&p_{i,j}^{i+1,j} = \frac{\rho_{h}}{2}, & &p_{i,j}^{i,j-1} = \frac{\rho_{v}}{4}, & &p_{i,j}^{i+1,j-1} = \frac{\rho_{d^{-}}}{4},
				\end{alignat*}
			\end{fleqn}
			where $\rho_{r} = 3(\rho_{xx} + \rho_{yy})/2$.
		\end{minipage}
		\begin{minipage}[c]{0.325\textwidth}
			\centering
			\vspace{0pt}
			\includegraphics[width=\textwidth]{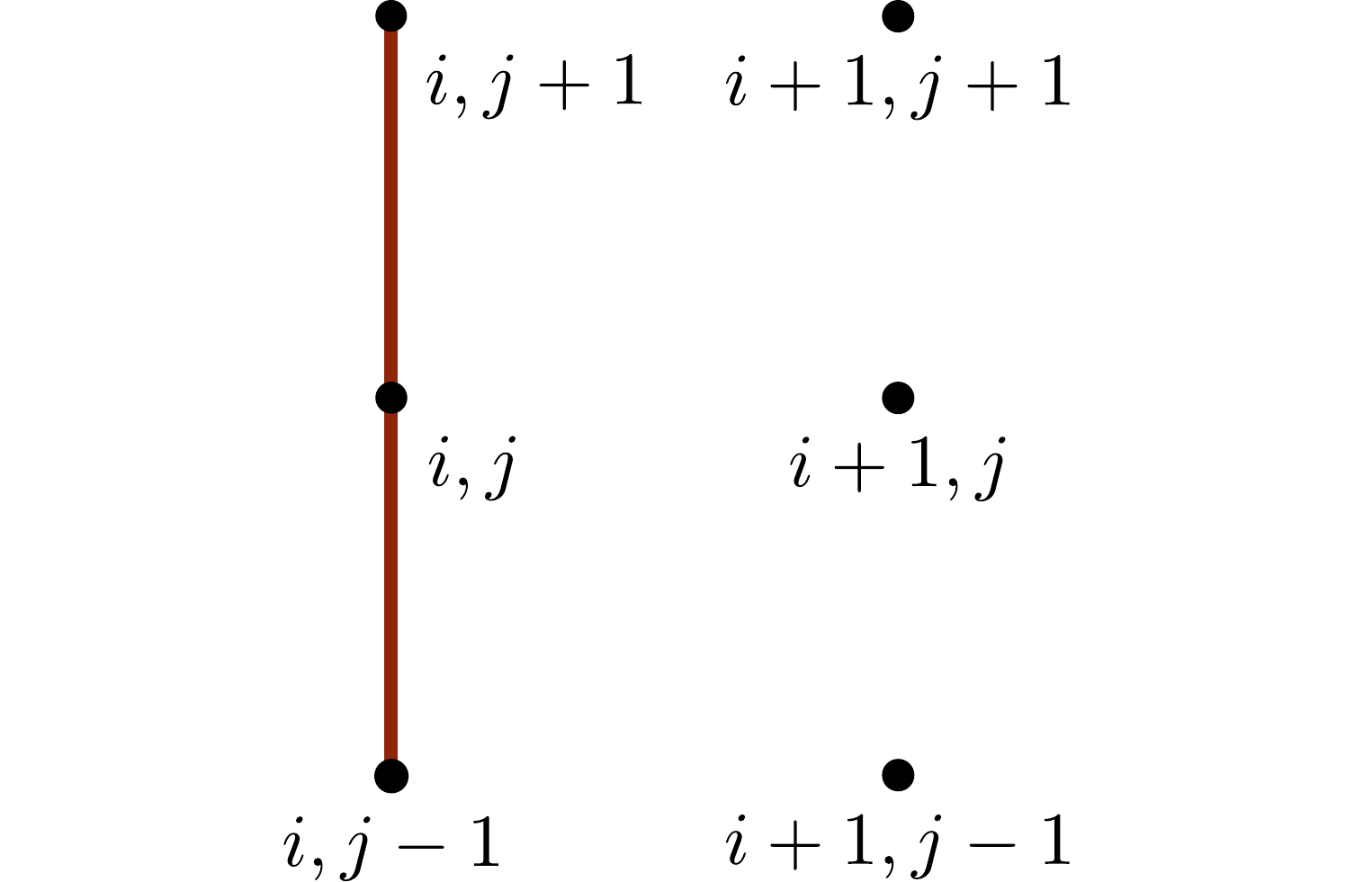}
		\end{minipage}
	\end{figure}
	\vfill
	\vspace{-\intextsep}\noindent\dotfill
	\vfill
	\begin{figure}[H]\ContinuedFloat
		\begin{minipage}[c]{0.65\textwidth}
			\vspace{0pt}
			{\bfseries (c) Southern edge ($i=2,\hdots,N_{x}-1$ and $j=1$)}:
			\begin{fleqn}[10pt]
				\begin{alignat*}{3}
					&p_{i,j}^{i-1,j+1} = \frac{\rho_{d^{-}}}{4}, & \quad & p_{i,j}^{i,j+1} = \frac{\rho_{v}}{2}, & \quad &p_{i,j}^{i+1,j+1} = \frac{\rho_{d^{+}}}{4}, \\ 
					&p_{i,j}^{i-1,j} = \frac{\rho_{h}}{4}, & &p_{i,j}^{i,j} = 1 - \rho_{r}, & &p_{i,j}^{i+1,j} = \frac{\rho_{h}}{4},
				\end{alignat*}
			\end{fleqn}
			where $\rho_{r} = 3(\rho_{xx} + \rho_{yy})/2$.
		\end{minipage}
		\begin{minipage}[c]{0.325\textwidth}
			\vspace{0pt}
			\includegraphics[width=\textwidth]{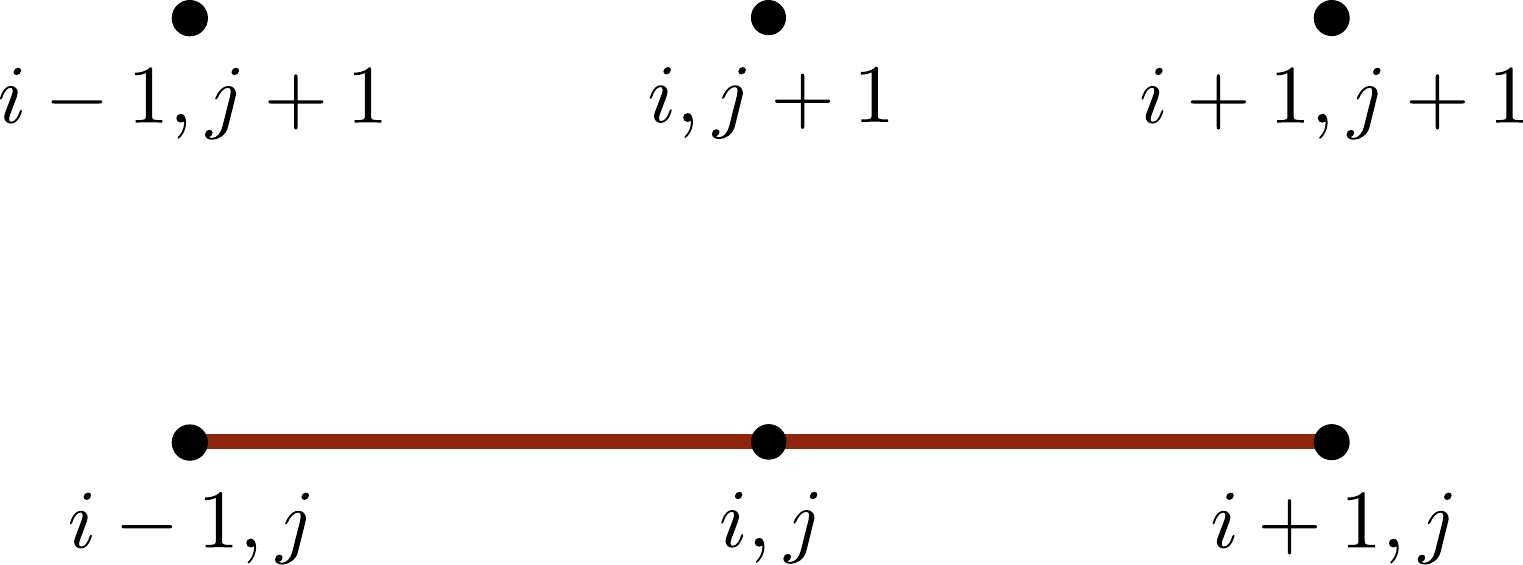}
		\end{minipage}
	\end{figure}
	\vfill
	\vspace{-\intextsep}\noindent\dotfill
	\vfill
	\begin{figure}[H]\ContinuedFloat
		\begin{minipage}[c]{0.65\textwidth}
			\vspace{0pt}
			{\bfseries (d) Eastern edge ($i=N_{x}$ and $j=2,\hdots,N_{y}-1$)}:
			\begin{fleqn}[10pt]
				\begin{alignat*}{3}
					&p_{i,j}^{i-1,j+1} = \frac{\rho_{d^{-}}}{4}, & \quad &p_{i,j}^{i,j+1} = \frac{\rho_{v}}{4}, & \quad &p_{i,j}^{i-1,j} = \frac{\rho_{h}}{2}, \\
					&p_{i,j}^{i,j} =  1 - \rho_{r}, & &p_{i,j}^{i-1,j-1} = \frac{\rho_{d^{+}}}{4}, & &p_{i,j}^{i,j-1} = \frac{\rho_{v}}{4},
				\end{alignat*}
			\end{fleqn}
			where $\rho_{r} = 3(\rho_{xx} + \rho_{yy})/2$.
		\end{minipage}
		\begin{minipage}[c]{0.325\textwidth}
			\centering
			\vspace{0pt}
			\includegraphics[width=\textwidth]{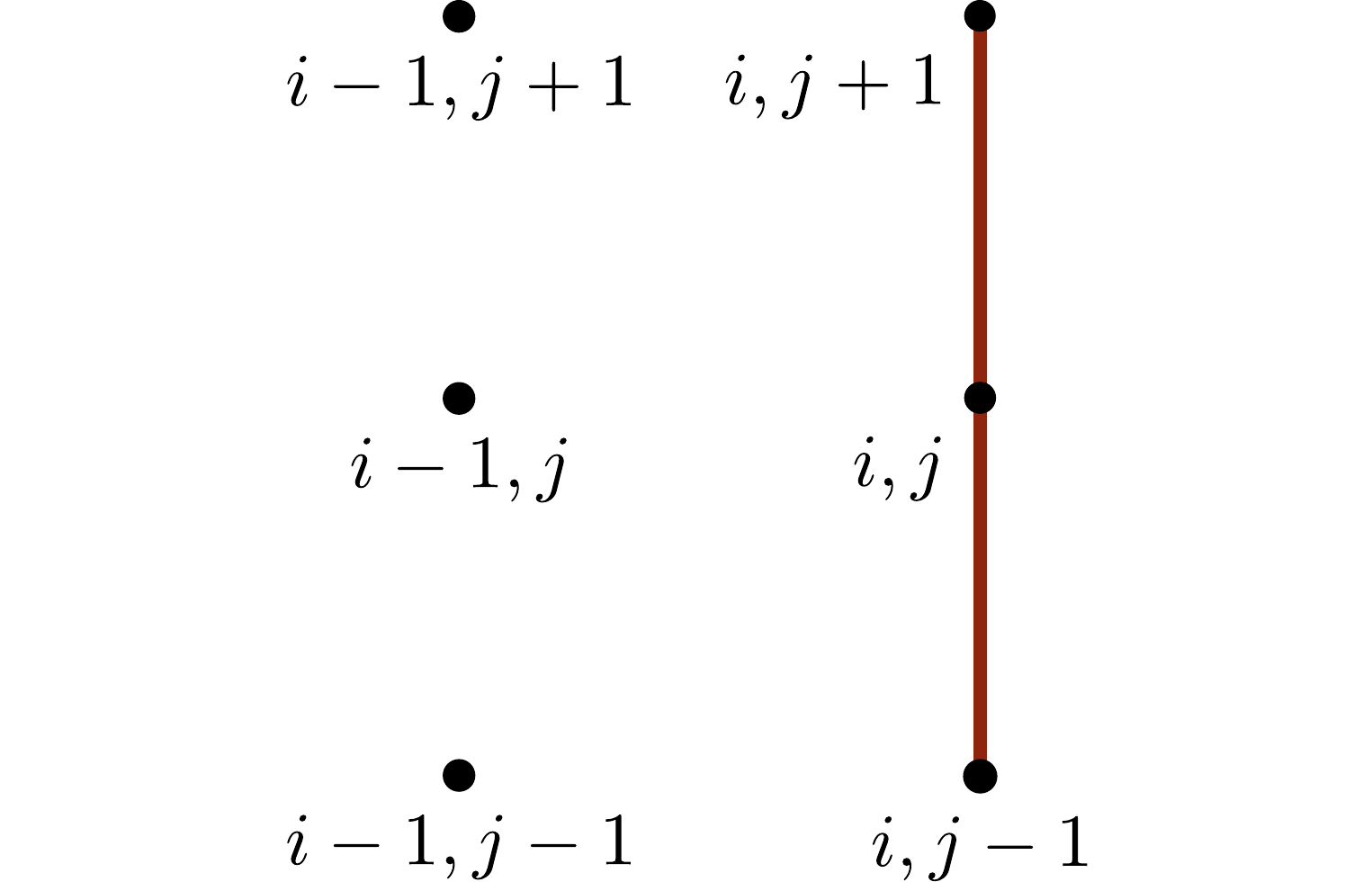}
		\end{minipage}
	\end{figure}
	\vfill
	\vspace{-\intextsep}\noindent\dotfill
	\vfill
	\begin{figure}[H]\ContinuedFloat
		\begin{minipage}[c]{0.65\textwidth}
			\vspace{0pt}
			{\bfseries (e) Northern edge ($i=2,\hdots,N_{x}-1$ and $j=N_{y}$)}:
			\begin{fleqn}[10pt]
				\begin{alignat*}{3}
					&p_{i,j}^{i-1,j} = \frac{\rho_{h}}{4}, & \quad &p_{i,j}^{i,j} = 1 - \rho_{r}, & \quad &p_{i,j}^{i+1,j} = \frac{\rho_{h}}{4}, \\ 
					&p_{i,j}^{i-1,j-1} = \frac{\rho_{d^{+}}}{4}, & &p_{i,j}^{i,j-1} = \frac{\rho_{v}}{2}, & &p_{i,j}^{i+1,j-1} = \frac{\rho_{d^{-}}}{4},
				\end{alignat*}
			\end{fleqn}
			where $\rho_{r} = 3(\rho_{xx} + \rho_{yy})/2$.
		\end{minipage}
		\begin{minipage}[c]{0.325\textwidth}
			\centering
			\vspace{0pt}
			\includegraphics[width=\textwidth]{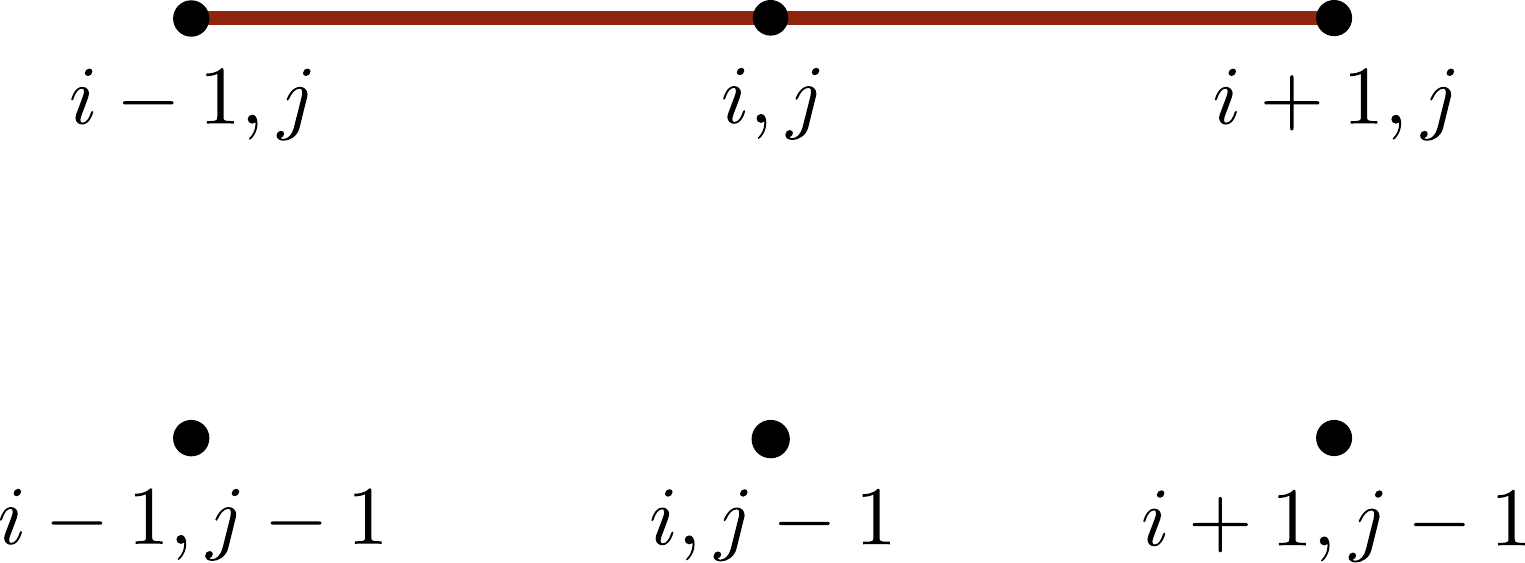}
		\end{minipage}
	\end{figure}
	\vfill
	\vspace{-\intextsep}\noindent\dotfill
	\vfill
	\begin{figure}[H]\ContinuedFloat
		\begin{minipage}[c]{0.65\textwidth}
			\vspace{0pt}
			{\bfseries (f) Southwest corner ($i=1$ and $j=1$)}:
			\begin{fleqn}[10pt]
				\begin{alignat*}{4}
					&p_{i,j}^{i,j+1} = \frac{\rho_{v}}{2}, &\quad &p_{i,j}^{i+1,j+1} = \frac{\rho_{d^{+}}}{2}, &\quad & p_{i,j}^{i,j} = 1 - \rho_{r}, & \quad &p_{i,j}^{i+1,j} = \frac{\rho_{h}}{2},
				\end{alignat*}
			\end{fleqn}
			where $\rho_{r} = 3(\rho_{xx} + \rho_{yy})/2 + 2\rho_{xy}$.
		\end{minipage}
		\begin{minipage}[c]{0.325\textwidth}
			\centering
			\vspace{0pt}
			\includegraphics[width=\textwidth]{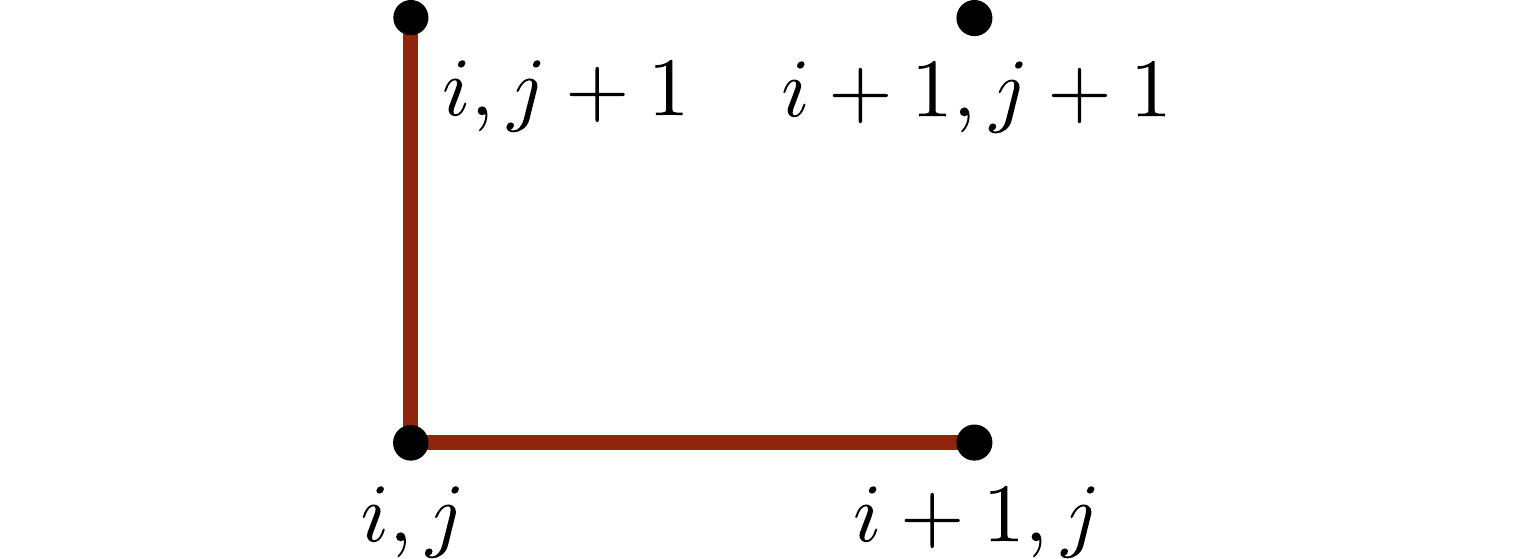}
		\end{minipage}
	\end{figure}
	\vfill
	\vspace{-\intextsep}\noindent\dotfill
	\vfill
	\begin{figure}[H]\ContinuedFloat
		\begin{minipage}[c]{0.65\textwidth}
			\vspace{0pt}
			{\bfseries (g) Southeast corner ($i=N_{x}$ and $j=1$)}:
			\begin{fleqn}[10pt]
				\begin{alignat*}{4}
					&p_{i,j}^{i-1,j+1} = \frac{\rho_{d^{-}}}{2}, &\quad &p_{i,j}^{i,j+1} = \frac{\rho_{v}}{2}, & \quad &p_{i,j}^{i-1,j} = \frac{\rho_{h}}{2}, & \quad &p_{i,j}^{i,j} = 1 - \rho_{r},
				\end{alignat*}
			\end{fleqn}
			where $\rho_{r} = 3(\rho_{xx} + \rho_{yy})/2 - 2\rho_{xy}$.
		\end{minipage}
		\begin{minipage}[c]{0.325\textwidth}
			\centering
			\vspace{0pt}
			\includegraphics[width=\textwidth]{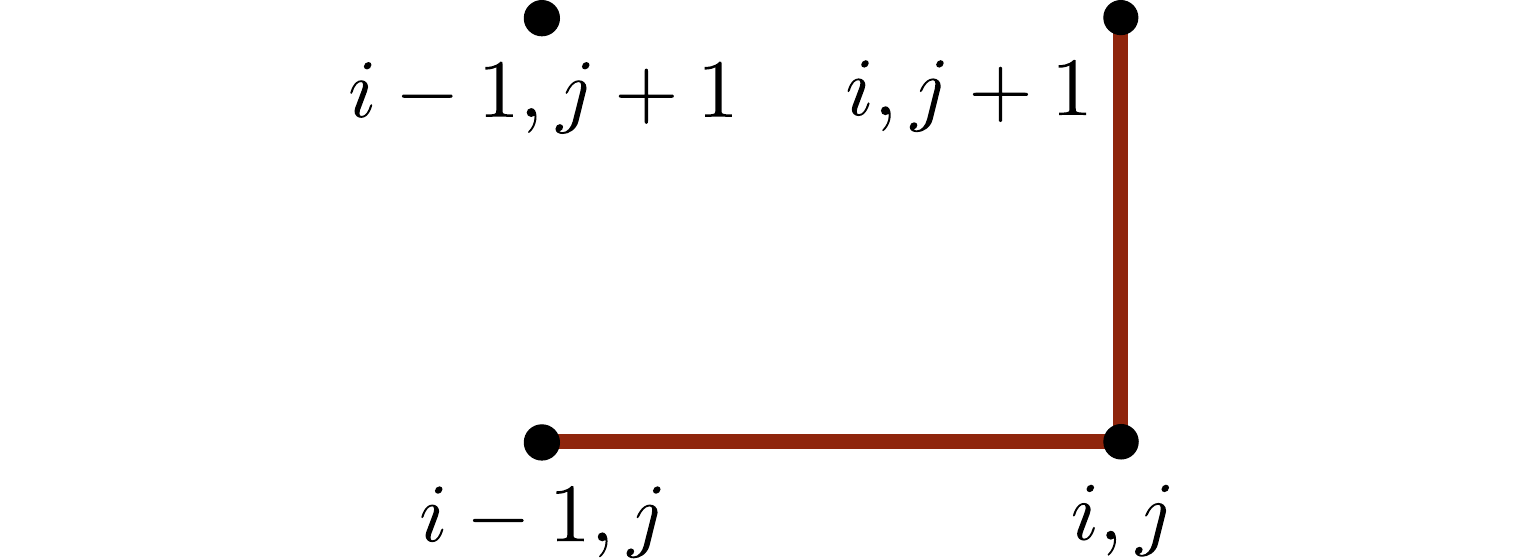}
		\end{minipage}
	\end{figure}
	\vfill
	\vspace{-\intextsep}\noindent\dotfill
	\vfill
	\begin{figure}[H]\ContinuedFloat
		\begin{minipage}[c]{0.65\textwidth}
			\vspace{0pt}
			{\bfseries (h) Northwest corner ($i=1$ and $j=N_{y}$)}:
			\begin{fleqn}[10pt]
				\begin{alignat*}{4}
					&p_{i,j}^{i,j} = 1 - \rho_{r}, &\quad &p_{i,j}^{i+1,j} = \frac{\rho_{h}}{2}, & \quad &p_{i,j}^{i,j-1} = \frac{\rho_{v}}{2}, & \quad &p_{i,j}^{i+1,j-1} = \frac{\rho_{d^{-}}}{2},
				\end{alignat*}
			\end{fleqn}
			where $\rho_{r} = 3(\rho_{xx} + \rho_{yy})/2 - 2\rho_{xy}$.
		\end{minipage}
		\begin{minipage}[c]{0.325\textwidth}
			\centering
			\vspace{0pt}
			\includegraphics[width=\textwidth]{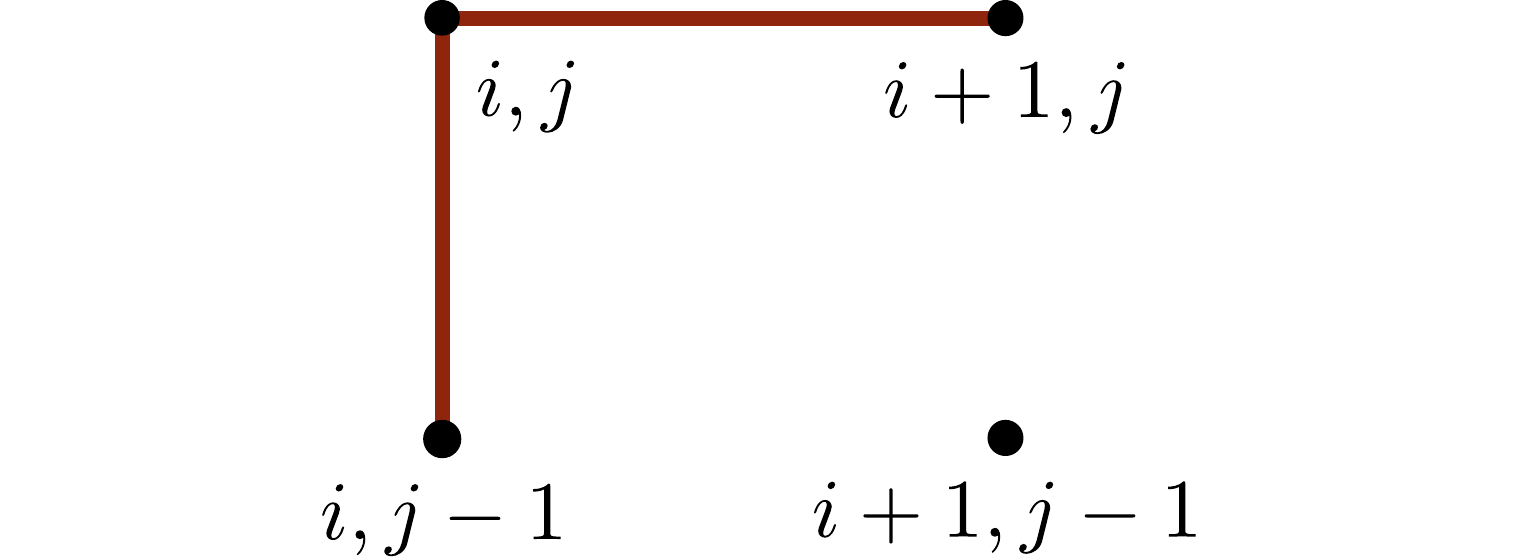}
		\end{minipage}
	\end{figure}    
	\vfill
	\vspace{-\intextsep}\noindent\dotfill
	\vfill
	\begin{figure}[H]\ContinuedFloat
		\begin{minipage}[c]{0.65\textwidth}
			\vspace{0pt}
			{\bfseries (i) Northeast corner ($i=N_{x}$ and $j=N_{y}$)}:
			\begin{fleqn}[10pt]
				\begin{alignat*}{4}
					&p_{i,j}^{i-1,j} = \frac{\rho_{h}}{2}, &\quad &p_{i,j}^{i,j} = 1 - \rho_{r}, & \quad &p_{i,j}^{i-1,j-1} = \frac{\rho_{d^{+}}}{2}, & \quad &p_{i,j}^{i,j-1} = \frac{\rho_{v}}{2},
				\end{alignat*}
			\end{fleqn}
			where $\rho_{r} = 3(\rho_{xx} + \rho_{yy})/2 + 2\rho_{xy}$.
		\end{minipage}
		\begin{minipage}[c]{0.325\textwidth}
			\centering
			\vspace{0pt}
			\includegraphics[width=\textwidth]{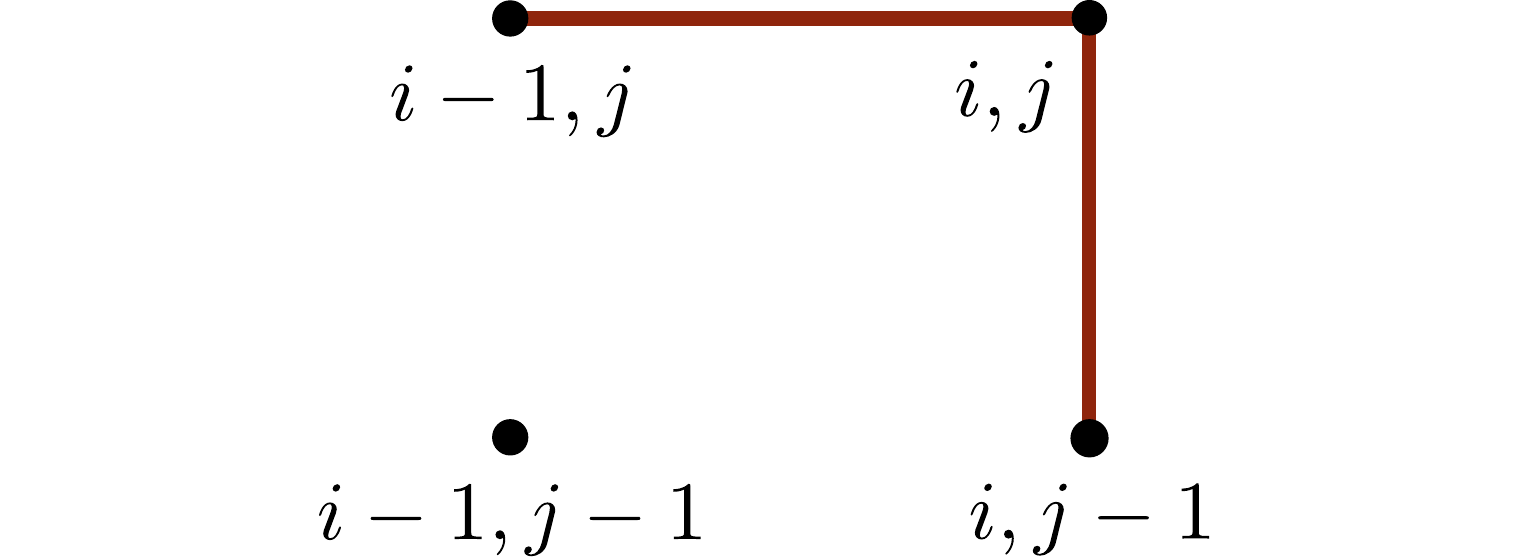}
		\end{minipage}
		\caption{\textbf{Transition probabilities governing particle movement on a rectangular lattice.} Transition probabilities defined by the stochastic matrix $\mathbf{P}=\mathbf{I}+\tau\mathbf{C}$ governing particle movement between (a) interior, (b)--(e) edge and (f)--(i) corner lattice sites in a rectangular configuration. Here, we let $p_{i,j}^{k,m}$ denote the probability that a particle located at $\mathbf{x}_{i,j} = (x_{i},y_{j})$ at time $t=t_{n-1}$ moves to $\mathbf{x}_{k,m} = (x_{k},y_{m})$ at time $t=t_{n}$. These probabilities are defined using expressions pertaining to horizontal ($\rho_{h}$), vertical ($\rho_{v}$) and diagonal ($\rho_{d^{+}}$ and $\rho_{d^{-}}$) movement (see expressions (\ref{eq:rec_horiz_vert}), (\ref{eq:rec_diag_plus}) and (\ref{eq:rec_diag_minus})). Additionally, the probability of a particle remaining in its current position (at $\mathbf{x}_{i,j}$) is defined as $p_{i,j}^{i,j}=1-\rho_{r}$, where $\rho_{r}$ is given by a linear combination of $\rho_{xx}$, $\rho_{yy}$ and/or $\rho_{xy}$ (see expressions in (\ref{eq:remain_expr})). For each case of lattice sites, we provide a schematic of the local lattice (central and neighbouring sites labelled with position indices).}
		\label{fig:rec_probs}
	\end{figure}}

	Studying the probabilities in Figure \ref{fig:rec_probs}, we see that $\mathbf{P}$ is a stochastic matrix when
	\begin{gather}\label{eq:rec_time_step}
		\tau \leq \frac{2\delta_{x}^2\delta_{y}^2}{3\delta_{y}^2D_{xx} + 4\delta_{x}\delta_{y}|D_{xy}| + 3\delta_{x}^2D_{yy}},
	\end{gather}
	and
	\begin{gather}
		3\delta_{y}^2D_{xx} \geq \delta_{x}^2D_{yy}, \label{eq:rec_cond1} \\
		3\delta_{x}^2D_{yy} \geq \delta_{y}^2D_{xx}, \label{eq:rec_cond2} \\
		\delta_{y}^2D_{xx} + \delta_{x}^2D_{yy} \geq 4\delta_{x}\delta_{y}|D_{xy}|, \label{eq:rec_cond3}
	\end{gather}
	where the constraints (\ref{eq:rec_cond1})--(\ref{eq:rec_cond3}) are required to ensure all transition probabilities are non-negative, and (\ref{eq:rec_time_step}) enforces $\rho_{r} \leq 1$ for each case (all probabilities are between zero and one). Inspecting the expressions in (\ref{eq:rec_horiz_vert}), we observe that particle movement occurs more frequently in the horizontal ($x$) or vertical ($y$) direction as the diffusivity $D_{xx}$ or $D_{yy}$ increases, respectively, assuming fixed values for other spatial parameters. Moreover, expressions (\ref{eq:rec_diag_plus}) and (\ref{eq:rec_diag_minus}) suggest that particles are more likely to move to the northeast or southwest (to $\mathbf{x}_{i+1,j+1}$ or $\mathbf{x}_{i-1,j-1}$) as opposed to the northwest or southeast (to $\mathbf{x}_{i-1,j+1}$ or $\mathbf{x}_{i+1,j-1}$) if $D_{xy} > 0$ and vice versa if $D_{xy} < 0$. Additionally, the frequency of this diagonal movement increases with larger absolute values of $D_{xy}$. These observations agree with the physical interpretation of the diffusion tensor $\mathbf{D}$.
	
	We now analyse and simplify the non-negativity constraints (\ref{eq:rec_cond1})--(\ref{eq:rec_cond3}). Firstly, the constraints (\ref{eq:rec_cond1}) and (\ref{eq:rec_cond2}) can be combined to give a condition on the spatial step $\delta_{y}$,
	\begin{gather}\label{eq:rec_interval_dy}
		\delta_{x}\sqrt{\frac{D_{yy}}{3D_{xx}}} \leq \delta_{y} \leq \delta_{x}\sqrt{\frac{3D_{yy}}{D_{xx}}},
	\end{gather}
	or, alternatively, $\delta_{x}$,
	\begin{gather}\label{eq:rec_interval_dx}
		\delta_{y}\sqrt{\frac{D_{xx}}{3D_{yy}}} \leq \delta_{x} \leq \delta_{y}\sqrt{\frac{3D_{xx}}{D_{yy}}},
	\end{gather}
	where both conditions are satisfied for any valid diffusion tensor as the lower bound is always less then the upper bound. Here, we use the constraint (\ref{eq:rec_interval_dy}) when $D_{xx} < D_{yy}$ and (\ref{eq:rec_interval_dx}) when $D_{xx} > D_{yy}$ to avoid being limited to a very small spatial step $\delta_{y}$ or $\delta_{x}$ when $D_{yy}/D_{xx}$ or $D_{xx}/D_{yy}$ is small, respectively. Moreover, the constraint (\ref{eq:rec_cond3}) can be considered as a quadratic in $\delta_{y}$ where the coefficient for $\delta_{y}^2$ is strictly positive ($D_{xx} > 0$). Thus, this constraint can be simplified by ensuring that the minimum of the quadratic is non-negative, which gives                                                                                                     
	\begin{gather}\label{eq:rec_det_cond}
		\det(\mathbf{D}) \geq 3D_{xy}^2.
	\end{gather}
	The constraint (\ref{eq:rec_det_cond}) on the diffusion tensor $\mathbf{D}$ suggests that a stochastic matrix, assuming model implementation on a rectangular lattice, can be obtained for isotropic or orthotropic particle diffusion ($D_{xy} = 0$) without conditions on the diffusion coefficients. For anisotropic media, this approach is, however, only suitable for a restricted range of diffusion tensors.
	
	\subsection{Flat-top hexagonal lattice}\label{sec:flat-top_probs}
	Secondly, we consider the transition matrix $\mathbf{P} = \mathbf{I} + \tau\mathbf{C}$ whose entries correspond to the probabilities of particle movement on a flat-top hexagonal lattice (see Figure \ref{fig:lattice_examples}). These entries are presented in Figure \ref{fig:flat-top_probs} for interior, edge and corner lattice sites. In each case, the probabilities of particle movement occurring along the coordinate axes in a horizontal (to $\mathbf{x}_{i-2,j}$ or $\mathbf{x}_{i+2,j}$) direction are defined in terms of the following expression:
	\begin{gather}\label{eq:flat-top_horiz}
		\rho_{h} = \frac{\tau(\delta_{y}^2D_{xx} - \delta_{x}^2D_{yy})}{\delta_{x}^2\delta_{y}^2}.
	\end{gather}
	Additionally, the expressions
	\begin{gather}\label{eq:flat-top_diag_plus}
		\rho_{d^{+}} = \frac{\tau(\delta_{x}D_{yy} + \delta_{y}D_{xy})}{\delta_{x}\delta_{y}^2},
	\end{gather}
	and 
	\begin{gather}\label{eq:flat-top_diag_minus}
		\rho_{d^{-}} = \frac{\tau(\delta_{x}D_{yy} - \delta_{y}D_{xy})}{\delta_{x}\delta_{y}^2},
	\end{gather}
	are used to define the probabilities of particle movement to the northeast or southwest sites (to $\mathbf{x}_{i+1,j+1}$ or $\mathbf{x}_{i-1,j-1}$) or to the northwest or southeast sites (to $\mathbf{x}_{i-1,j+1}$ or $\mathbf{x}_{i+1,j-1}$), respectively. Finally, we utilise a linear combination of the expressions in (\ref{eq:remain_expr}) to define the probability that a particle remains in its current position. For each case of lattice sites, we provide a schematic of the local lattice (central and neighbouring sites labelled with position indices) to ease visual association between each lattice site and the corresponding probability of a particle moving to, or remaining at, that location (see Figure \ref{fig:flat-top_probs}). Note that the corresponding row and column positions in the matrix $\mathbf{P}$ for the probability $p_{i,j}^{k,m}$ are given by $m_{i,j}$ and $m_{k,m}$, respectively, where $m_{i,j} = (j-1)N_{x}/2 + \lceil i/2 \rceil$ is the mapping function from (\ref{eq:map}). Additionally, we remind the reader that lattice sites in a hexagonal configuration are only defined if the sum of the position indices is even (i.e $\mathbf{x}_{i,j} = (x_{i},y_{j})$ is only defined if $i + j$ is even).
	\afterpage{
	\begin{figure}[H]
		\begin{minipage}[c]{0.65\textwidth}
			\vspace{0pt}
			{\bfseries (a) Interior ($i=3,\hdots,N_{x}-2$ and $j=2,\hdots,N_{y}-1$)}:
			\begin{fleqn}[10pt]
				\begin{alignat*}{3}
					&p_{i,j}^{i-1,j+1} = 2\rho_{d^{-}}, &\quad &p_{i,j}^{i+1,j+1} = 2\rho_{d^{+}}, &\quad & \\
					&p_{i,j}^{i-2,j} = \rho_{h}, & &p_{i,j}^{i,j} = 1 - \rho_{r}, & &p_{i,j}^{i+2,j} = \rho_{h}, \\
					&p_{i,j}^{i-1,j-1} = 2\rho_{d^{+}}, & &p_{i,j}^{i+1,j-1} = 2\rho_{d^{-}}, & &
				\end{alignat*}
			\end{fleqn}
			where $\rho_{r} = 2(\rho_{xx} + 3\rho_{yy})$.
		\end{minipage}
		\begin{minipage}[c]{0.325\textwidth}
			\centering
			\vspace{0pt}
			\includegraphics[width=\textwidth]{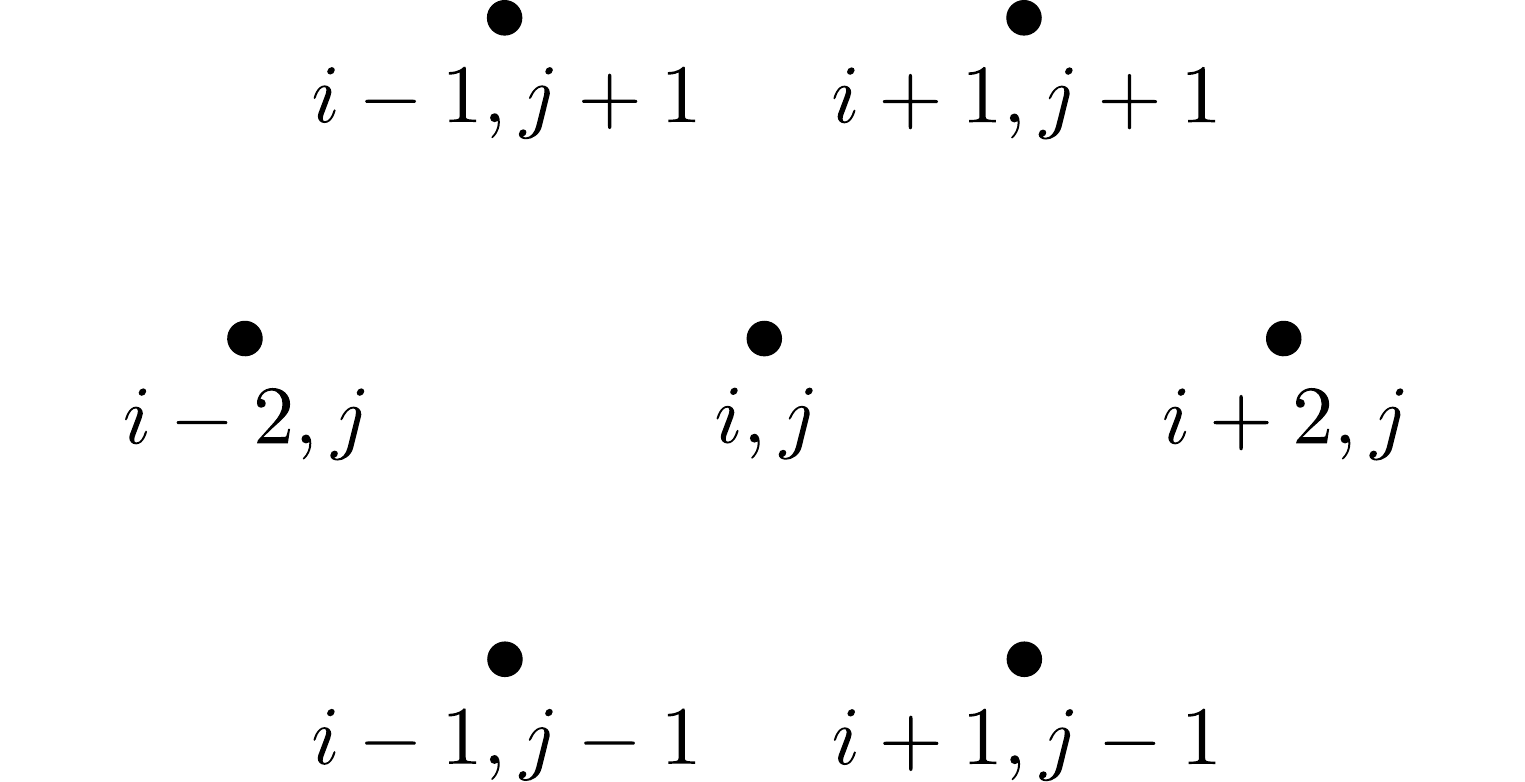}
		\end{minipage}
		\phantomcaption
	\end{figure}
	\vfill
	\vspace{-\intextsep}\noindent\dotfill
	\vfill
	\begin{figure}[H]\ContinuedFloat
		\begin{minipage}[c]{0.65\textwidth}
			\vspace{0pt}
			{\bfseries (b) Western concave edge ($i=1$ and $j = 2,\hdots,N_{y}-1$)}:
			\begin{fleqn}[10pt]
				\begin{alignat*}{2}
					&p_{i,j}^{i+1,j+1} = 3\rho_{d^{+}}, &\quad &p_{i,j}^{i,j} = 1 - \rho_{r}, \\
					&p_{i,j}^{i+2,j} = 3\rho_{h}, & &p_{i,j}^{i+1,j-1} = 3\rho_{d^{-}},
				\end{alignat*}
			\end{fleqn}
			where $\rho_{r} = 3(\rho_{xx} + \rho_{yy})$. 
		\end{minipage}
		\begin{minipage}[c]{0.325\textwidth}
			\centering
			\vspace{0pt}
			\includegraphics[width=\textwidth]{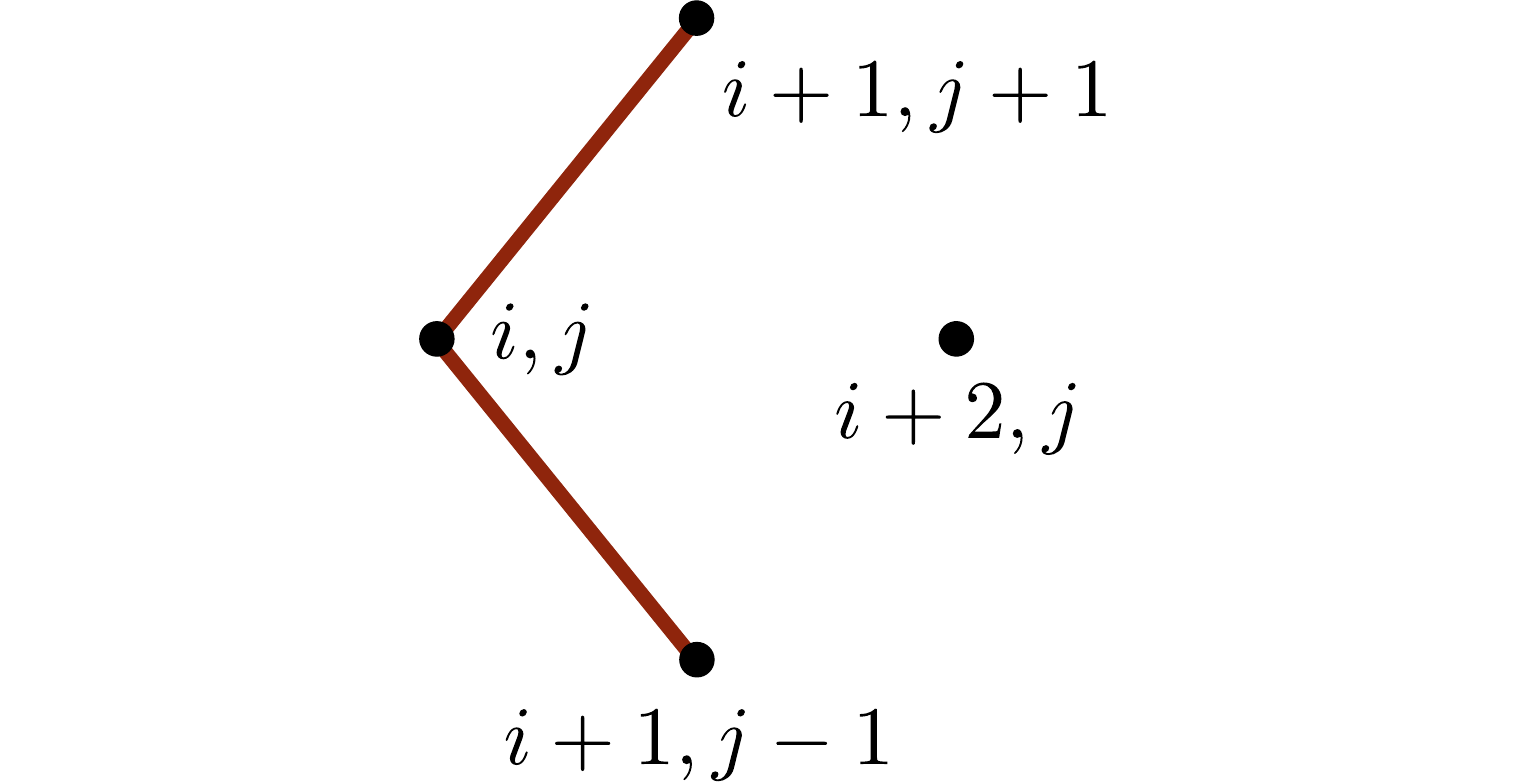}
		\end{minipage}
	\end{figure}
	\vfill
	\vspace{-\intextsep}\noindent\dotfill
	\vfill
	\begin{figure}[H]\ContinuedFloat
		\begin{minipage}[c]{0.65\textwidth}
			\vspace{0pt}
			{\bfseries (c) Western convex edge ($i=2$ and $j=2,\hdots,N_{y}-1$):}
			\begin{fleqn}[10pt]
				\begin{alignat*}{3}
					&p_{i,j}^{i-1,j+1} = \frac{3\rho_{d^{-}}}{2}, &\quad &p_{i,j}^{i+1,j+1} = 3\rho_{d^{+}}, &\quad &p_{i,j}^{i,j} = 1 - \rho_{r}, \\
					&p_{i,j}^{i+2,j} = \frac{3\rho_{h}}{2}, & &p_{i,j}^{i-1,j-1} = \frac{3\rho_{d^{+}}}{2}, & &p_{i,j}^{i+1,j-1} = 3\rho_{d^{-}},
				\end{alignat*}
			\end{fleqn}
			where $\rho_{r} = 3(\rho_{xx} + 5\rho_{yy})/2$.
		\end{minipage}
		\begin{minipage}[c]{0.325\textwidth}
			\centering
			\vspace{0pt}
			\includegraphics[width=\textwidth]{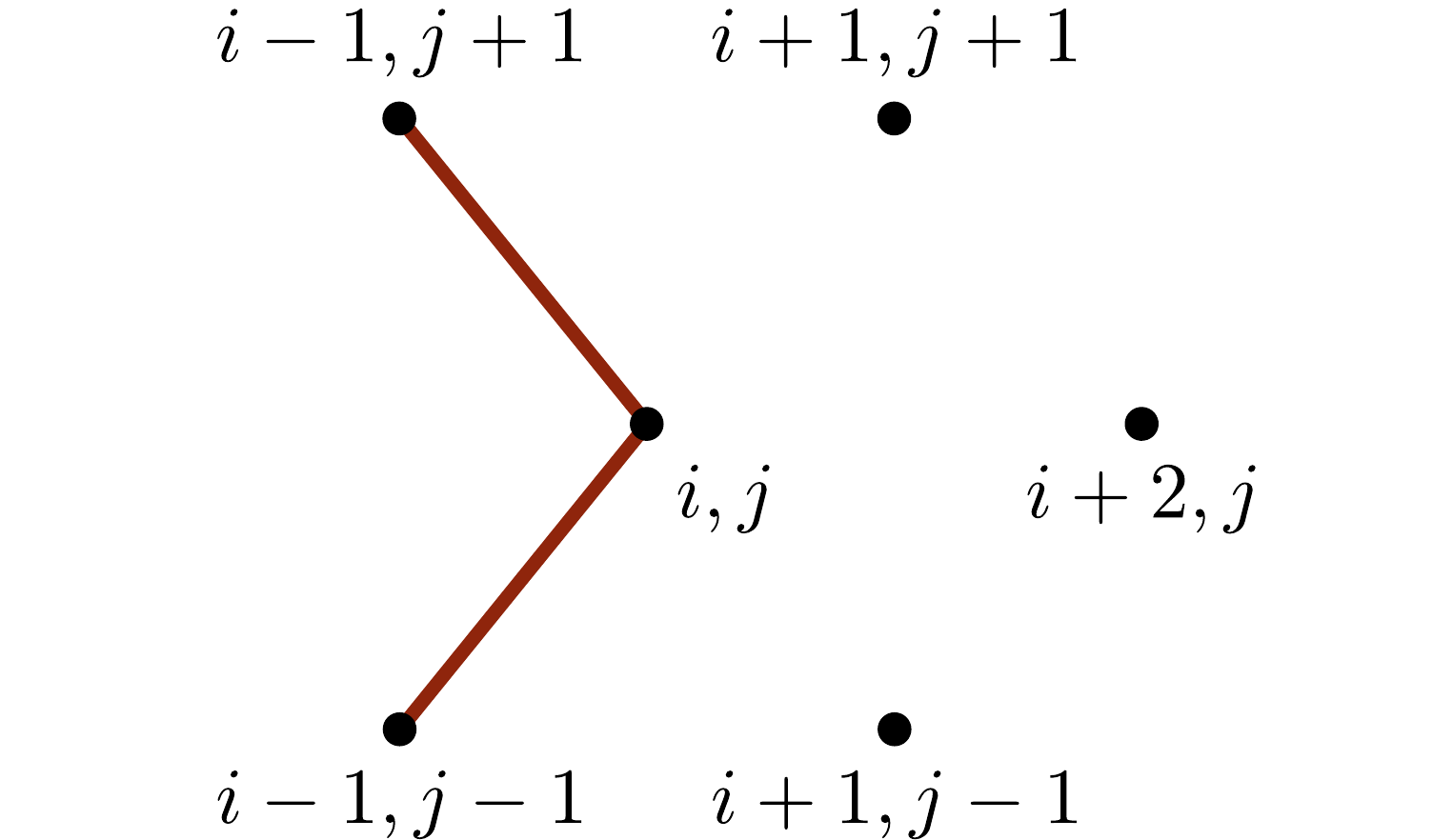}
		\end{minipage}
	\end{figure}
	\vfill
	\vspace{-\intextsep}\noindent\dotfill
	\vfill
	\begin{figure}[H]\ContinuedFloat
		\begin{minipage}[c]{0.65\textwidth}
			\vspace{0pt}
			{\bfseries (d) Southern edge ($i=3,\hdots,N_{x}-2$ and $j=1$)}:
			\begin{fleqn}[10pt]
				\begin{alignat*}{3}
					&p_{i,j}^{i-1,j+1} = 4\rho_{d^{-}}, &\quad &p_{i,j}^{i+1,j+1} = 4\rho_{d^{+}}, &\quad & \\
					&p_{i,j}^{i-2,j} = \rho_{h}, & &p_{i,j}^{i,j} = 1 - \rho_{r}, & &p_{i,j}^{i+2,j} = \rho_{h},
				\end{alignat*}
			\end{fleqn}
			where $\rho_{r} = 2(\rho_{xx} + 3\rho_{yy})$.
		\end{minipage}
		\begin{minipage}[c]{0.325\textwidth}
			\centering
			\vspace{0pt}
			\includegraphics[width=\textwidth]{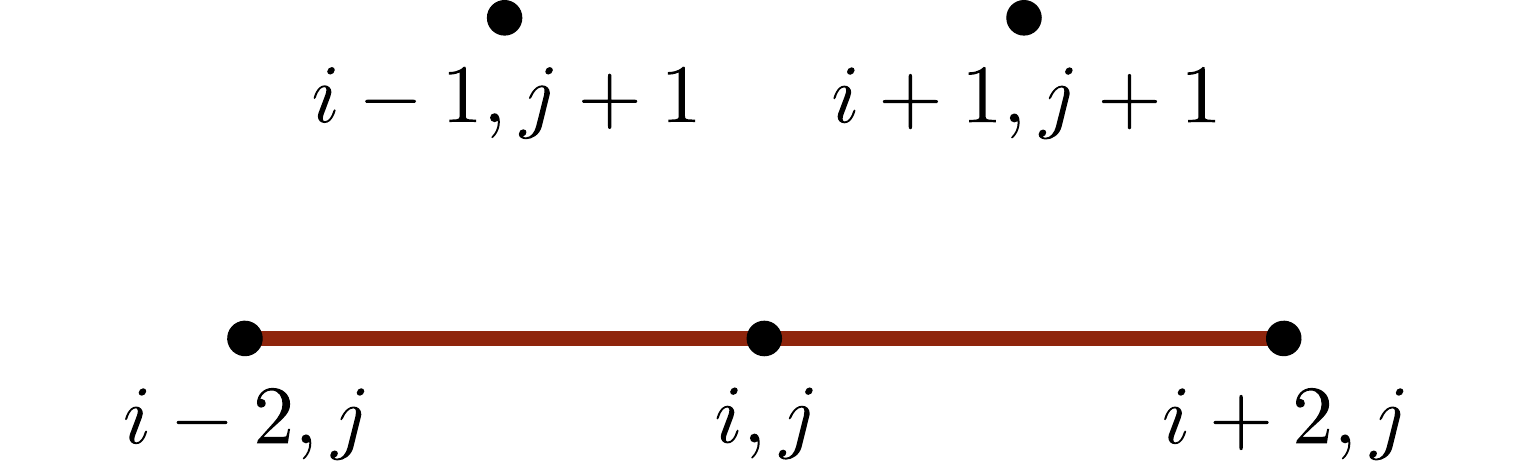}
		\end{minipage}
	\end{figure}
	\vfill
	\vspace{-\intextsep}\noindent\dotfill
	\vfill
	\begin{figure}[H]\ContinuedFloat
		\begin{minipage}[c]{0.65\textwidth}
			\vspace{0pt}
			{\bfseries (e) Eastern convex edge ($i=N_{x}-1$ and $j=2,\hdots,N_{y}-1$)}:
			\begin{fleqn}[10pt]
				\begin{alignat*}{3}
					&p_{i,j}^{i-1,j+1} = 3\rho_{d^{-}}, &\qquad &p_{i,j}^{i+1,j+1} = \frac{3\rho_{d^{+}}}{2}, &\qquad &p_{i,j}^{i-2,j} = \frac{3\rho_{h}}{2}, \\
					&p_{i,j}^{i,j} = 1 - \rho_{r}, & &p_{i,j}^{i-1,j-1} = 3\rho_{d^{+}}, & &p_{i,j}^{i+1,j-1} = \frac{3\rho_{d^{-}}}{2},
				\end{alignat*}
			\end{fleqn}
			where $\rho_{r} = 3(\rho_{xx} + 5\rho_{yy})/2$.
		\end{minipage}
		\begin{minipage}[c]{0.325\textwidth}
			\centering
			\vspace{0pt}
			\includegraphics[width=\textwidth]{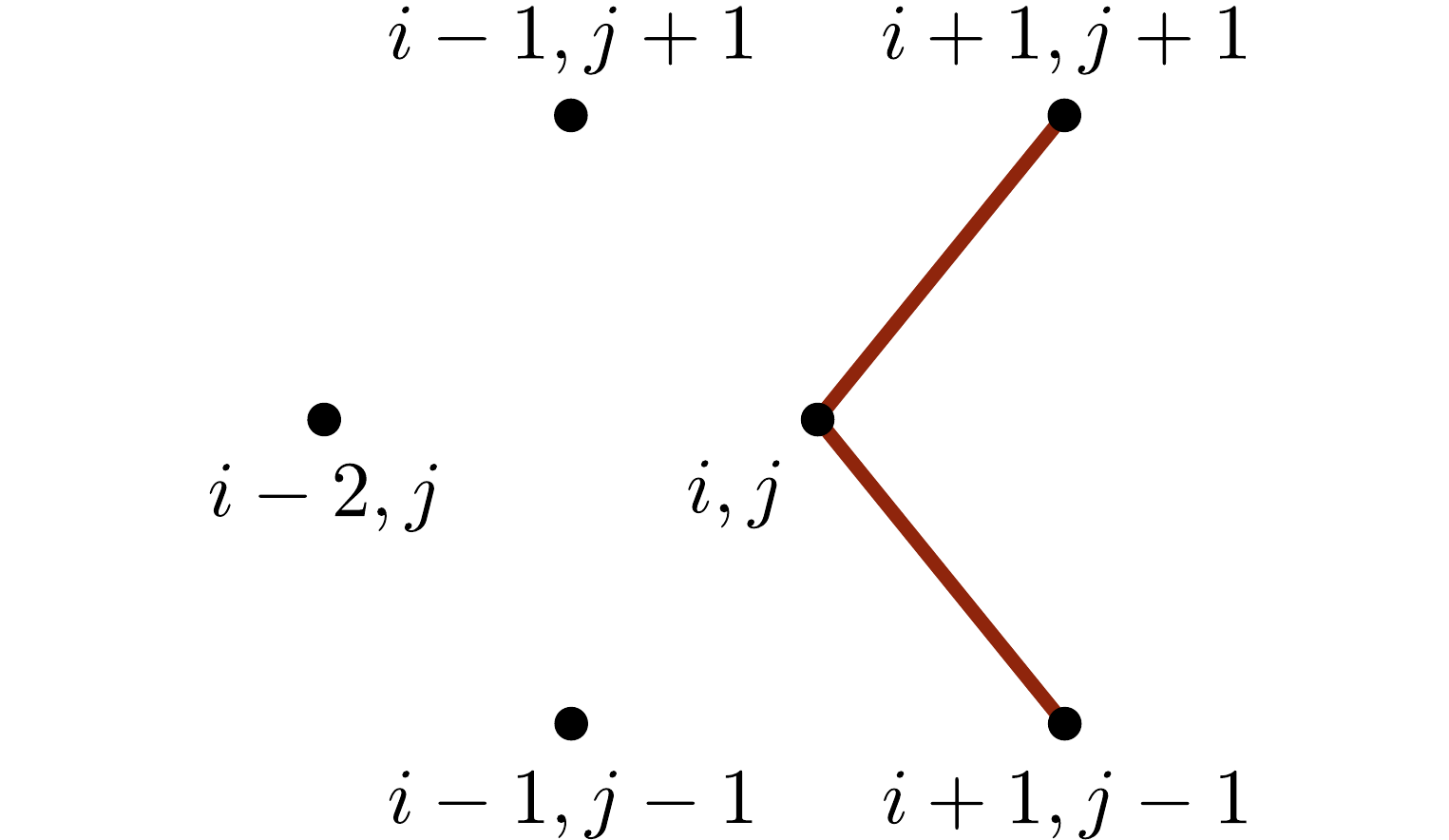}
		\end{minipage}
	\end{figure}
	\vfill
	\vspace{-\intextsep}\noindent\dotfill
	\vfill
	\begin{figure}[H]\ContinuedFloat
		\begin{minipage}[c]{0.65\textwidth}
			\vspace{0pt}
			{\bfseries (f) Eastern concave edge ($i=N_{x}$ and $j=2,\hdots,N_{y}-1$)}:
			\begin{fleqn}[10pt]
				\begin{alignat*}{2}
					&p_{i,j}^{i-1,j+1} = 3\rho_{d^{-}}, &\qquad &p_{i,j}^{i-2,j} = 3\rho_{h}, \\
					&p_{i,j}^{i,j} = 1 - \rho_{r}, & &p_{i,j}^{i-1,j-1} = 3\rho_{d^{+}},
				\end{alignat*}
			\end{fleqn}
			where $\rho_{r} = 3(\rho_{xx} + \rho_{yy})$.
		\end{minipage}
		\begin{minipage}[c]{0.325\textwidth}
			\centering
			\vspace{0pt}
			\includegraphics[width=\textwidth]{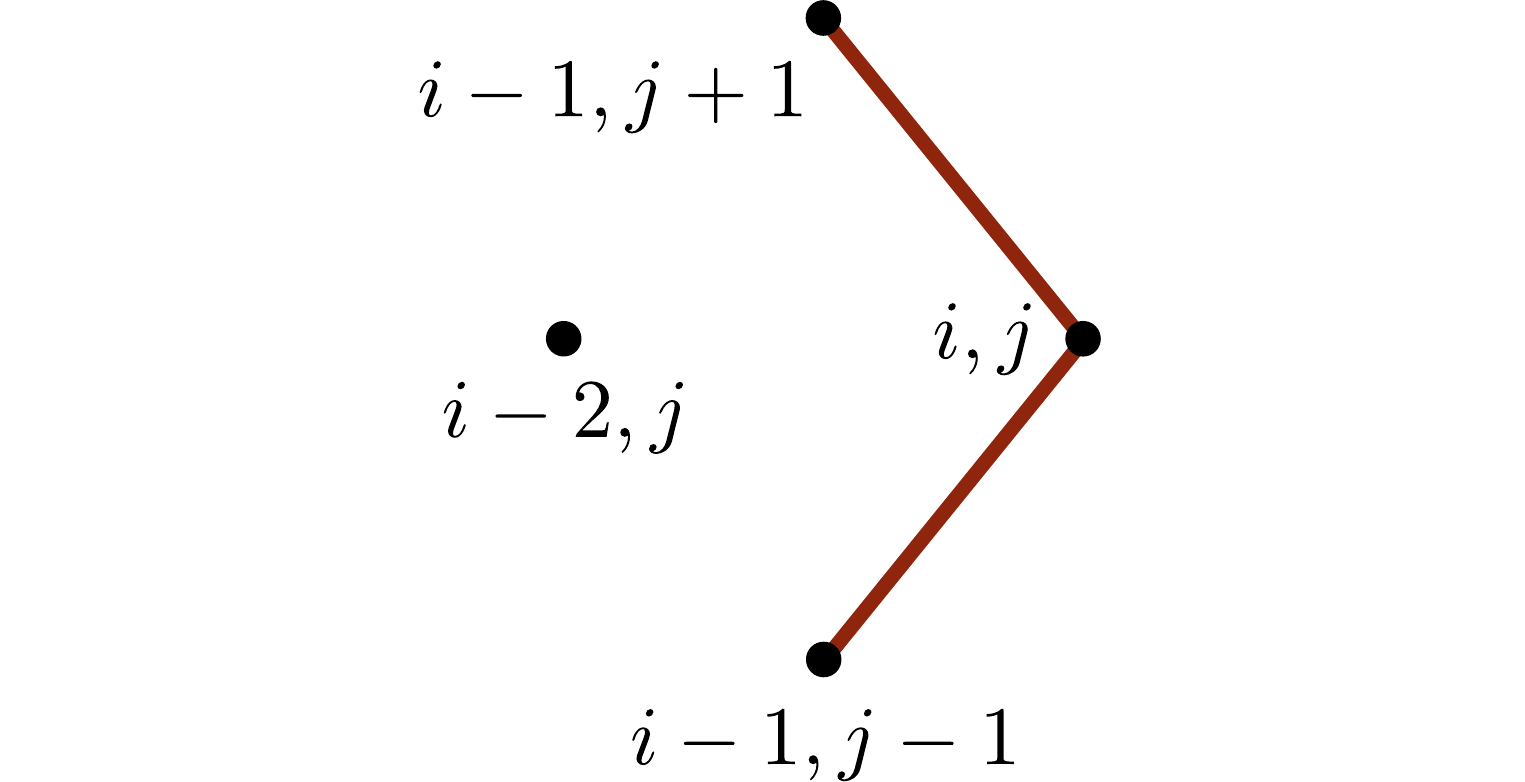}
		\end{minipage}
	\end{figure}
	\vfill
	\vspace{-\intextsep}\noindent\dotfill
	\vfill
	\begin{figure}[H]\ContinuedFloat
		\begin{minipage}[c]{0.65\textwidth}
			\vspace{0pt}
			{\bfseries (g) Northern edge ($i=3,\hdots,N_{x}-2$ and $j=N_{y}$)}:
			\begin{fleqn}[10pt]
				\begin{alignat*}{3}
					&p_{i,j}^{i-2,j} = \rho_{h}, &\qquad &p_{i,j}^{i,j} = 1 - \rho_{r}, &\qquad &p_{i,j}^{i+2,j} = \rho_{h}, \\
					&p_{i,j}^{i-1,j-1} = 4\rho_{d^{+}}, & &p_{i,j}^{i+1,j-1} = 4\rho_{d^{-}}, & &
				\end{alignat*}
			\end{fleqn}
			where $\rho_{r} = 2(\rho_{xx} + 3\rho_{yy})$.
		\end{minipage}
		\begin{minipage}[c]{0.325\textwidth}
			\centering
			\vspace{0pt}
			\includegraphics[width=\textwidth]{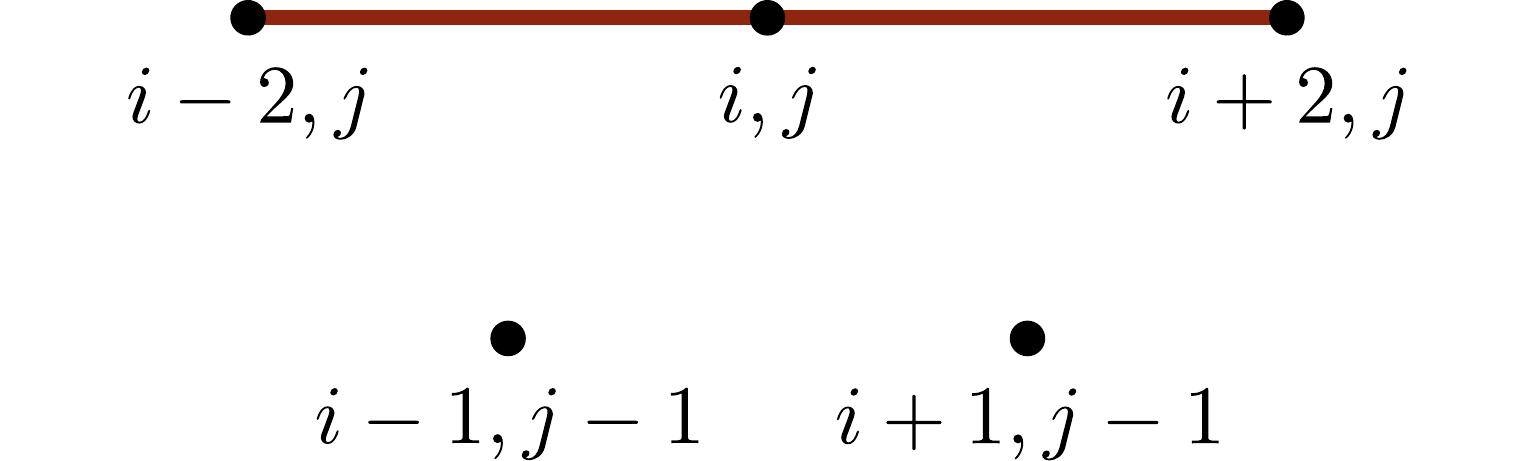}
		\end{minipage}
	\end{figure}
	\vfill
	\vspace{-\intextsep}\noindent\dotfill
	\vfill
	\begin{figure}[H]\ContinuedFloat
		\begin{minipage}[c]{0.65\textwidth}
			\vspace{0pt}
			{\bfseries (h) Southwest concave corner ($i=1$ and $j=1$)}:
			\begin{fleqn}[10pt]
				\begin{alignat*}{3}
					&p_{i,j}^{i+1,j+1} = 6\rho_{d^{+}}, &\quad &p_{i,j}^{i,j} = 1 - \rho_{r}, &\quad &p_{i,j}^{i+2,j} = 3\rho_{h},
				\end{alignat*}
			\end{fleqn}
			where $\rho_{r} = 3(\rho_{xx} + 2\rho_{xy} + \rho_{yy})$.
		\end{minipage}
		\begin{minipage}[c]{0.325\textwidth}
			\centering
			\vspace{0pt}
			\includegraphics[width=\textwidth]{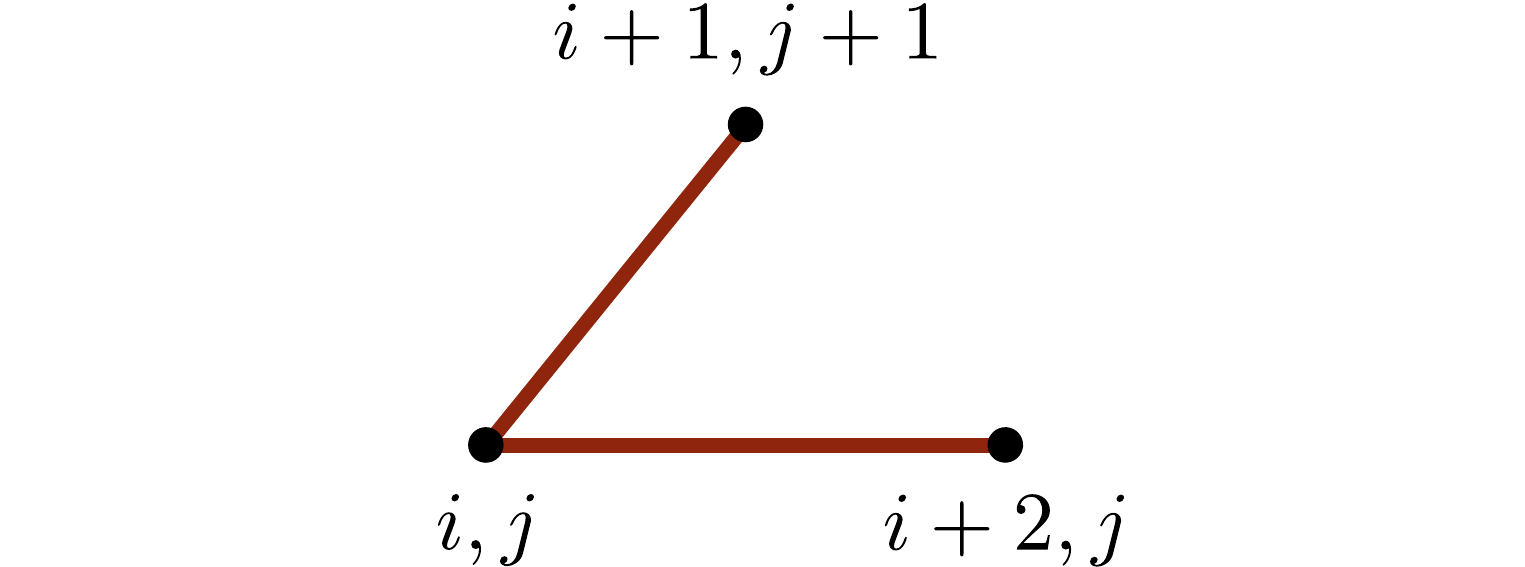}
		\end{minipage}
	\end{figure}
	\vfill
	\vspace{-\intextsep}\noindent\dotfill
	\vfill
	\begin{figure}[H]\ContinuedFloat
		\begin{minipage}[c]{0.65\textwidth}
			\vspace{0pt}
			{\bfseries (i) Southeast convex corner ($i=N_{x}-1$ and $j=1$)}:
			\begin{fleqn}[10pt]
				\begin{alignat*}{4}
					&p_{i,j}^{i-1,j+1} = 6\rho_{d^{-}}, & \quad &p_{i,j}^{i+1,j+1} = 3\rho_{d^{+}}, & \quad &p_{i,j}^{i-2,j} = \frac{3\rho_{h}}{2}, &\quad &p_{i,j}^{i,j} = 1 - \rho_{r},
				\end{alignat*}
			\end{fleqn}
			where $\rho_{r} = 3(\rho_{xx} + 5\rho_{yy})/2 - 3\rho_{xy}$.
		\end{minipage}
		\begin{minipage}[c]{0.325\textwidth}
			\centering
			\vspace{0pt}
			\includegraphics[width=\textwidth]{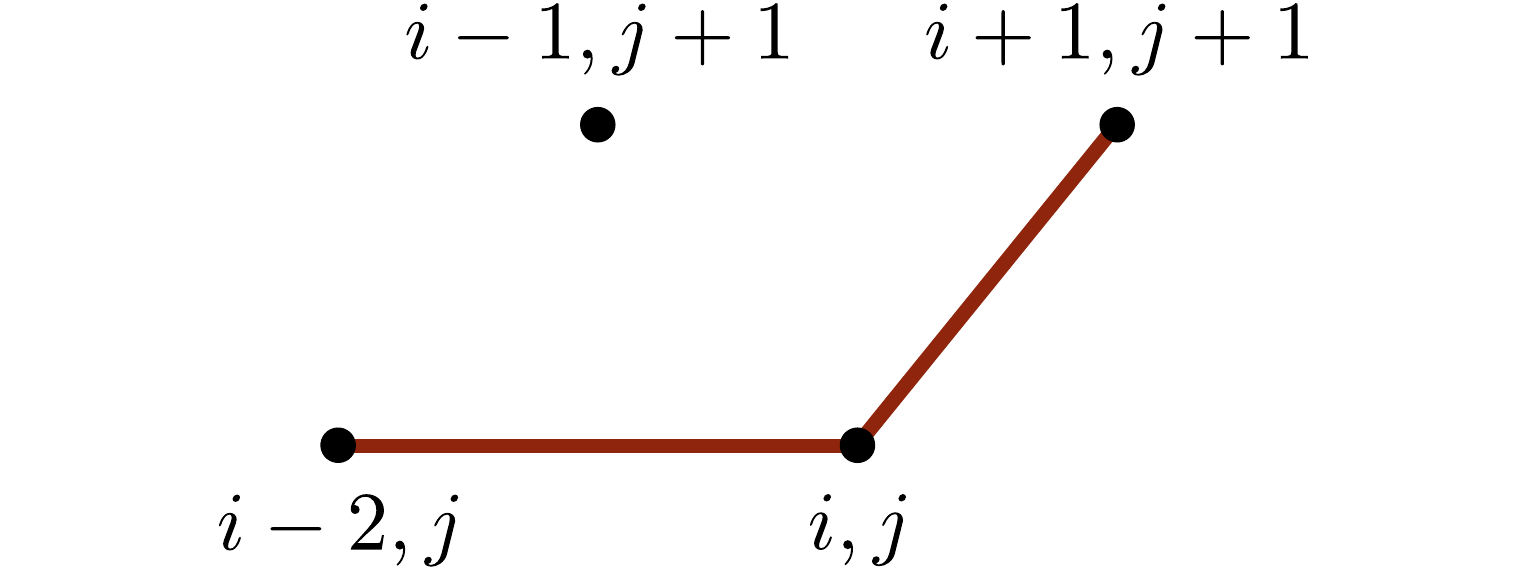}
		\end{minipage}
	\end{figure}
	\vfill
	\vspace{-\intextsep}\noindent\dotfill
	\vfill
	\begin{figure}[H]\ContinuedFloat
		\begin{minipage}[c]{0.65\textwidth}
			\vspace{0pt}
			{\bfseries (j) Northwest concave corner ($i=1$ and $j=N_{y}$)}:
			\begin{fleqn}[10pt]
				\begin{alignat*}{3}
					&p_{i,j}^{i,j} = 1 - \rho_{r}, & \quad &p_{i,j}^{i+2,j} = 3\rho_{h}, & \quad &p_{i,j}^{i+1,j-1} = 6\rho_{d^{-}},
				\end{alignat*}
			\end{fleqn}
			where $\rho_{r} = 3(\rho_{xx} - 2\rho_{xy} + \rho_{yy})$.
		\end{minipage}
		\begin{minipage}[c]{0.325\textwidth}
			\centering
			\vspace{0pt}
			\includegraphics[width=\textwidth]{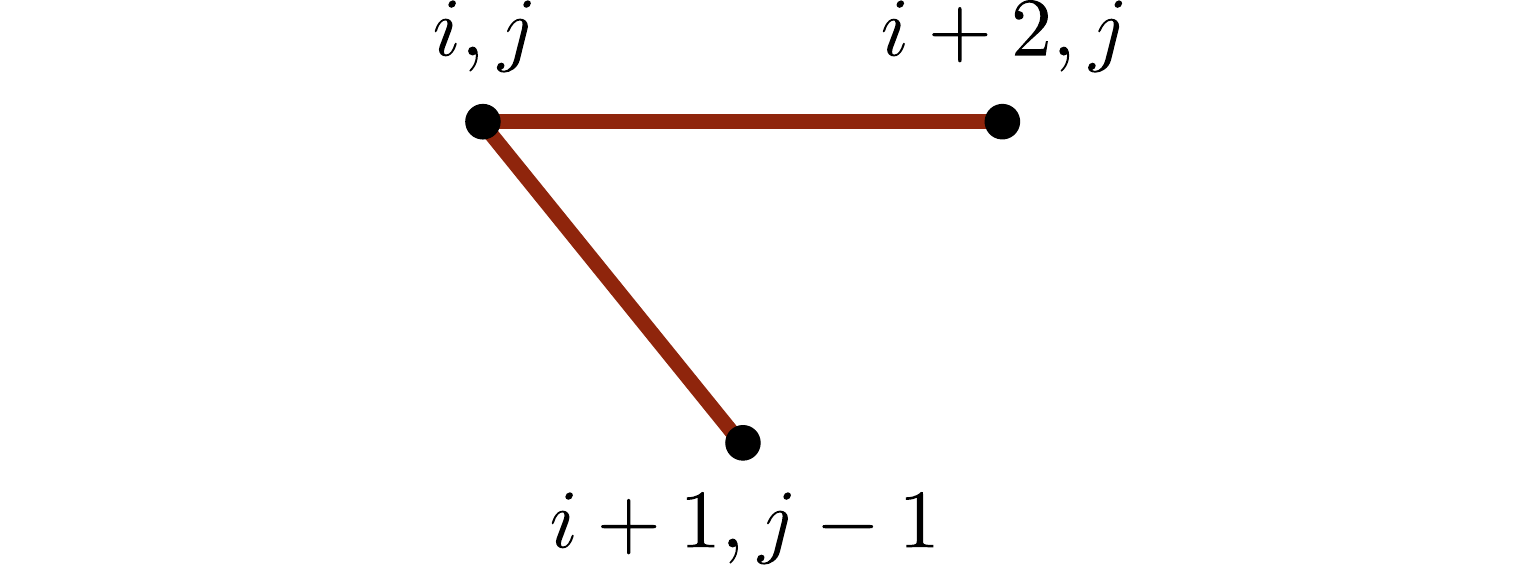}
		\end{minipage}
	\end{figure}
	\vfill
	\vspace{-\intextsep}\noindent\dotfill
	\vfill
	\begin{figure}[H]\ContinuedFloat
		\begin{minipage}[c]{0.65\textwidth}
			\vspace{0pt}
			{\bfseries (k) Northeast convex corner ($i = N_{x}-1$ and $j = N_{y}$)}
			\begin{fleqn}[10pt]
				\begin{alignat*}{4}
					&p_{i,j}^{i-2,j} = \frac{3\rho_{h}}{2}, & \quad &p_{i,j}^{i,j} = 1 - \rho_{r}, & \quad & p_{i,j}^{i-1,j-1} = 6\rho_{d^{+}}, & \quad &p_{i,j}^{i-1,j+1} = 3\rho_{d^{-}},
				\end{alignat*}
			\end{fleqn}
			where $\rho_{r} = 3(\rho_{xx} + 5\rho_{yy})/2 + 3\rho_{xy}$. 
		\end{minipage}
		\begin{minipage}[c]{0.325\textwidth}
			\centering
			\vspace{0pt}
			\includegraphics[width=\textwidth]{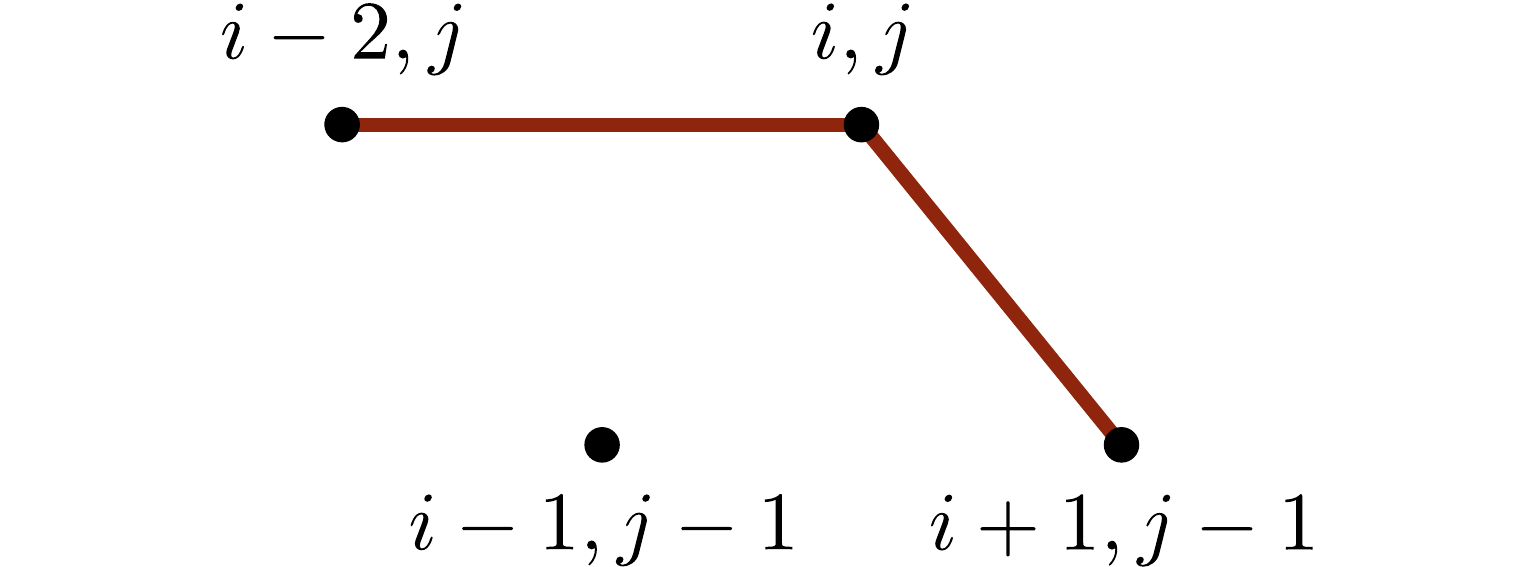}
		\end{minipage}
		\caption{\textbf{Transition probabilities governing particle movement on a flat-top hexagonal lattice.} Transition probabilities defined by the stochastic matrix $\mathbf{P}=\mathbf{I}+\tau\mathbf{C}$ governing particle movement between (a) interior, (b)--(g) edge and (h)--(k) corner lattice sites in a flat-top hexagonal lattice configuration. Here, we let $p_{i,j}^{k,m}$ denote the probability that a particle located at $\mathbf{x}_{i,j} = (x_{i},y_{j})$ at time $t=t_{n-1}$ moves to $\mathbf{x}_{k,m} = (x_{k},y_{m})$ at time $t=t_{n}$. These probabilities are defined using expressions pertaining to horizontal ($\rho_{h}$) and diagonal ($\rho_{d^{+}}$ and $\rho_{d^{-}}$) movement (see expressions (\ref{eq:flat-top_horiz}), (\ref{eq:flat-top_diag_plus}) and (\ref{eq:flat-top_diag_minus})). Additionally, the probability of a particle remaining in its current position (at $\mathbf{x}_{i,j}$) is defined as $p_{i,j}^{i,j}=1-\rho_{r}$, where $\rho_{r}$ is given by a linear combination of $\rho_{xx}$, $\rho_{yy}$ and/or $\rho_{xy}$ (see expressions in (\ref{eq:remain_expr})). For each case of lattice sites, we provide a schematic of the local lattice (central and neighbouring sites labelled with position indices).}
		\label{fig:flat-top_probs}
	\end{figure}}
	
	Studying the probabilities in Figure \ref{fig:flat-top_probs}, we see that $\mathbf{P}$ is a stochastic matrix when
	\begin{gather}\label{eq:flat-top_time_cond}
		\tau \leq \min\left\{\frac{\delta_{x}^2\delta_{y}^2}{2(\delta_{y}^2D_{xx} + 3\delta_{x}^2D_{yy})},\frac{\delta_{x}^2\delta_{y}^2}{3(\delta_{y}^2D_{xx} + 2\delta_{x}\delta_{y}|D_{xy}| + \delta_{x}^2D_{yy})},\frac{2\delta_{x}^2\delta_{y}^2}{3(\delta_{y}^2D_{xx} + 2\delta_{x}\delta_{y}|D_{xy}| + 5\delta_{x}^2D_{yy})}\right\}, \
	\end{gather}
	and
	\begin{gather}
		\delta_{y}^2D_{xx} \geq \delta_{x}^2D_{yy}, \label{eq:flat-top_cond_1} \\
		\delta_{x}D_{yy} \geq \delta_{y}|D_{xy}|, \label{eq:flat-top_cond_2}
	\end{gather}
	where the constraints (\ref{eq:flat-top_cond_1})--(\ref{eq:flat-top_cond_2}) are required to ensure all transition probabilities are non-negative, and (\ref{eq:flat-top_time_cond}) enforces $\rho_{r} \leq 1$ for each case (all probabilities are between zero and one). Inspecting the expressions (\ref{eq:flat-top_horiz})--(\ref{eq:flat-top_diag_minus}), we see that analogous observations to those made in section \ref{sec:rec_probs} (rectangular lattice) apply to horizontal and diagonal particle movement on a flat-top hexagonal lattice. The constraints (\ref{eq:flat-top_cond_1}) and (\ref{eq:flat-top_cond_2}) can be combined and simplified to give a condition on the spatial step $\delta_{y}$,
	\begin{gather}\label{eq:flat-top_interval_dy}
		\delta_{x}\sqrt{\frac{D_{yy}}{D_{xx}}} \leq \delta_{y} \leq \delta_{x}\frac{D_{yy}}{|D_{xy}|},
	\end{gather}
	or, alternatively, $\delta_{x}$,
	\begin{gather}\label{eq:flat-top_interval_dx}
		\delta_{y}\frac{|D_{xy}|}{D_{yy}} \leq \delta_{x} \leq \delta_{y}\sqrt{\frac{D_{xx}}{D_{yy}}}.
	\end{gather}
	Here, we consider the constraint (\ref{eq:flat-top_interval_dy}) when $D_{xx} < D_{yy}$ and (\ref{eq:flat-top_interval_dx}) when $D_{xx} > D_{yy}$ to avoid being limited to a very small spatial step $\delta_{y}$ or $\delta_{x}$ when $D_{yy}/|D_{xy}|$ or $D_{xx}/D_{yy}$ is small, respectively. For either interval to exist, we require the lower bound to not exceed the upper bound, which gives the following constraint on the diffusion tensor $\mathbf{D}$:
	\begin{gather}\label{eq:hex_det_cond}
		\det(\mathbf{D}) \geq 0.
	\end{gather}
	Given that $\mathbf{D}$ is symmetric positive definite, the condition (\ref{eq:hex_det_cond}) is always satisfied. Thus, a stochastic matrix governing particle movement on a flat-top hexagonal lattice can be always be obtained for any valid diffusion tensor.
	
	\subsection{Pointy-top hexagonal lattice}\label{sec:pointy-top_probs}
	Finally, we consider the transition matrix $\mathbf{P} = \mathbf{I} + \tau\mathbf{C}$ whose entries correspond to the probabilities of particle movement on a pointy-top hexagonal lattice (see Figure \ref{fig:lattice_examples}). These entries are presented in Figure \ref{fig:pointy-top_probs} for interior, edge and corner lattice sites. In each case, the probabilities of particle movement occurring along the coordinate axes in a vertical (to $\mathbf{x}_{i,j-2}$ or $\mathbf{x}_{i,j+2}$) direction are defined in terms of the following expression:
	\begin{gather}\label{eq:pointy-top_vert}
		\rho_{v} = \frac{\tau(\delta_{x}^2D_{yy} - \delta_{y}^2D_{xx})}{\delta_{x}^2\delta_{y}^2}.
	\end{gather}
	Additionally, the expressions
	\begin{gather}\label{eq:pointy-top_diag_plus}
		\rho_{d^{+}} = \frac{\tau(\delta_{y}D_{xx} + \delta_{x}D_{xy})}{\delta_{x}^2\delta_{y}},
	\end{gather}
	and 
	\begin{gather}	\label{eq:pointy-top_diag_minus}
		\rho_{d^{-}} = \frac{\tau(\delta_{y}D_{xx} - \delta_{y}D_{xy})}{\delta_{x}^2\delta_{y}},
	\end{gather}
	are used to define the probabilities of particle movement to the northeast or southwest sites (to $\mathbf{x}_{i+1,j+1}$ or $\mathbf{x}_{i-1,j-1}$) or to the northwest or southeast sites (to $\mathbf{x}_{i-1,j+1}$ or $\mathbf{x}_{i+1,j-1}$), respectively. Finally, we utilise a linear combination of the expressions in (\ref{eq:remain_expr}) to define the probability that a particle remains in its current position. For each case of lattice sites, we provide a schematic of the local lattice (central and neighbouring sites labelled with position indices) to ease visual association between each lattice site and the corresponding probability of a particle moving to, or remaining at, that location (see Figure \ref{fig:pointy-top_probs}). Note that the corresponding row and column positions in the matrix $\mathbf{P}$ for the probability $p_{i,j}^{k,m}$ are given by $m_{i,j}$ and $m_{k,m}$, respectively, where $m_{i,j} = (i-1)N_{y}/2 + \lceil j/2 \rceil$ is the mapping function from (\ref{eq:map}). Additionally, we remind the reader that lattice sites in a hexagonal configuration are only defined if the sum of position indices is even (i.e. $\mathbf{x}_{i,j} = (x_{i},y_{j})$ is only defined if $i + j$ is even).
	\afterpage{\centering
	\begin{figure}[H]
		\setstretch{0.75}
		\begin{minipage}[c]{0.65\textwidth}
			{\bfseries (a) Interior ($i=2,\hdots,N_{x}-1$ and $j=3,\hdots,N_{y}-2$)}:
			\begin{fleqn}[10pt]
				\begin{alignat*}{3}
					&p_{i,j}^{i,j+2} = \rho_{v}, &\quad &p_{i,j}^{i-1,j+1} = 2\rho_{d^{-}}, &\quad &p_{i,j}^{i+1,j+1} = 2\rho_{d^{+}}, \\
					&p_{i,j}^{i,j} = 1 - \rho_{r}, & &p_{i,j}^{i-1,j-1} = 2\rho_{d^{+}}, & & \\
					&p_{i,j}^{i+1,j-1} = 2\rho_{d^{-}}, & &p_{i,j}^{i,j-2} = \rho_{v}, & &
				\end{alignat*}
			\end{fleqn}
			where $\rho_{r} = 2(3\rho_{xx} + \rho_{yy})$.
		\end{minipage}
		\hfill
		\begin{minipage}[c]{0.325\textwidth}
			\vspace{0pt}
			\includegraphics[width=\textwidth]{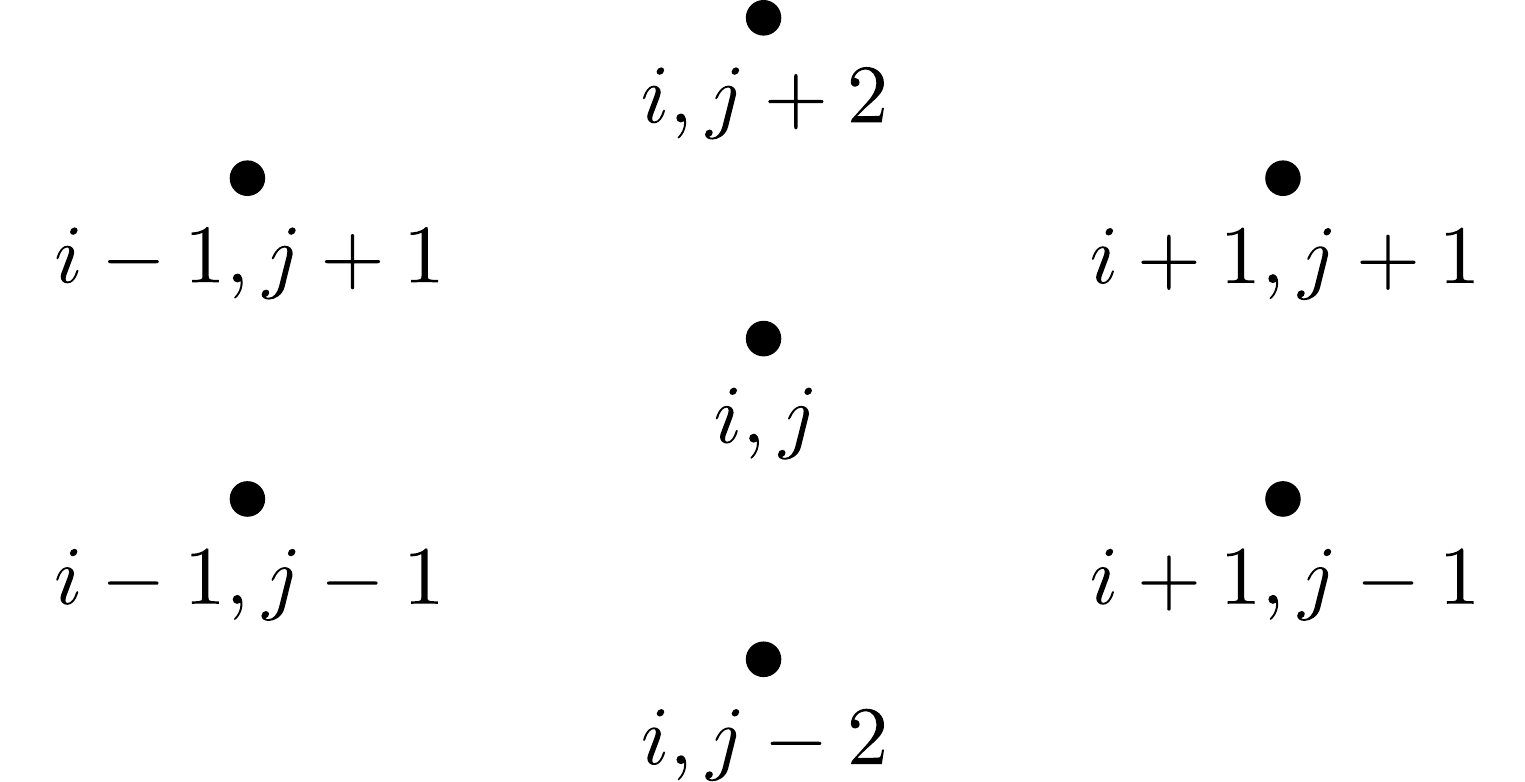}
		\end{minipage}
		\phantomcaption
	\end{figure}
	\vfill
	\vspace{-\intextsep}\noindent\dotfill
	\vfill
	\begin{figure}[H]\ContinuedFloat
		\setstretch{0.75}
		\begin{minipage}[c]{0.65\textwidth}
			\vspace{0pt}
			{\bfseries (b) Western edge ($i=1$ and $j=3,\hdots,N_{y}-2$)}:
			\begin{fleqn}[10pt]
				\begin{alignat*}{3}
					&p_{i,j}^{i,j+2} = \rho_{v}, &\quad &p_{i,j}^{i+1,j+1} = 4\rho_{d^{+}}, &\quad &p_{i,j}^{i,j} = 1 - \rho_{r}, \\
					&p_{i,j}^{i+1,j-1} = 4\rho_{d^{-}}, & &p_{i,j}^{i,j-2} = \rho_{v}, & &
				\end{alignat*}
			\end{fleqn}
			where $\rho_{r} = 2(3\rho_{xx} + \rho_{yy})$.
		\end{minipage}
		\hfill
		\begin{minipage}[c]{0.325\textwidth}
			\vspace{0pt}
			\includegraphics[width=\textwidth]{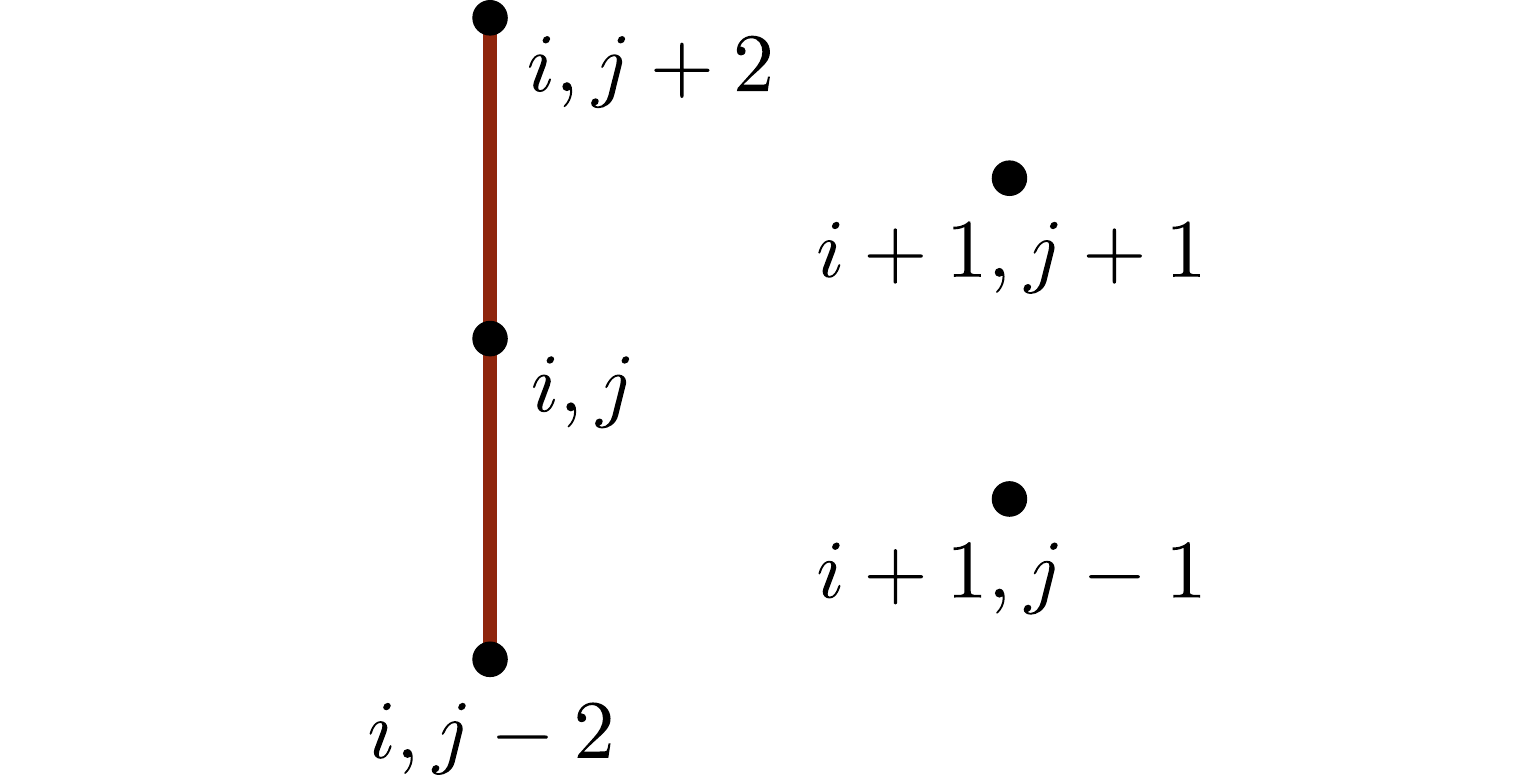}
		\end{minipage}
	\end{figure}
	\vfill
	\vspace{-\intextsep}\noindent\dotfill
	\vfill
	\begin{figure}[H]\ContinuedFloat
		\setstretch{0.75}
		\begin{minipage}[c]{0.65\textwidth}
			\vspace{0pt}
			{\bfseries (c) Southern concave edge ($i=2,\hdots,N_{x}-1$ and $j=1$)}:
			\begin{fleqn}[10pt]
				\begin{alignat*}{2}
					&p_{i,j}^{i,j+2} = 3\rho_{v}, &\quad &p_{i,j}^{i-1,j+1} = 3\rho_{d^{-}}, \\
					&p_{i,j}^{i+1,j+1} = 3\rho_{d^{+}}, & &p_{i,j}^{i,j} = 1 - \rho_{r},
				\end{alignat*}
			\end{fleqn}
			where $\rho_{r} = 3(\rho_{xx} + \rho_{yy})$.
		\end{minipage}
		\hfill
		\begin{minipage}[c]{0.325\textwidth}
			\vspace{0pt}
			\centering
			\includegraphics[width=\textwidth]{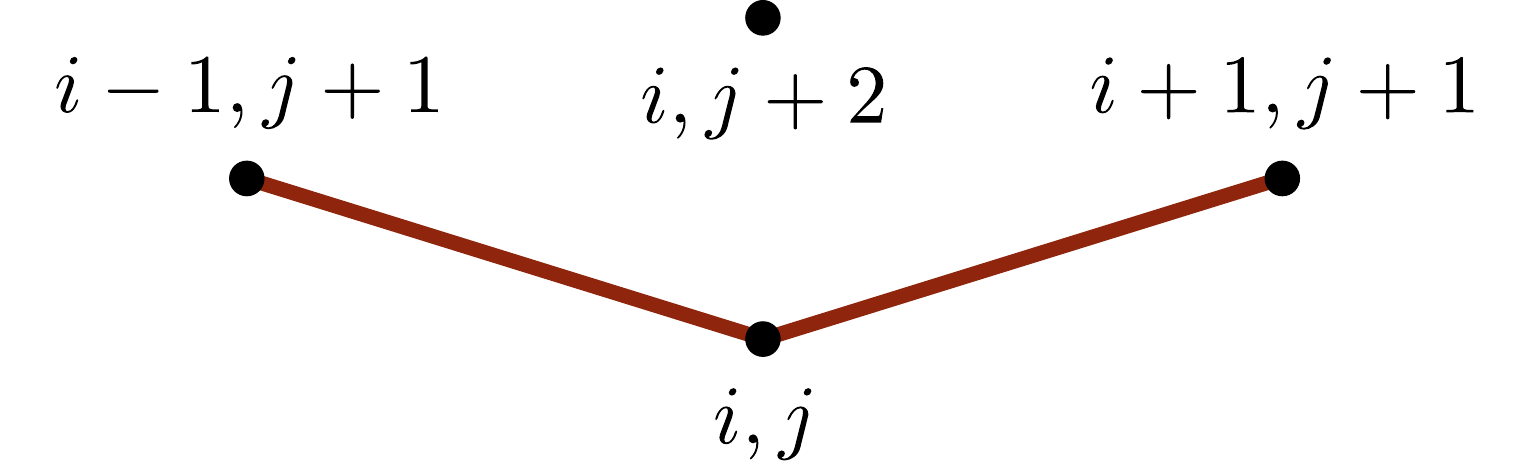}
		\end{minipage}
	\end{figure}
	\vfill
	\vspace{-\intextsep}\noindent\dotfill
	\vfill
	\begin{figure}[H]\ContinuedFloat
		\setstretch{0.75}
		\begin{minipage}[c]{0.65\textwidth}
			\vspace{0pt}
			{\bfseries (d) Southern convex edge ($i=2,\hdots,N_{x}-1$ and $j=2$)}:
			\begin{fleqn}[10pt]
				\begin{alignat*}{3}
					&p_{i,j}^{i,j+2} = \frac{3\rho_{v}}{2}, &\quad &p_{i,j}^{i-1,j+1} = 3\rho_{d^{-}}, &\quad &p_{i,j}^{i+1,j+1} = 3\rho_{d^{+}}, \\
					&p_{i,j}^{i,j} = 1 - \rho_{r}, & &p_{i,j}^{i-1,j-1} = \frac{3\rho_{d^{+}}}{2}, & &p_{i,j}^{i+1,j-1} = \frac{3\rho_{d^{-}}}{2},
				\end{alignat*}
			\end{fleqn}
			where $\rho_{r} = 3(5\rho_{xx} + \rho_{yy})/2$.
		\end{minipage}
		\hfill
		\begin{minipage}[c]{0.325\textwidth}
			\vspace{0pt}
			\includegraphics[width=\textwidth]{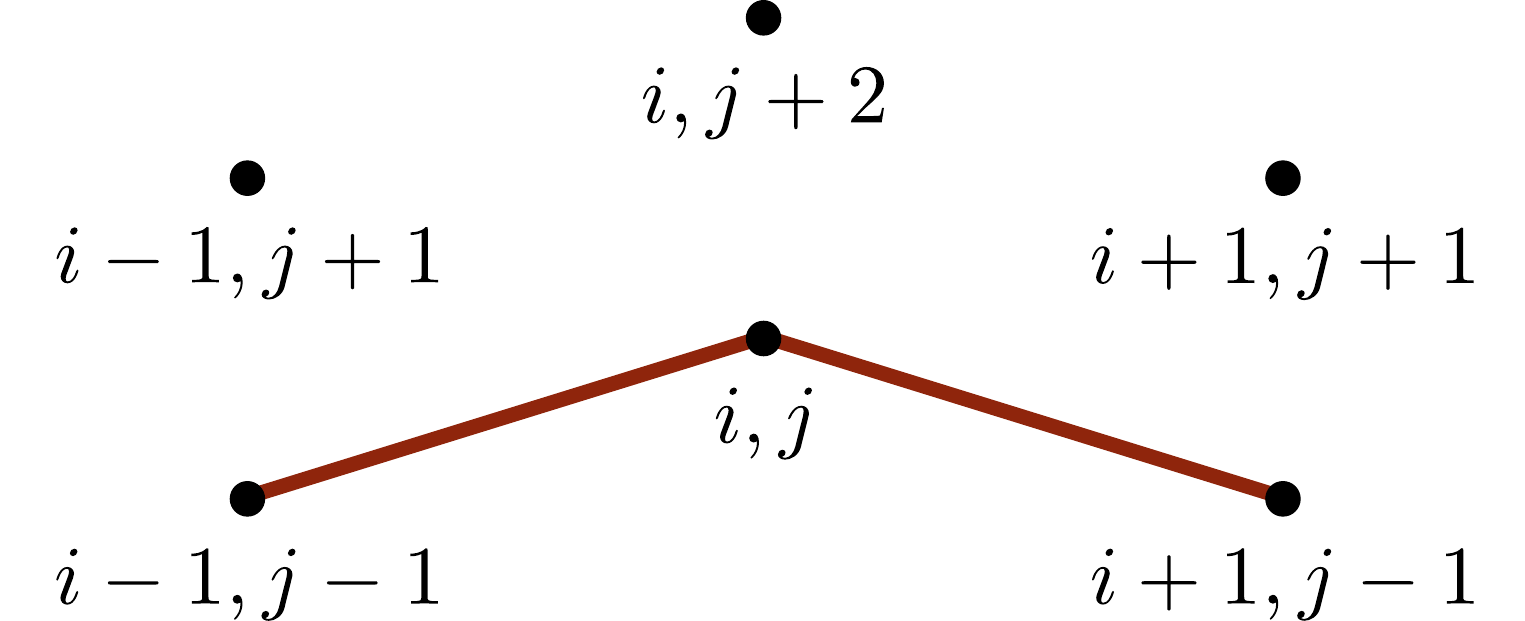}
		\end{minipage}
	\end{figure}
	\vfill
	\vspace{-\intextsep}\noindent\dotfill
	\vfill
	\begin{figure}[H]\ContinuedFloat
		\setstretch{0.75}
		\begin{minipage}[c]{0.65\textwidth}
			\vspace{0pt}
			{\bfseries (e) Eastern edge ($i = N_{x}$ and $j=3,\hdots,N_{y}-2$)}:
			\begin{fleqn}[10pt]
				\begin{alignat*}{3}
					&p_{i,j}^{i,j+2} = \rho_{v}, &\quad &p_{i,j}^{i-1,j+1} = 4\rho_{d^{-}}, &\quad & \\
					&p_{i,j}^{i,j} = 1 - \rho_{r}, & &p_{i,j}^{i-1,j-1} = 4\rho_{d^{+}}, & &p_{i,j}^{i,j-2} = \rho_{v},
				\end{alignat*}
			\end{fleqn}
			where $\rho_{r} = 2(3\rho_{xx} + \rho_{yy})$.
		\end{minipage}
		\hfill
		\begin{minipage}[c]{0.325\textwidth}
			\vspace{0pt}
			\centering
			\includegraphics[width=\textwidth]{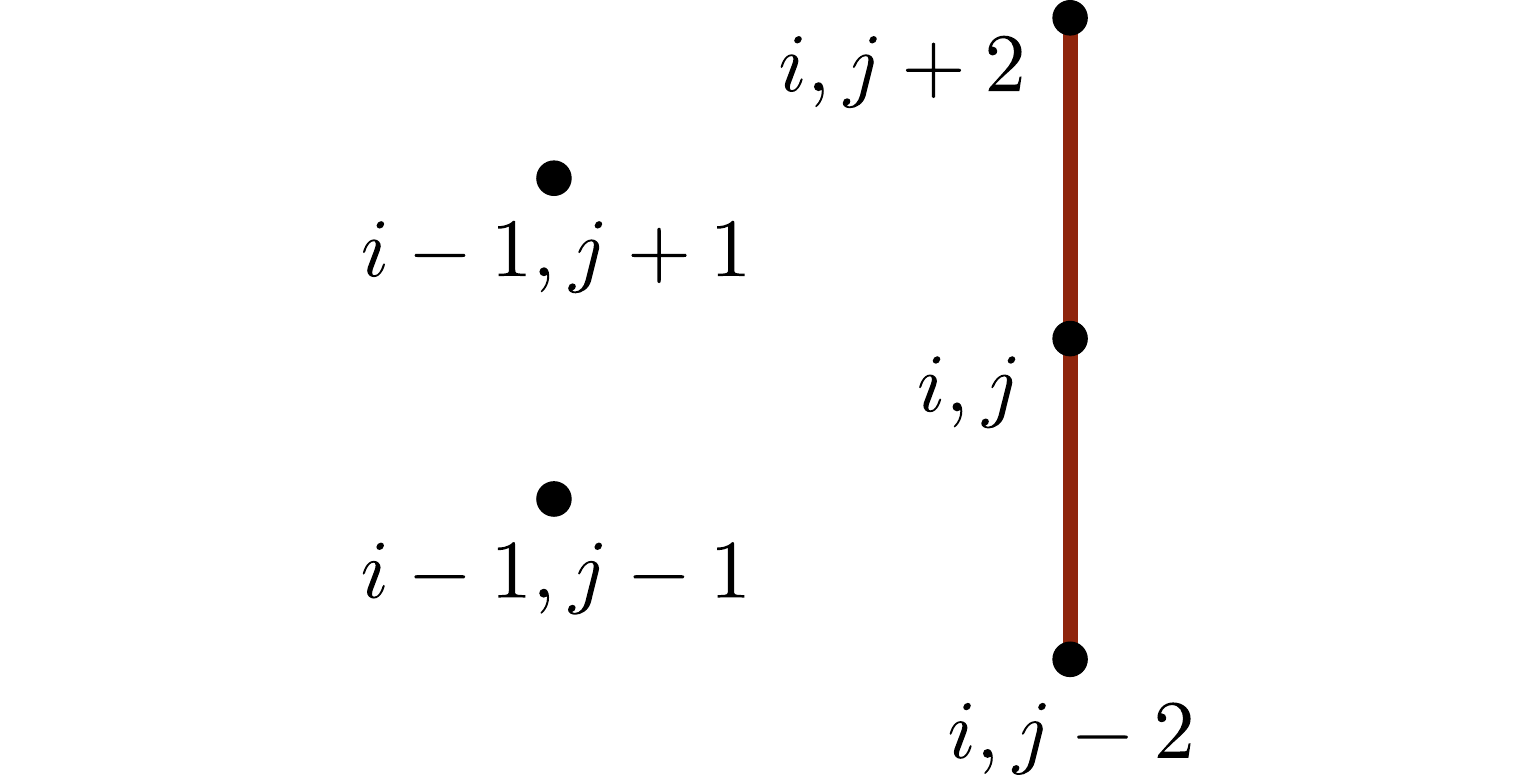}
		\end{minipage}
	\end{figure}
	\vfill
	\vspace{-\intextsep}\noindent\dotfill
	\vfill
	\begin{figure}[H]\ContinuedFloat
		\setstretch{0.75}
		\begin{minipage}[c]{0.65\textwidth}
			\vspace{0pt}
			{\bfseries (f) Northern convex edge ($i=2,\hdots,N_{x}-1$ and $j=N_{y}-1$)}:
			\begin{fleqn}[10pt]
				\begin{alignat*}{3}
					&p_{i,j}^{i-1,j+1} = \frac{3\rho_{d^{-}}}{2}, &\quad &p_{i,j}^{i+1,j+1} = \frac{3\rho_{d^{+}}}{2}, &\quad &p_{i,j}^{i,j} = 1 - \rho_{r}, \\
					&p_{i,j}^{i-1,j-1} = 3\rho_{d^{+}}, & &p_{i,j}^{i+1,j-1} = 3\rho_{d^{-}}, & &p_{i,j}^{i,j-2} = \frac{3\rho_{v}}{2},
				\end{alignat*}
				where $\rho_{r} = 3(5\rho_{xx} + \rho_{yy})/2$.
			\end{fleqn}
		\end{minipage}
		\hfill
		\begin{minipage}[c]{0.325\textwidth}
			\vspace{0pt}
			\centering
			\includegraphics[width=\textwidth]{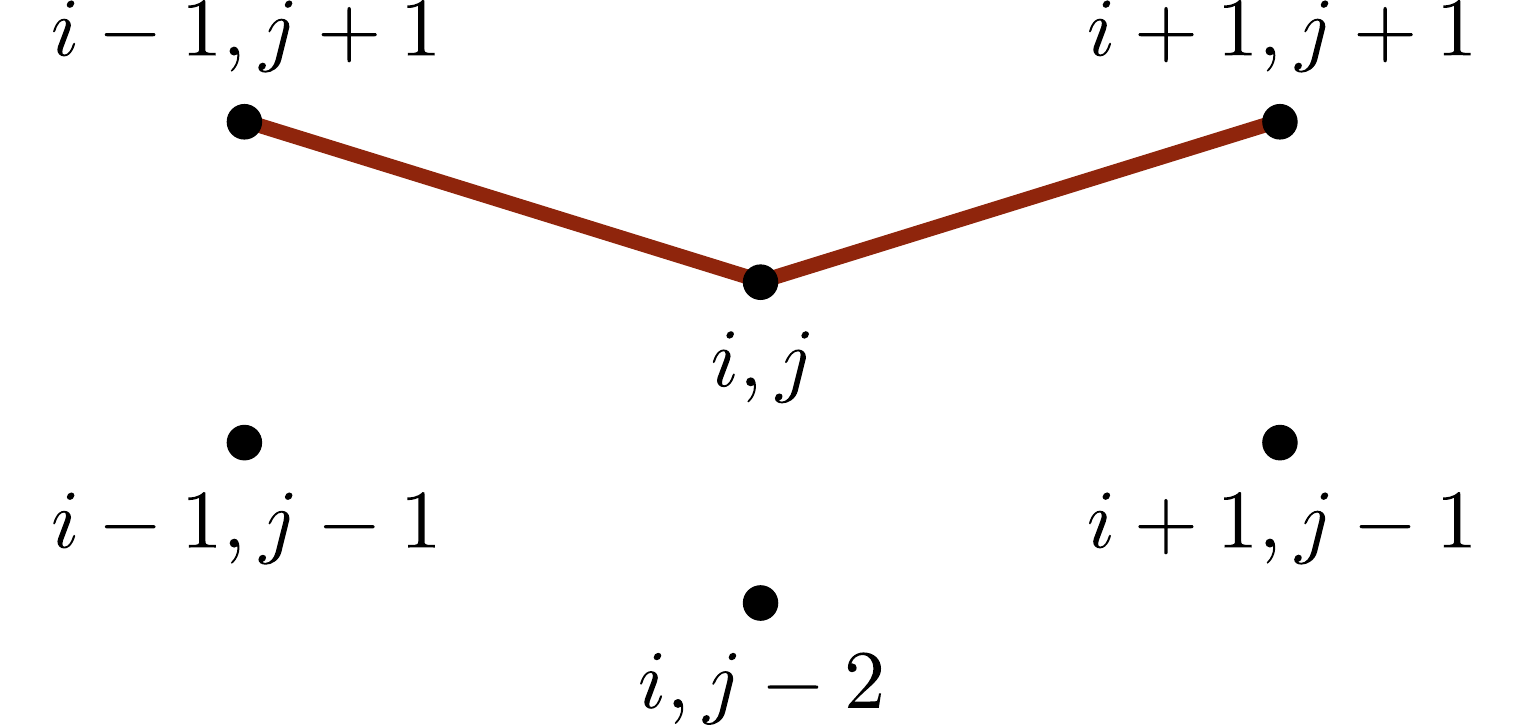}
		\end{minipage}
	\end{figure}
	\vfill
	\vspace{-\intextsep}\noindent\dotfill
	\vfill
	\begin{figure}[H]\ContinuedFloat
		\setstretch{0.75}
		\begin{minipage}[c]{0.65\textwidth}
			\vspace{0pt}
			{\bfseries (g) Northern concave edge ($i=2,\hdots,N_{x}-1$ and $j=N_{y}$)}:
			\begin{fleqn}[10pt]
				\begin{alignat*}{2}
					&p_{i,j}^{i,j} = 1 - \rho_{r}, &\quad &p_{i,j}^{i-1,j-1} = 3\rho_{d^{+}}, \\
					&p_{i,j}^{i+1,j-1} = 3\rho_{d^{-}}, & &p_{i,j}^{i,j-2} = 3\rho_{v},
				\end{alignat*}
			\end{fleqn}
			where $\rho_{r} = 3(\rho_{xx} + \rho_{yy})$.
		\end{minipage}
		\hfill
		\begin{minipage}[c]{0.325\textwidth}
			\vspace{0pt}
			\centering
			\includegraphics[width=\textwidth]{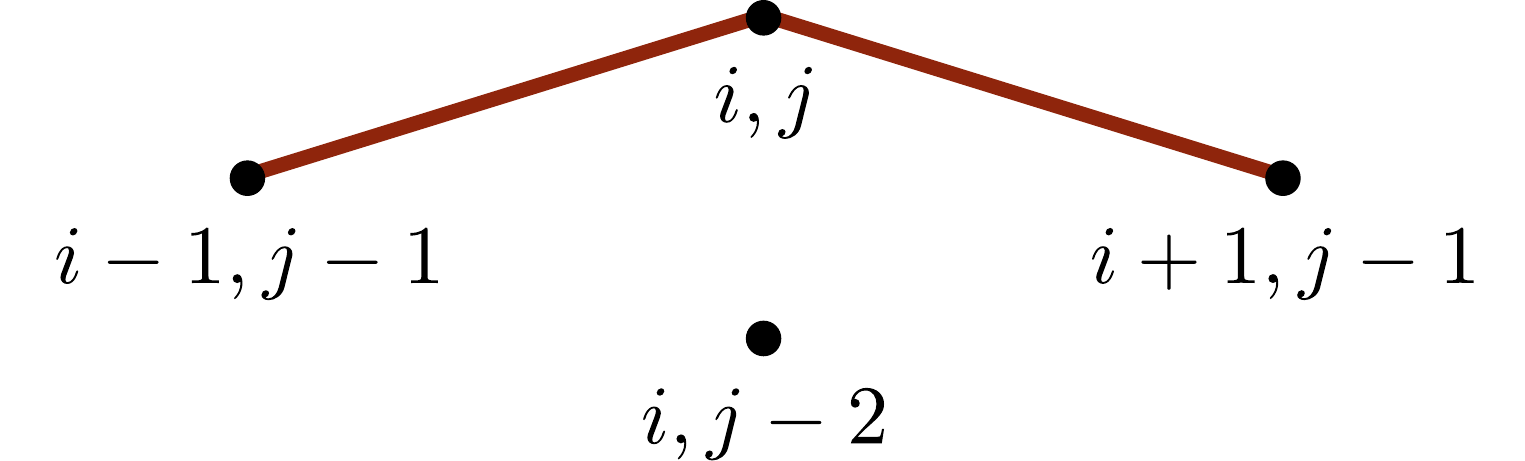}
		\end{minipage}
	\end{figure}
	\vfill
	\vspace{-\intextsep}\noindent\dotfill
	\vfill
	\begin{figure}[H]\ContinuedFloat
		\setstretch{0.75}
		\begin{minipage}[c]{0.65\textwidth}
			\vspace{0pt}
			{\bfseries (h) Southwest concave corner ($i = 1$ and $j = 1$)}:
			\begin{fleqn}[10pt]
				\begin{alignat*}{3}
					&p_{i,j}^{i,j+2} = 3\rho_{v}, & \quad &p_{i,j}^{i+1,j+1} = 6\rho_{d^{+}}, & \quad &p_{i,j}^{i,j} = 1 - \rho_{r},
				\end{alignat*}
			\end{fleqn}
			where $\rho_{r} = 3(\rho_{xx} + 2\rho_{xy} + \rho_{yy})$.
		\end{minipage}
		\hfill
		\begin{minipage}[c]{0.325\textwidth}
			\vspace{0pt}
			\centering
			\includegraphics[width=\textwidth]{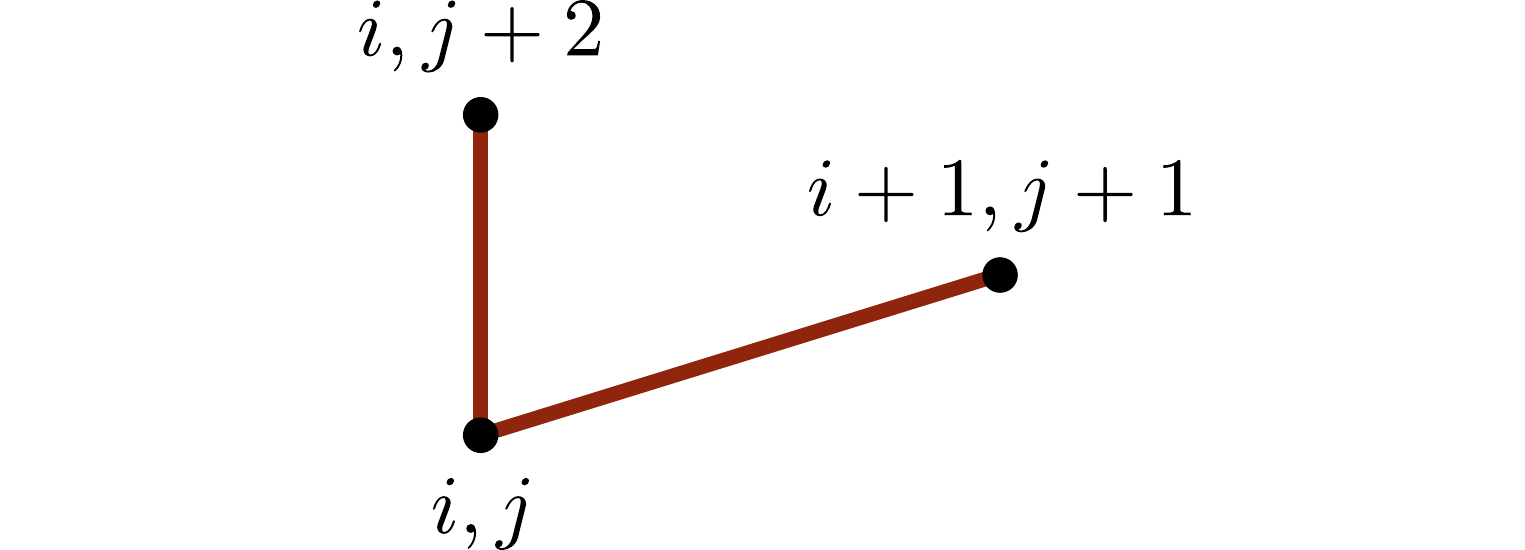}
		\end{minipage}
	\end{figure}
	\vfill
	\vspace{-\intextsep}\noindent\dotfill
	\vfill
	\begin{figure}[H]\ContinuedFloat
		\setstretch{0.75}
		\begin{minipage}[c]{0.65\textwidth}
			\vspace{0pt}
			{\bfseries (i) Southeast concave corner ($i = N_{x}$ and $j = 1$)}:
			\begin{fleqn}[10pt]
				\begin{alignat*}{3}
					&p_{i,j}^{i,j+2} = 3\rho_{v}, & \quad &p_{i,j}^{i-1,j+1} = 6\rho_{d^{-}}, & \quad &p_{i,j}^{i,j} = 1 - \rho_{r},
				\end{alignat*}
			\end{fleqn}
			where $\rho_{r} = 3(\rho_{xx} - 2\rho_{xy} + \rho_{yy})$.
		\end{minipage}
		\hfill
		\begin{minipage}[c]{0.325\textwidth}
			\vspace{0pt}
			\centering
			\includegraphics[width=\textwidth]{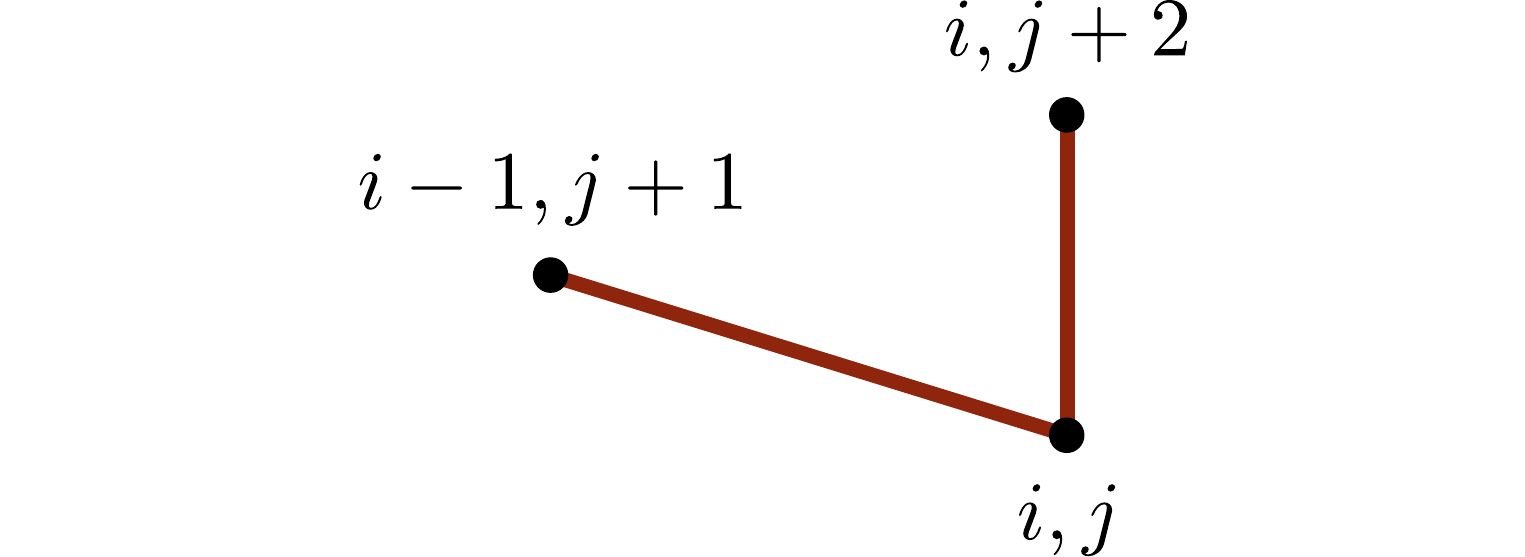}
		\end{minipage}
		\vfill
	\end{figure}
	\vfill
	\vspace{-\intextsep}\noindent\dotfill
	\vfill
	\begin{figure}[H]\ContinuedFloat
		\setstretch{0.75}
		\begin{minipage}[c]{0.65\textwidth}
			\vspace{0pt}
			{\bfseries (j) Northwest convex corner ($i = 1$ and $j = N_{y}-1$)}:
			\begin{fleqn}[10pt]
				\begin{alignat*}{4}
					&p_{i,j}^{i+1,j+1} = 3\rho_{d^{+}}, & \quad &p_{i,j}^{i,j} = 1 - \rho_{r}, & \quad &p_{i,j}^{i+1,j-1} = 6\rho_{d^{-}}, & \quad &p_{i,j}^{i,j-2} = \frac{3\rho_{v}}{2},
				\end{alignat*}
			\end{fleqn}
			where $\rho_{r} = 3(5\rho_{xx} + \rho_{yy})/2 - 3\rho_{xy}$.
		\end{minipage}
		\hfill
		\begin{minipage}[c]{0.325\textwidth}
			\vspace{0pt}
			\centering
			\includegraphics[width=\textwidth]{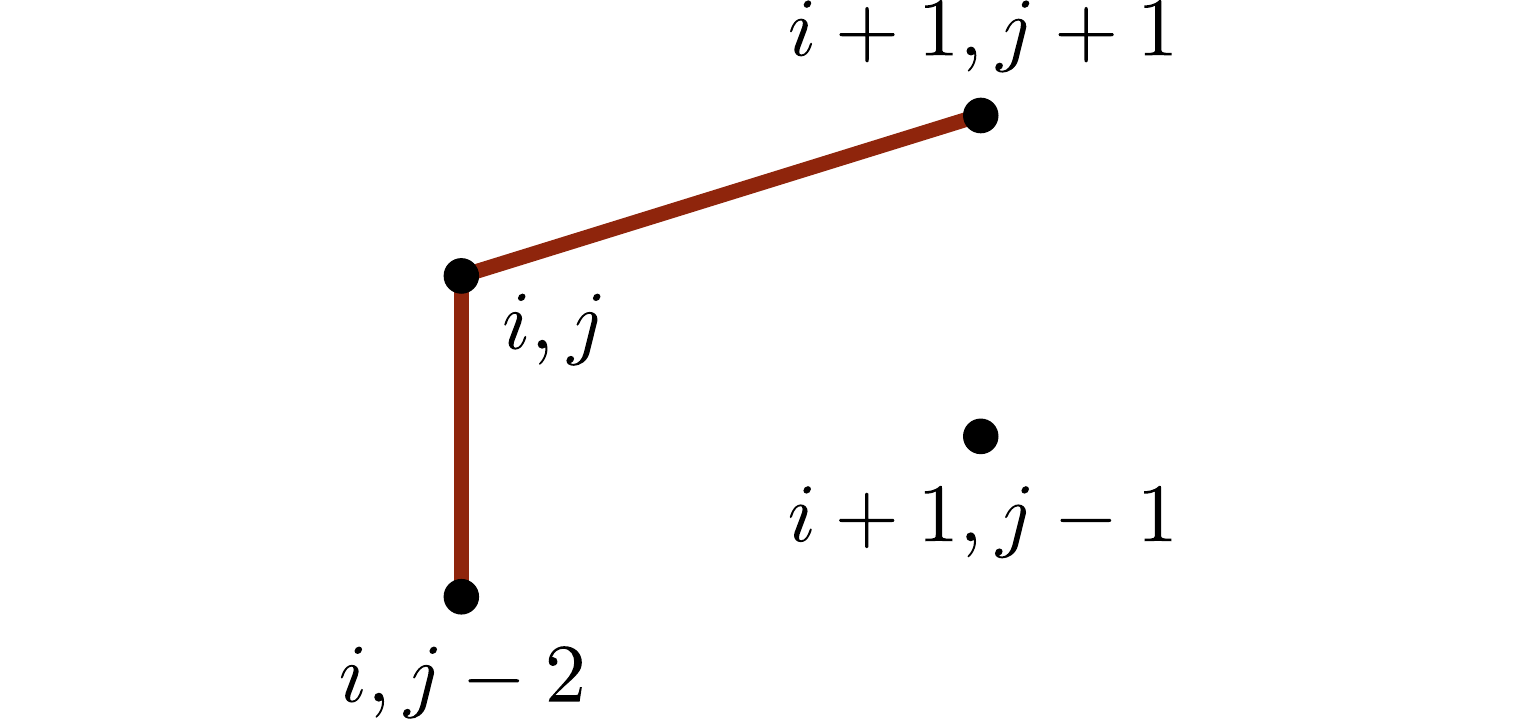}
		\end{minipage}
	\end{figure}
	\vfill
	\vspace{-\intextsep}\noindent\dotfill
	\vfill
	\begin{figure}[H]\ContinuedFloat
		\setstretch{0.75}
		\begin{minipage}[c]{0.65\textwidth}
			\vspace{0pt}
			{\bfseries (k) Northeast convex corner ($i = N_{x}$ and $j = N_{y}-1$)}:
			\begin{fleqn}[10pt]
				\begin{alignat*}{4}
					&p_{i,j}^{i-1,j+1} = 3\rho_{d^{-}}, &\quad &p_{i,j}^{i,j} = 1 - \rho_{r}, & \quad &p_{i,j}^{i-1,j-1} = 6\rho_{d^{+}}, & \quad &p_{i,j}^{i,j-2} = \frac{3\rho_{v}}{2}.
				\end{alignat*}
			\end{fleqn}
			where $\rho_{r} = 3(5\rho_{xx} + \rho_{yy})/2 + 3\rho_{xy}$,
		\end{minipage}
		\hfill
		\begin{minipage}[c]{0.325\textwidth}
			\vspace{0pt}
			\centering
			\includegraphics[width=\textwidth]{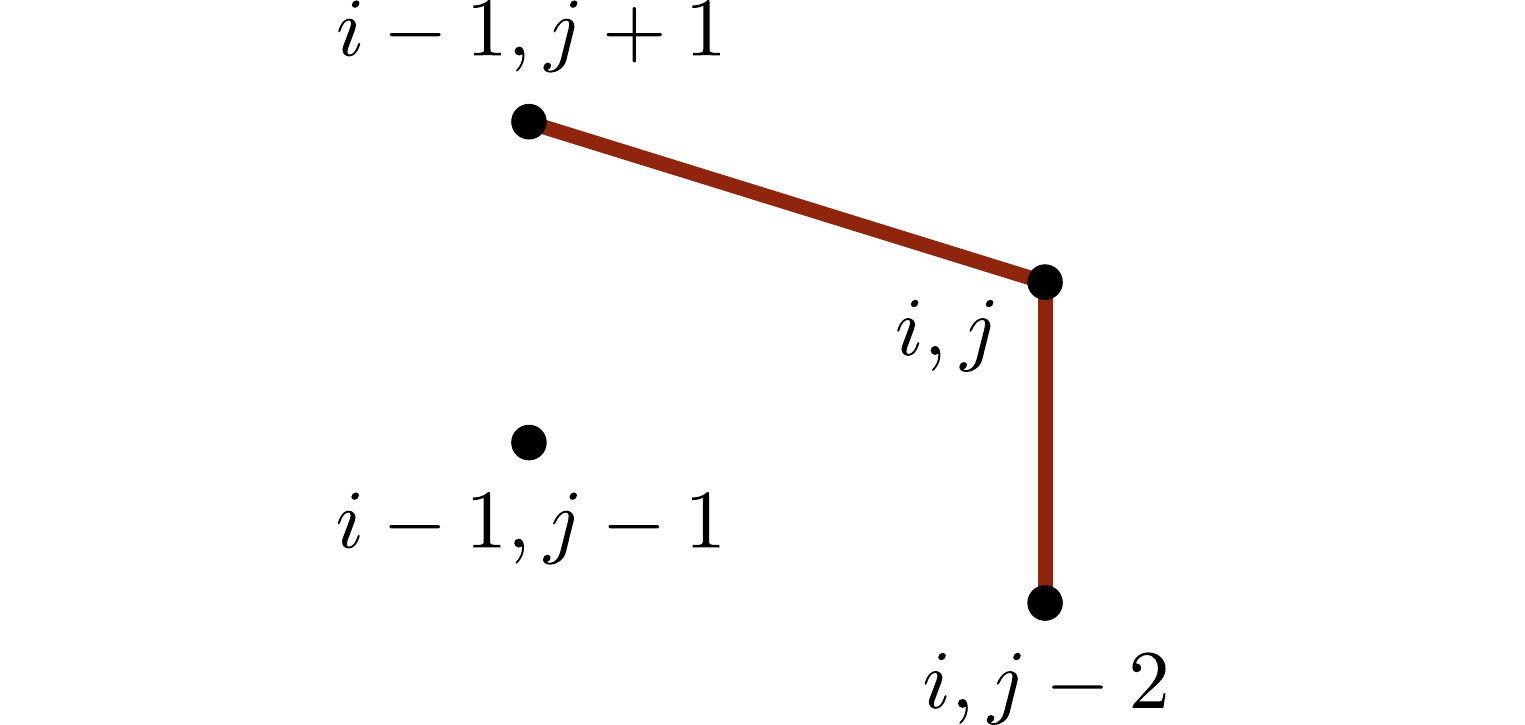}
		\end{minipage}
		\caption{\textbf{Transition probabilities governing particle movement on a pointy-top hexagonal lattice.} Transition probabilities defined by the stochastic matrix $\mathbf{P}=\mathbf{I}+\tau\mathbf{C}$ governing particle movement between (a) interior, (b)--(g) edge and (h)--(k) corner lattice sites in a pointy-top hexagonal lattice configuration. Here, we let $p_{i,j}^{k,m}$ denote the probability that a particle located at $\mathbf{x}_{i,j} = (x_{i},y_{j})$ at time $t=t_{n-1}$ moves to $\mathbf{x}_{k,m} = (x_{k},y_{m})$ at time $t=t_{n}$. These probabilities are defined using expressions pertaining to vertical ($\rho_{v}$) and diagonal ($\rho_{d^{+}}$ and $\rho_{d^{-}}$) movement (see expressions (\ref{eq:pointy-top_vert}), (\ref{eq:pointy-top_diag_plus}) and (\ref{eq:pointy-top_diag_minus})). Additionally, the probability of a particle remaining in its current position (at $\mathbf{x}_{i,j}$) is defined as $p_{i,j}^{i,j}=1-\rho_{r}$, where $\rho_{r}$ is given by a linear combination of $\rho_{xx}$, $\rho_{yy}$ and/or $\rho_{xy}$ (see expressions in (\ref{eq:remain_expr})). For each case of lattice sites, we provide a schematic of the local lattice (central and neighbouring sites labelled with position indices).}
		\label{fig:pointy-top_probs}
	\end{figure}
	}

	Studying the probabilities in Figure \ref{fig:pointy-top_probs}, we see that $\mathbf{P}$ is a stochastic matrix when
	\begin{gather}\label{eq:pointy-top_time_cond}
		\tau \leq \min\left\{\frac{\delta_{x}^2\delta_{y}^2}{2(3\delta_{y}^2D_{xx} + \delta_{x}^2D_{yy})},\frac{\delta_{x}^2\delta_{y}^2}{3(\delta_{y}^2D_{xx} + 2\delta_{x}\delta_{y}|D_{xy}| + \delta_{x}^2D_{yy})},\frac{2\delta_{x}^2\delta_{y}^2}{3(5\delta_{y}^2D_{xx} + 2\delta_{x}\delta_{y}|D_{xy}| + \delta_{x}^2D_{yy})}\right\}, \
	\end{gather}
	and
	\begin{gather}
		\delta_{x}^2D_{yy} \geq \delta_{y}^2D_{xx}, \label{eq:pointy-top_cond_1} \\
		\delta_{y}D_{xx} \geq \delta_{x}|D_{xy}|, \label{eq:pointy-top_cond_2}
	\end{gather}
	where the constraints (\ref{eq:pointy-top_cond_1})--(\ref{eq:pointy-top_cond_2}) are required to ensure all transition probabilities are non-negative, and (\ref{eq:pointy-top_time_cond}) enforces $\rho_{r} \leq 1$ for each case (all probabilities are between zero and one). Inspecting the expressions (\ref{eq:pointy-top_vert})--(\ref{eq:pointy-top_diag_minus}), we see that analogous observations to those made in section \ref{sec:rec_probs} (rectangular lattice) apply to vertical and diagonal particle movement on a pointy-top hexagonal lattice. The constraints (\ref{eq:pointy-top_cond_1}) and (\ref{eq:pointy-top_cond_2}) can be combined to give a condition on the spatial step $\delta_{y}$,
	\begin{gather}\label{eq:pointy-top_interval_dy}
		\delta_{x}\frac{|D_{xy}|}{D_{xx}} \leq \delta_{y} \leq \delta_{x}\sqrt{\frac{D_{yy}}{D_{xx}}},
	\end{gather}
	or, alternatively, $\delta_{x}$,
	\begin{gather}\label{eq:pointy-top_interval_dx}
		\delta_{y}\sqrt{\frac{D_{xx}}{D_{yy}}} \leq \delta_{x} \leq \delta_{y}\frac{D_{xx}}{|D_{xy}|}.
	\end{gather}
	Here, we consider the constraint (\ref{eq:pointy-top_interval_dy}) when $D_{xx} < D_{yy}$ and (\ref{eq:pointy-top_interval_dx}) when $D_{xx} > D_{yy}$ to avoid being limited to a very small spatial step $\delta_{y}$ or $\delta_{x}$ when $D_{yy}/D_{xx}$ or $D_{xx}/|D_{xy}|$ is small, respectively. For either interval (\ref{eq:pointy-top_interval_dy}) or (\ref{eq:pointy-top_interval_dx}) to exist, we require the lower bound to not exceed the upper bound, which gives the constraint (\ref{eq:hex_det_cond}) on the diffusion tensor $\mathbf{D}$ obtained in section \ref{sec:flat-top_probs}:
	\begin{gather*}
		\det(\mathbf{D}) \geq 0.
	\end{gather*}
	As previously discussed, this condition is always satisfied for a symmetric positive definite diffusion tensor $\mathbf{D}$. Thus, a stochastic matrix governing particle movement on a flat-top or pointy-top hexagonal lattice can always be obtained for any valid diffusion tensor.
	
	\section{Discussion of simulations and results}\label{sec:discussion}
\begin{figure}[H]
		\begin{subfigure}{0.47\textwidth}
			\centering
			\includegraphics[width=\textwidth]{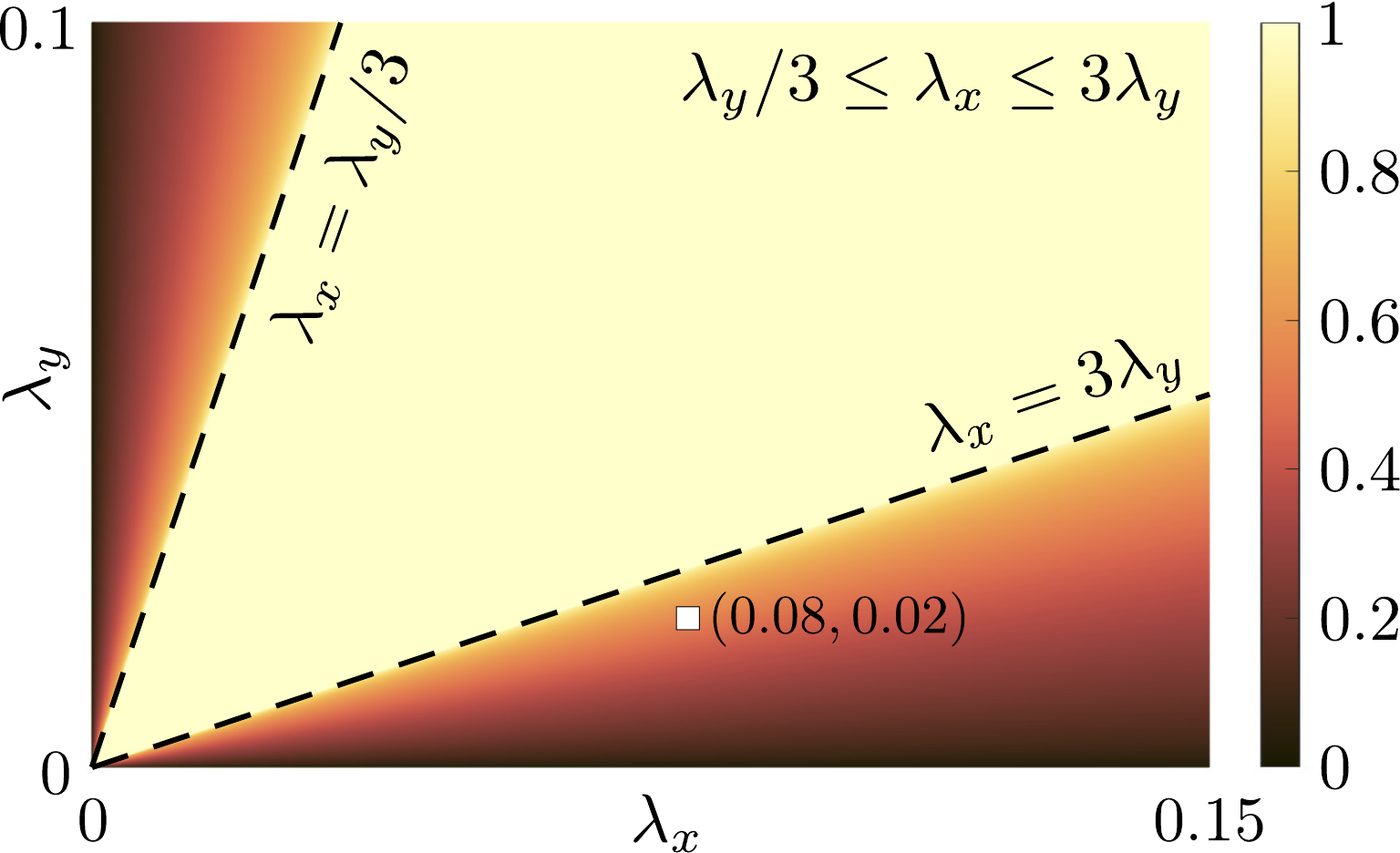}
			\caption{Proportion of unit semicircle satisfying (\ref{eq:rec_theta})}
			\label{fig:eigenvalue_sweep}
		\end{subfigure}
		\hfill
		\begin{subfigure}{0.47\textwidth}
			\centering
			\includegraphics[width=\textwidth,trim=0cm -1.35cm 0cm 0cm, clip=true]{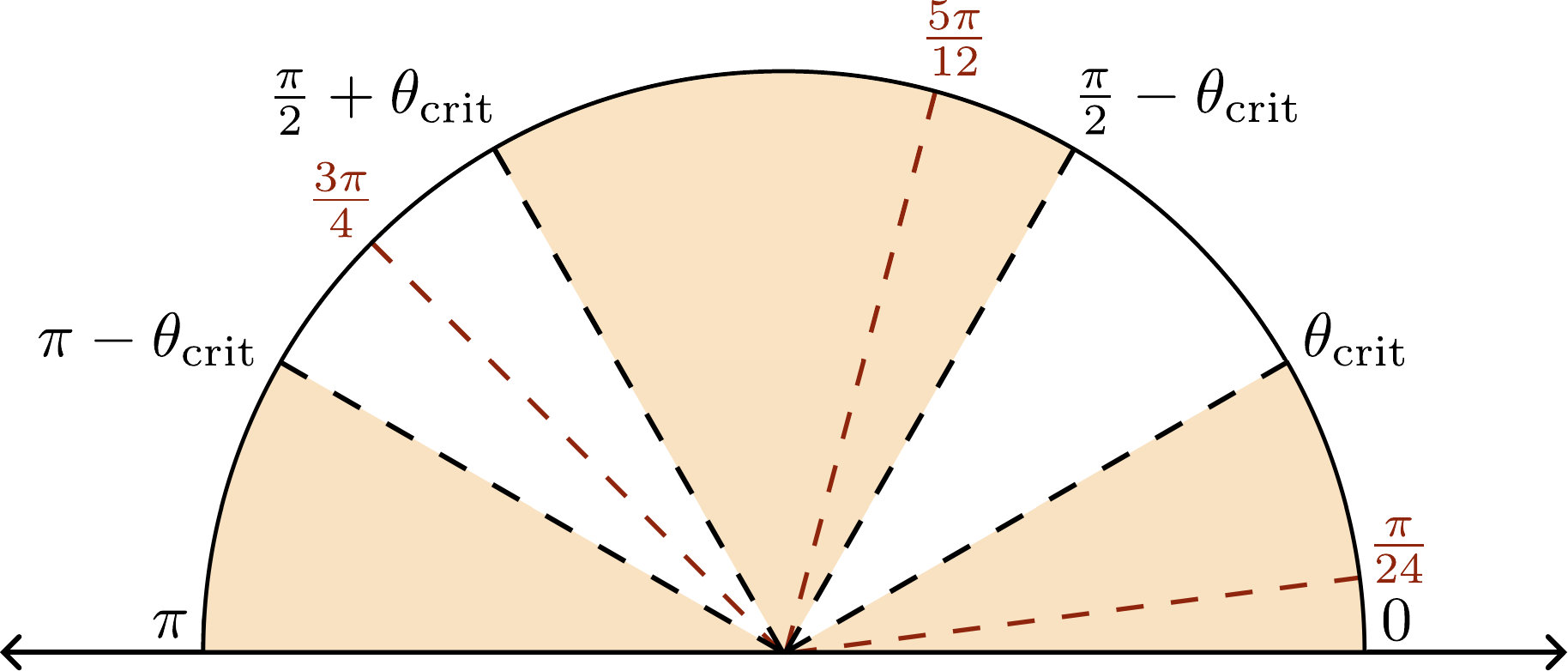}
			\caption{Angular regions for $\lambda_{x} = 0.08$ and $\lambda_{y} = 0.02$}
			\label{fig:test_case_regions}
		\end{subfigure}
		\caption{\textbf{Prototype tensor constraints for particle diffusion on a rectangular lattice.} (a) Proportion (between zero and one) of the unit semicircle within which the angle of rotation $0 < \theta < \pi$ satisfies (\ref{eq:rec_theta}) for implementation of the deterministic and random walk models (with the prototype diffusion tensor (\ref{eq:prototype_tensor})) on a rectangular lattice. If $\lambda_{y}/3 \leq \lambda_{x} \leq 3\lambda_{y}$, where $\lambda_{x}$ and $\lambda_{y}$ are the tensor eigenvalues, there is no restriction on $\theta$ (labelled region with bounds denoted by dashed black lines). Otherwise, the proportion of the unit semicircle (range of appropriate angles) decreases as $\lambda_{x}$ and $\lambda_{y}$ diverge away from each other. (b) Angular regions (shaded) of the unit semicircle within which $\theta$ satisfies (\ref{eq:rec_theta}) for eigenvalues $\lambda_{x} = 0.08$ and $\lambda_{y} = 0.02$ (labelled on (a)). The three prototype diffusion tensors used in this work are generated using these eigenvalues and the respective angles $\theta = \pi/24$, $\theta = 5\pi/12$ and $\theta=3\pi/4$ (dashed red lines). The latter angle does not satisfy the constraint (\ref{eq:rec_theta}).}
		\label{fig:rec_angle_restrictions}
	\end{figure}

	In this section, we provide visual and quantitative evidence to support the equivalence of the deterministic and random walk models. To elaborate, we (i) present two-dimensional comparisons of the particle densities obtained from the deterministic and random walk models, (ii) compare and demonstrate the agreement between these particle densities for a one-dimensional slice of the domain, and (iii) quantitatively validate the equivalence of the models for several test cases. For each test case, we generate a prototype diffusion tensor using rotation matrices,
	\begin{gather}\label{eq:prototype_tensor}
		\mathbf{D} = \mathbf{R}\boldsymbol{\Lambda}\mathbf{R}^{\mathsf{T}}, \quad \mathbf{R} = \begin{bmatrix}\cos\theta & -\sin\theta \\ \sin\theta & \cos\theta \end{bmatrix}, \quad \boldsymbol{\Lambda} = \begin{bmatrix} \lambda_{x} & 0 \\ 0 & \lambda_{y} \end{bmatrix} , 
	\end{gather}
	where $0 < \theta < \pi$ is the (anti-clockwise) angle of rotation, and $\lambda_{x} > 0$ and $\lambda_{y} > 0$ are the eigenvalues.  This yields the following expressions for the diffusion tensor components:
	\begin{gather}\label{eq:D_components}
		D_{xx} = \lambda_{x}\cos^2\theta + \lambda_{y}\sin^2\theta, \quad D_{xy} = (\lambda_{x} - \lambda_{y})\sin\theta\cos\theta, \quad D_{yy} = \lambda_{x}\sin^2\theta + \lambda_{y}\cos^2\theta.
	\end{gather}
	For deterministic and random walk models implemented on the hexagonal lattices, there are no restrictions on $\lambda_{x}, \lambda_{y}$ or $\theta$, as implied by the condition $\det(\mathbf{D}) \geq 0$ (see sections \ref{sec:flat-top_probs} and \ref{sec:pointy-top_probs}). On the other hand, implementation of the models on a rectangular lattice can, depending on the choice of eigenvalues, introduce a restriction on the angle of rotation. Recalling the constraint $\det(\mathbf{D}) \geq 3D_{xy}^2$ on the diffusion tensor (see section \ref{sec:rec_probs}), substitution of the expressions in (\ref{eq:D_components}) into the constraint yields
	\begin{gather}\label{eq:sin_cond}
		|\sin(2\theta)| \leq \frac{2}{|\lambda_{x} - \lambda_{y}|}\sqrt{\frac{\lambda_{x}\lambda_{y}}{3}}, \quad \lambda_{x} \neq \lambda_{y}.
	\end{gather}
	For $\lambda_{y}/3 \leq \lambda_{x} \leq 3\lambda_{y}$, the condition (\ref{eq:sin_cond}) is satisfied for any choice of $\theta$, as the upper bound is greater than or equal to one (exceeds the maximum absolute value of $\sin(2\theta)$). However, for $\lambda_{x} < \lambda_{y}/3$ or $\lambda_{x} > 3\lambda_{y}$, the condition (\ref{eq:sin_cond}) yields the following constraint:
	\begin{gather}\label{eq:rec_theta}
		\theta \in (0,\theta_{\rm crit}) \cup (\pi/2 - \theta_{\rm crit},\pi/2 + \theta_{\rm crit}) \cup (\pi - \theta_{\rm crit},\pi), \quad \theta_{\rm crit} = \frac{1}{2}\arcsin\left(\frac{2}{|\lambda_{x}-\lambda_{y}|}\sqrt{\frac{\lambda_{x}\lambda_{y}}{3}}\right),
	\end{gather}
	where $0 < \theta_{\rm crit} \leq \pi/4$. The constraint (\ref{eq:rec_theta}) implies the range of suitable values for the angle of rotation depends on the choice of eigenvalues (see Figure \ref{fig:eigenvalue_sweep}). 
	
\begin{figure}[H]
		\begin{subfigure}{0.49\textwidth}
			\centering
			\includegraphics[width=\textwidth, trim=0cm -0.25cm 0cm 0.5cm, clip=true]{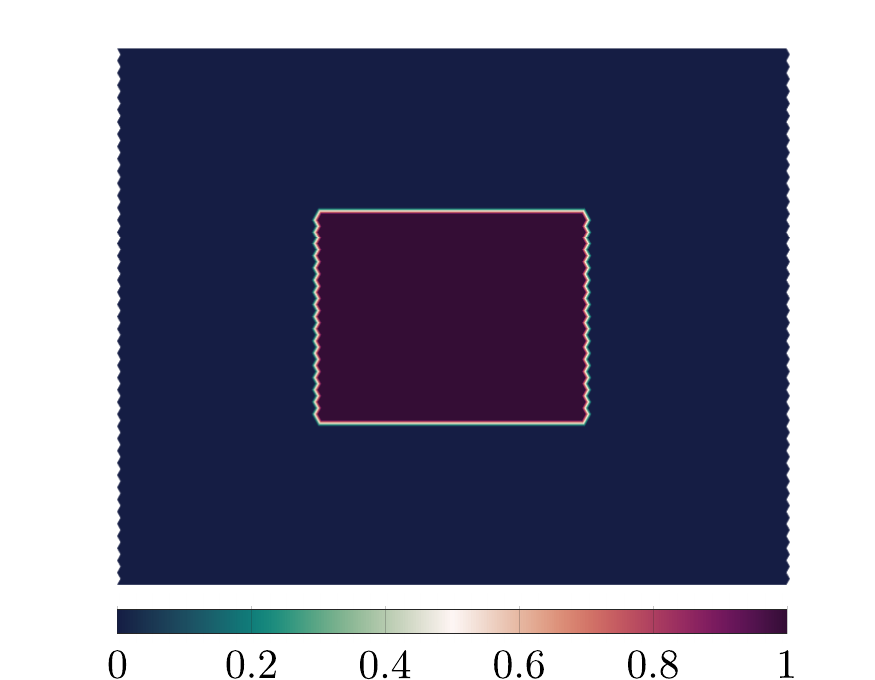}
			\caption{Simple initial condition}
			\label{fig:initial_simple}
		\end{subfigure}
		\hfill
		\begin{subfigure}{0.49\textwidth}
			\centering
			\includegraphics[width=\textwidth, trim=0cm -0.25cm 0cm 0.5cm, clip=true]{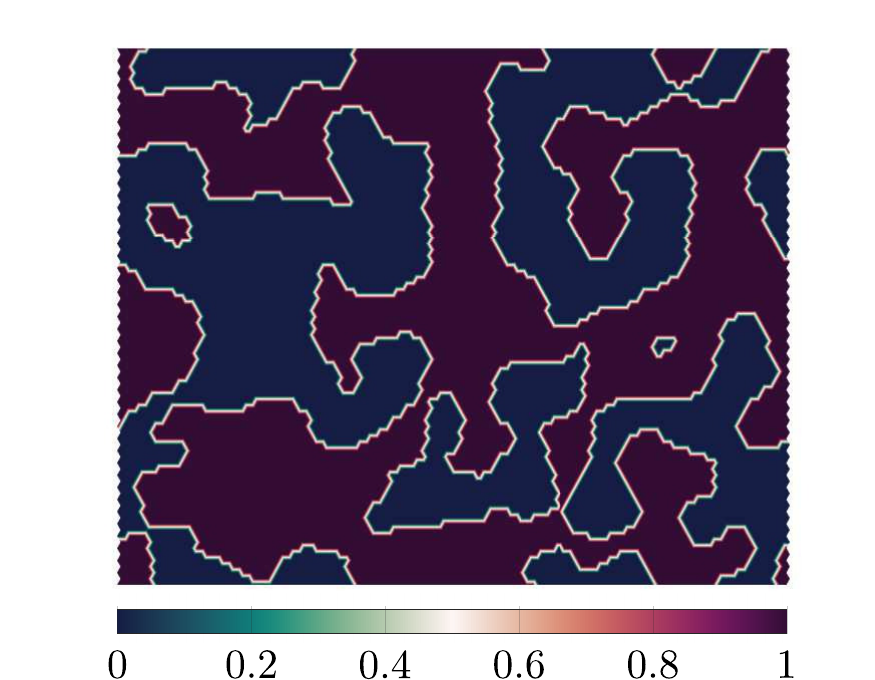}
			\caption{Complex initial condition}
			\label{fig:initial_complex}
		\end{subfigure}
		\caption{\textbf{Simple and complex initial conditions.} Two-dimensional (a) simple and (b) complex initial particle densities (at time $t = 0$) for the deterministic and random walk models implemented on a flat-top hexagonal lattice. The simple and complex initial conditions are defined in (\ref{eq:f_simple}) and (\ref{eq:f_complex}), respectively, where the latter utilises the density aggregation algorithm (\ref{eq:flat-top_complex}). Simulation results (two-dimensional model comparisons) using these initial conditions are presented in Figures \ref{fig:2D_flat-top_simple} and \ref{fig:2D_flat-top_complex}, respectively. Parameters: $L_{x} = 1$, $L_{y} = 0.8$, $N_{x} = 202$, $N_{y} = 89$, $\delta_{x} = 0.01$, $\delta_{y} \approx 0.0182$, $\kappa = 25$, $\gamma = 0.5$.}
		\label{fig:initial_conditions}
	\end{figure}

\begin{figure}[p!]
		\centering
		\includegraphics[width=\textwidth]{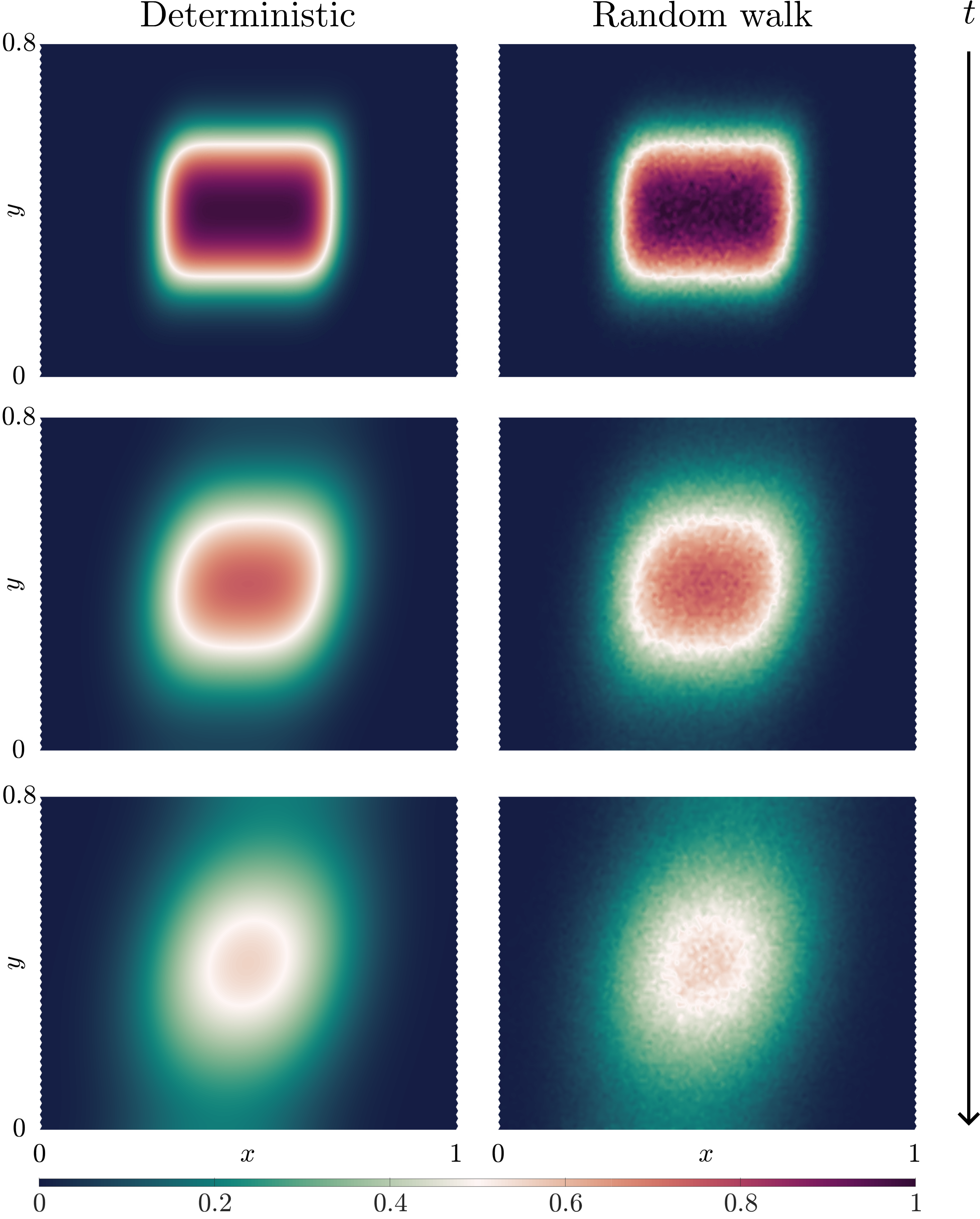}
		\caption{\textbf{Two-dimensional comparisons of anisotropic diffusion models (simple initial condition)}. Two-dimensional comparisons of particle density obtained from the deterministic (left column of panels) and random walk (right column of panels) models implemented on a flat-top hexagonal lattice. The solutions are plotted for times $t = 0.03, 0.13, 0.25$ (corresponding to each row of panels in descending order) where the deterministic particle densities are obtained from the solution of the spatially-discretised model (\ref{eq:density_ODEs}), and the stochastic particle densities are averaged across five simulations of the random walk model (see Algorithm \ref{alg:random_walk}). A simple (rectangular) initial condition (\ref{eq:f_simple}) is used: $f(\mathbf{x}_{i,j}) = 1$ if $3L_{x}/10 \leq x_{i} \leq 7L_{x}/10$ and $3L_{y}/10 \leq y_{j} \leq 7L_{y}/10$, otherwise $f(\mathbf{x}_{i,j}) = 0$. Parameters: $L_{x} = 1$, $L_{y} = 0.8$, $N_{x}=202$, $N_{y}=89$, $\delta_{x} = 0.01$, $\delta_{y} \approx 0.0182$, $N_{p}^{*}=449189$, $N_{s}=5$, $N_{t}=2332$, $T=1$, $\tau \approx 0.00043$, $\lambda_{x} = 0.08$, $\lambda_{y}=0.02$, $\theta = 5\pi/12$, $D_{xx} \approx 0.024$, $D_{xy}=0.015$ and $D_{yy} \approx 0.076$. Colour map sourced from \href{https://matplotlib.org/cmocean/}{https://matplotlib.org/cmocean/} \cite{thyng_et_al_2016}.}
		\label{fig:2D_flat-top_simple}
	\end{figure}

\begin{figure}[p!]
		\centering
		\includegraphics[width=\textwidth]{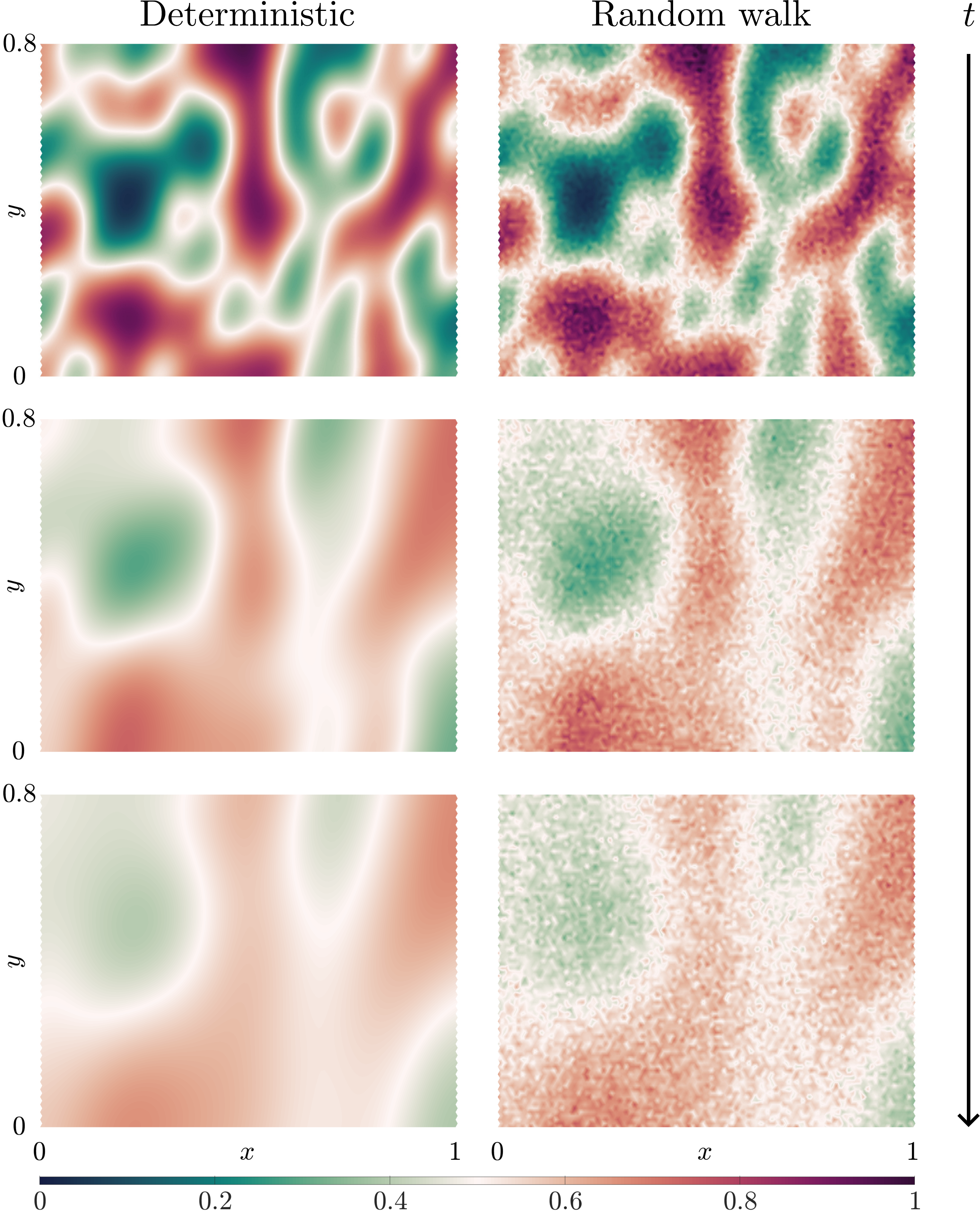}
		\caption{\textbf{Two-dimensional comparisons of anisotropic diffusion models (complex initial condition)}. Two-dimensional comparisons of particle density obtained from the deterministic (left column of panels) and random walk (right column of panels) models implemented on a pointy-top hexagonal lattice. The solutions are plotted for times $t = 0.03, 0.13, 0.25$ (corresponding to each row of panels in descending order) where the deterministic particle densities are obtained from the solution of the spatially-discretised model (\ref{eq:density_ODEs}), and the stochastic particle densities are averaged across five simulations of the random walk model (see Algorithm \ref{alg:random_walk}). A complex initial condition (\ref{eq:f_complex}) is used with 25 iterations of the density aggregation algorithm (\ref{eq:flat-top_complex}). Parameters: $L_{x} = 1$, $L_{y} = 0.8$, $N_{x}=202$, $N_{y}=89$, $\delta_{x} = 0.01$, $\delta_{y} \approx 0.0182$, $N_{p}^{*}=451197$, $N_{s}=5$, $N_{t}=2332$, $T=1$, $\tau \approx 0.00043$, $\lambda_{x} = 0.08$, $\lambda_{y}=0.02$, $\theta = 5\pi/12$, $D_{xx} \approx 0.024$, $D_{xy}=0.015$, $D_{yy} \approx 0.076$, $\kappa = 25$ and $\gamma = 0.5$. Colour map sourced from \href{https://matplotlib.org/cmocean/}{https://matplotlib.org/cmocean/} \cite{thyng_et_al_2016}.}
		\label{fig:2D_flat-top_complex}
	\end{figure}

	In this work, we compare the deterministic and random walk models for three prototype diffusion tensors generated using the eigenvalues $\lambda_{x} = 0.08$ and $\lambda_{y} = 0.02$ (labelled on Figure \ref{fig:eigenvalue_sweep}) and respective angles $\theta=\pi/24$, $\theta=5\pi/12$ and $\theta=3\pi/4$. Note that implementation of the models on a rectangular lattice is not suitable for the angle $\theta = 3\pi/4$ (see Figure \ref{fig:test_case_regions}), as it violates the constraint (\ref{eq:rec_theta}). Moreover, we compare the models for two types of initial conditions. We utilise a simple (rectangular) initial condition, 
	\begin{gather}\label{eq:f_simple}
	 	f_{i,j} = f(\mathbf{x}_{i,j}) = \begin{cases} 1, & \text{if}\hspace{0.15cm} 3L_{x}/10 \leq x_{i} \leq 7L_{x}/10 \hspace{0.15cm} \text{and} \hspace{0.15cm} 3L_{y}/10 \leq y_{j} \leq 7L_{y}/10, \\ 0, & \text{otherwise}, \end{cases}
	\end{gather}
	and construct a complex initial condition by generating random regions of aggregated particle density. Here, we generate a uniformly distributed random value (between zero and one) for each lattice site in the specified configuration, denoted by $f_{i,j}^{(0)} = f^{(0)}(\mathbf{x}_{i,j})$. We then aggregate particle density by performing iterations for a rectangular lattice \cite{march_et_al_2021},
	\begin{gather}\label{eq:rec_complex}
		f_{i,j}^{(k)} = \frac{4}{9}f_{i,j}^{(k-1)} + \frac{1}{9}\left(f_{i,j-1}^{(k-1)} + f_{i-1,j}^{(k-1)} + f_{i+1,j}^{(k-1)} + f_{i,j+1}^{(k-1)}\right) + \frac{1}{36}\left(f_{i-1,j-1}^{(k-1)} + f_{i+1,j-1}^{(k-1)} + f_{i-1,j+1}^{(k-1)} + f_{i+1,j+1}^{(k-1)}\right), 
	\end{gather}
	flat-top hexagonal lattice,
	\begin{gather}\label{eq:flat-top_complex}
		f_{i,j}^{(k)} = \frac{1}{2}f_{i,j}^{(k-1)} + \frac{1}{12}\left(f_{i-1,j-1}^{(k-1)} + f_{i+1,j-1}^{(k-1)} + f_{i-2,j}^{(k-1)} + f_{i+2,j}^{(k-1)} + f_{i-1,j+1}^{(k-1)}+ 	f_{i+1,j+1}^{(k-1)}\right),
	\end{gather}
	or pointy-top hexagonal lattice,
	\begin{gather}\label{eq:pointy-top_complex}
		f_{i,j}^{(k)} = \frac{1}{2}f_{i,j}^{(k-1)} + \frac{1}{12}\left(f_{i,j-2}^{(k-1)} + f_{i-1,j-1}^{(k-1)} + f_{i+1,j-1}^{(k-1)} + f_{i-1,j+1}^{(k-1)} + f_{i+1,j+1}^{(k-1)}+ 	f_{i,j+2}^{(k-1)}\right),
	\end{gather}
	where $f_{i,j}^{(k)} = f^{(k)}(\mathbf{x}_{i,j})$ for $k=1,\hdots,\kappa$. The weights in the aggregation algorithms (\ref{eq:rec_complex})--(\ref{eq:pointy-top_complex}) are inspired by those used in lattice Boltzmann methods \cite{blaak_sloot_2000}, and periodicity is assumed when aggregating densities on the domain boundaries. Finally, the particle density at each lattice site is assigned a value of zero or one depending on its magnitude,
	\begin{gather}\label{eq:f_complex}
		f_{i,j} = f(\mathbf{x}_{i,j}) = \begin{cases} 1, & \text{if}\hspace{0.15cm} f_{i,j}^{(\kappa)} > (1-\gamma)\max\left\{f_{i,j}^{(\kappa)}\right\} + \gamma\min\left\{f_{i,j}^{(\kappa)}\right\}, \\ 0, & \text{otherwise}, \end{cases}
	\end{gather}
	where $0 < \gamma < 1$. Example simple (\ref{eq:f_simple}) and complex (\ref{eq:f_complex}) initial conditions are presented in Figures \ref{fig:initial_simple} and \ref{fig:initial_complex}, respectively, for a flat-top hexagonal lattice.
	In Figures \ref{fig:2D_flat-top_simple} and \ref{fig:2D_flat-top_complex}, we present two-dimensional comparisons of the deterministic and random walk models implemented on a flat-top hexagonal lattice with simple and complex initial conditions, respectively. These results correspond to particle diffusion governed by a prototype tensor (\ref{eq:prototype_tensor}) with diffusivities $D_{xx} \approx 0.024$, $D_{xy} = 0.015$ and $D_{yy} \approx 0.076$, spatial steps $\delta_{x} = 0.01$ and $\delta_{y} \approx 0.0182$ and a time step duration of $\tau = 1/2332$ $(\approx 0.00043)$. These parameters collectively satisfy the constraints (\ref{eq:flat-top_time_cond}) and (\ref{eq:flat-top_interval_dy}) for $\tau$ and $\delta_{y}$, respectively, which ensure all transition probabilities are between zero and one. Furthermore, the choice for $\delta_{y}$ is the smallest value satisfying (\ref{eq:flat-top_interval_dy}) that divides evenly into $2L_{y}$ (see Appendix \ref{sec:appendix_alt_constraints}), and the number of time steps $N_{t}=2332$ gives the largest possible value for $\tau$ satisfying (\ref{eq:flat-top_time_cond}). For the random walk models, we specify a desired number of $N_{p} = 50N_{\ell} = 449450$ particles (prior to scaling) and average the particle densities obtained from five simulations of Algorithm \ref{alg:random_walk}. 
	\begin{figure}[p!]
		\centering
		\includegraphics[width=\textwidth]{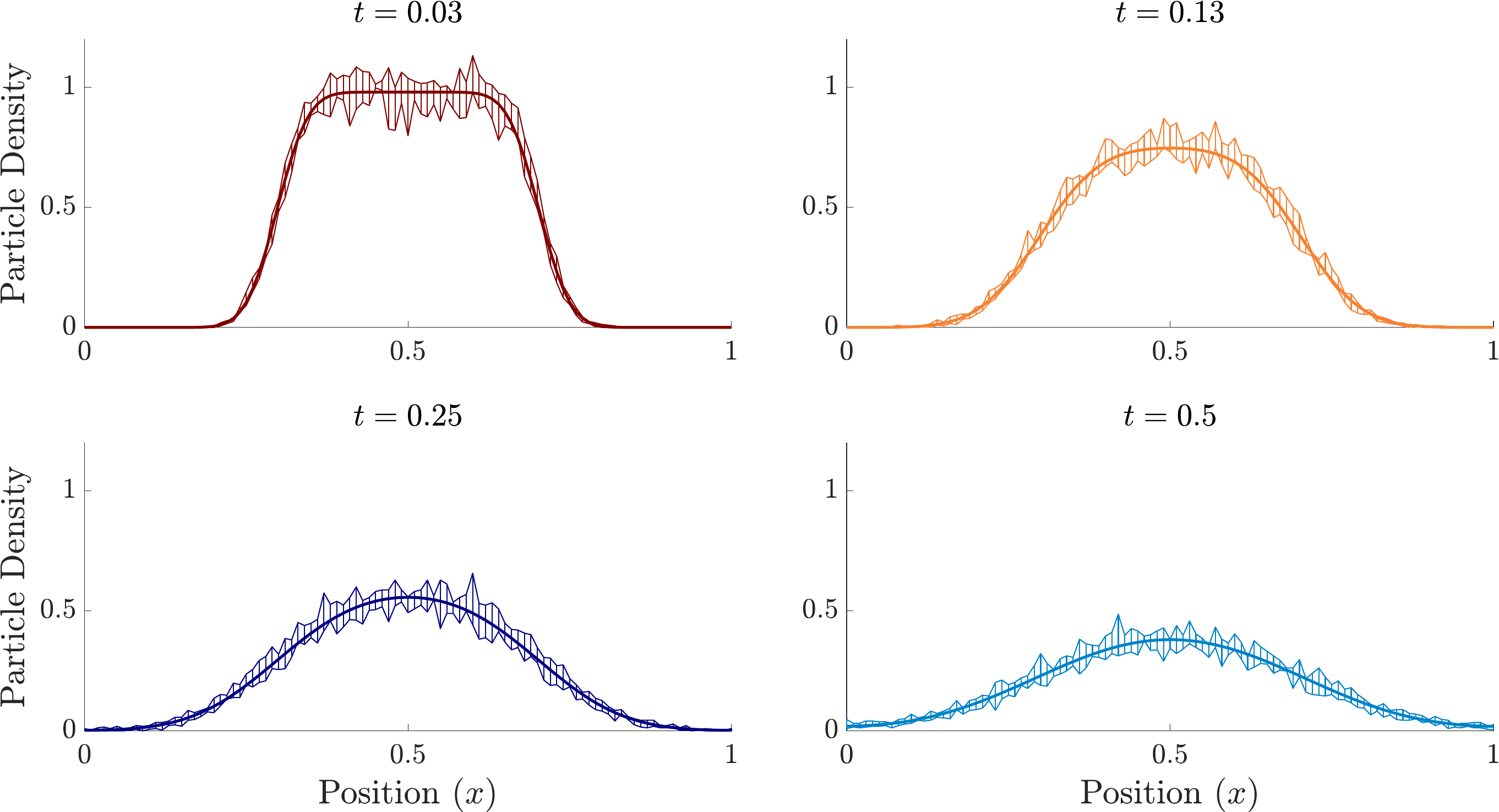}
		\caption{\textbf{One-dimensional comparisons of anisotropic diffusion models (simple initial condition)}. One-dimensional slices of particle density obtained from the deterministic (continuous line obtained from the solution of the spatially-discretised model (\ref{eq:density_ODEs})) and random walk (hatched regions bounded by 2.5\% and 97.5\% quantiles enveloping the particle density obtained from five simulations of Algorithm \ref{alg:random_walk}) models implemented on a flat-top hexagonal lattice. The solutions are plotted for times $t = 0.03, 0.13, 0.25, 0.5$ for the row of lattice sites located at $y=L_{y}/2$ and are obtained from the simulation results presented in Figure \ref{fig:2D_flat-top_simple}. A simple (rectangular) initial condition (\ref{eq:f_simple}) is used: $f(\mathbf{x}_{i,j}) = 1$ if $3L_{x}/10 \leq x_{i} \leq 7L_{x}/10$ and $3L_{y}/10 \leq y_{j} \leq 7L_{y}/10$, otherwise $f(\mathbf{x}_{i,j}) = 0$. Refer to Figure \ref{fig:2D_flat-top_simple} for relevant parameters.}
		\label{fig:1D_flat-top_simple}
		
		\vskip20pt%
		
		\includegraphics[width=\textwidth]{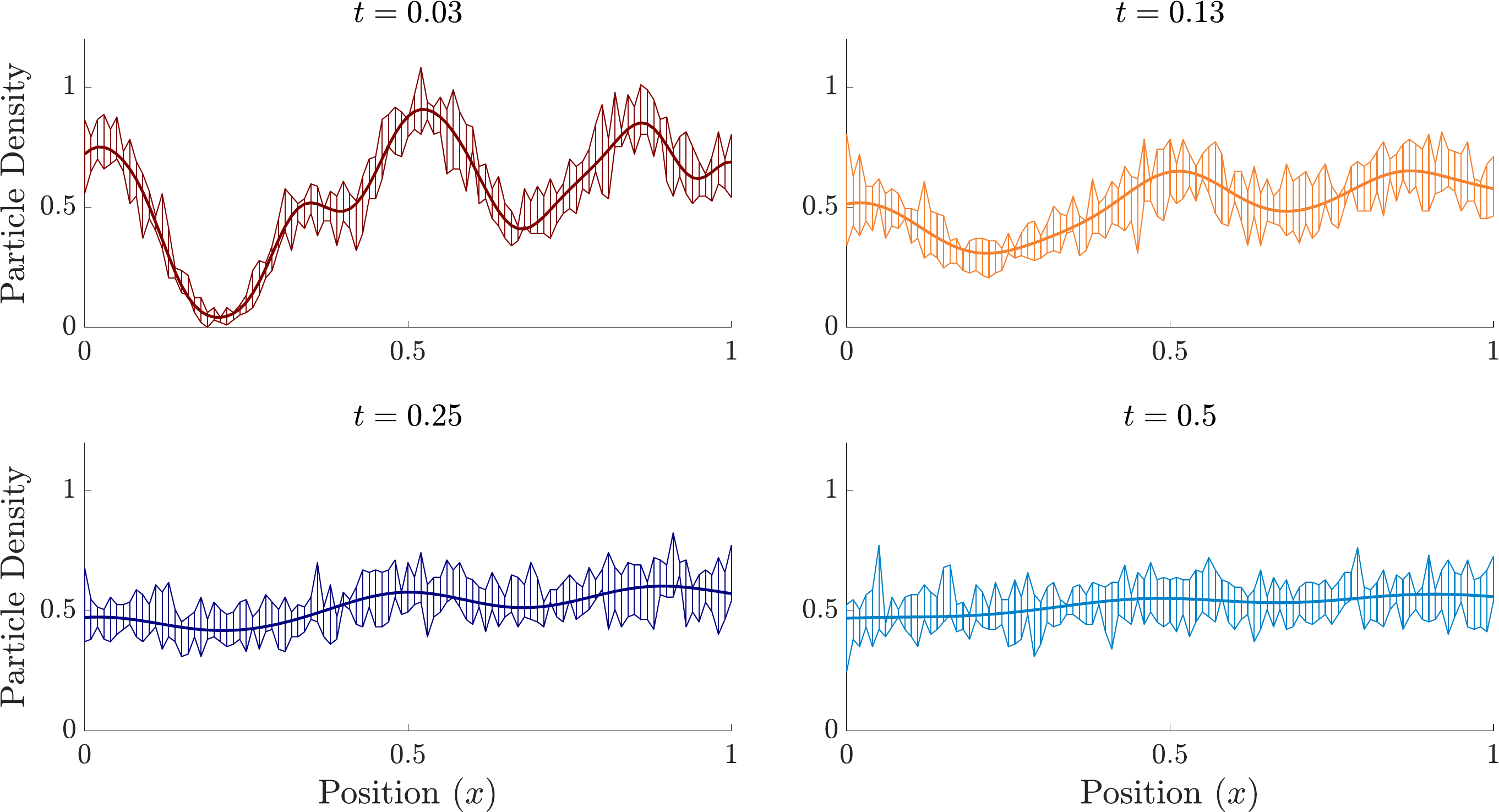}
		\caption{\textbf{One-dimensional comparisons of anisotropic diffusion models (complex initial condition)}. One-dimensional comparisons of particle density obtained from the deterministic (continuous line obtained from the solution of the spatially-discretised model (\ref{eq:density_ODEs})) and random walk (hatched regions bounded by 2.5\% and 97.5\% quantiles enveloping the particle density obtained from five simulations of Algorithm \ref{alg:random_walk}) models implemented on a flat-top hexagonal lattice. The solutions are plotted for times $t = 0.03, 0.13, 0.25, 0.5$ for the row of lattice sites located at $y=L_{y}/2$ and are obtained from the simulation results presented in Figure \ref{fig:2D_flat-top_complex}. A complex initial condition (\ref{eq:f_complex}) is used with 25 iterations of the density aggregation algorithm (\ref{eq:flat-top_complex}). Refer to Figure \ref{fig:2D_flat-top_complex} for relevant parameters.}
		\label{fig:1D_flat-top_complex}
	\end{figure}
	
	From these results, we observe clear agreement between the random walk simulations and deterministic model solution for both initial conditions. Furthermore, both models exhibit behaviour consistent with anisotropic diffusion described by the specified tensor. Particles are shown to diffuse more rapidly in the vertical direction as opposed to the horizontal direction, and there is clear positive correlation between the axial directions of diffusion (agrees with the physical interpretation of $D_{xx} < D_{yy}$ and $D_{xy} > 0$). This agreement can be further demonstrated by comparing the models for a one-dimensional slice (single row or column of lattice sites) of the domain. In Figures \ref{fig:1D_flat-top_simple} and \ref{fig:1D_flat-top_complex}, we compare the deterministic model with 2.5\% and 97.5\% quantiles obtained from the random walk simulations presented in Figures \ref{fig:2D_flat-top_simple} and \ref{fig:2D_flat-top_complex}, respectively, for the row of lattice sites located at $y = L_{y}/2=0.4$. 
	For both initial conditions, the deterministic model is consistently captured within these quantiles, thus demonstrating strong evidence of equivalence between the deterministic model and random walk simulations. Similar results (omitted from this work) are obtained for other parameter choices and lattice configurations.
	
	Finally, we provide quantitative evidence to support the visual comparisons presented in Figures \ref{fig:2D_flat-top_simple}--\ref{fig:1D_flat-top_complex} and validate the equivalence of the deterministic model and random walk simulations for other parameter choices and lattice configurations. The discrepancy between the solution of the spatially-discretised deterministic model (\ref{eq:density_ODEs}), obtained using a forward Euler discretisation in time, and random walk simulations is quantified using the mean-squared error (MSE),
	\begin{gather}\label{eq:MSE}
		{\rm MSE} = \frac{1}{N_{t}N_{\ell}}\sum_{n=1}^{N_{t}}\sum_{k=1}^{N_{\ell}}(U_{k,n}^{d} - U_{k,n}^{s})^2,
	\end{gather}
	where $U_{k,n}^{d}$ and $U_{k,n}^{s}$ represent the deterministic and random walk (averaged over $N_{s}$ simulations) particle densities at $\mathbf{X}_{k}$ and time $t_{n} = n\tau$. A smaller value of the MSE (\ref{eq:MSE}) is representative of closer agreement between the models. In Table \ref{tab:metrics}, we present the MSE between models implemented on the rectangular and hexagonal lattices for both types of initial conditions with particle diffusion described by the aforementioned prototype tensors. Recall that model implementation on a rectangular lattice is only suitable for two of the three prototype tensors ($\theta = \pi/24$ and $\theta = 5\pi/12$ only). For each test case, we perform simulations of the random walk model using four successive choices for the total number of particles (prior to scaling): $N_{p} = k N_{\ell}$ for $k = 10,25,50,100$. Additionally, we use the smallest suitable choices (satisfying the relevant conditions) for the spatial step $\delta_{x}$ or $\delta_{y}$ (see Appendix \ref{sec:appendix_alt_constraints}) and number of time steps $N_{t}$ (largest suitable $\tau$) for each test case. For the model comparisons shown in Figures \ref{fig:2D_flat-top_simple}--\ref{fig:1D_flat-top_complex} (flat-top hexagonal lattice), we highlight the corresponding entries for the MSE in Table \ref{tab:metrics} to provide a useful link between the visual observations and quantitative results. 
	\begin{table}[p!]	
		\caption{\textbf{Quantitative agreement between anisotropic diffusion models.} Quantitative agreement between the deterministic (solution of the spatially-discretised model (\ref{eq:density_ODEs})) and random walk (particle densities averaged over five simulations of Algorithm \ref{alg:random_walk}) models implemented on the (a) rectangular, (b) flat-top and (c) pointy-top lattices. The mean-squared error (\ref{eq:MSE}) is used to compare the models for simple (see expression (\ref{eq:f_simple})) and complex (see expressions (\ref{eq:rec_complex})--(\ref{eq:f_complex})) initial conditions and three prototype diffusion tensors (see expression (\ref{eq:prototype_tensor})) generated using eigenvalues $\lambda_{x}=0.08$ and $\lambda_{y}=0.02$ and respective angles $\theta=\pi/24$, $\theta=5\pi/12$ and $\theta=3\pi/4$. For each test case (specified initial condition and diffusion tensor), the results are presented for four choices of the number of particles $N_{p}$ (initialised as 10, 25, 50 and 100 times the number of lattice sites $N_{\ell}$ and adjusted according to Algorithm \ref{alg:random_walk}). Results in (a) for the rectangular lattice are omitted for the third prototype tensor, as the angle $\theta = 3\pi/4$ does not satisfy (\ref{eq:rec_theta}). The mean-squared error (\ref{eq:MSE}) for the results presented in Figures \ref{fig:2D_flat-top_simple}--\ref{fig:1D_flat-top_complex} are highlighted in (b). \textbf{Parameters (all configurations):} $L_{x} = 1$, $L_{y} = 0.8$, $\lambda_{x} = 0.08$, $\lambda_{y}=0.02$, $T=1$, $N_{s} = 5$, $\kappa=25$, $\gamma = 0.5$. \textbf{Angle 1:} $\theta=\pi/24$, $D_{xx} \approx 0.079$, $D_{xy} \approx 0.008$, $D_{yy} \approx 0.021$. \textbf{Angle 2:} $\theta = 5\pi/12$, $D_{xx} \approx 0.024$, $D_{xy} = 0.015$, $D_{yy} \approx 0.076$. \textbf{Angle 3:} $\theta = 3\pi/4$, $D_{xx} = 0.05$, $D_{xy} = -0.03$, $D_{yy} = 0.05$.}
		\label{tab:metrics}
		\renewcommand{\arraystretch}{1.15}
		\sbox0{$4.97 \times 10^{-4}$}
		\sbox1{1,131,680}
		\begin{center}
			\begin{tabular}{@{}l c r@{}}
				\hline \multicolumn{1}{c}{Angle 1: $\theta = \pi/24$} & Angle 2: $\theta = 5\pi/12$ & \multicolumn{1}{c}{Angle 3: $\theta = 3\pi/4$} \\ \hline & & \\[-10pt] \hline
				\begin{tabular}{c c c} a) & MSE & $N_{p}^{*}$ \\[1pt] \hline \multirow{4}*{\begin{turn}{90}\large Simple\end{turn}} & $4.97 \times 10^{-4}$ & 112,640 \\ & $1.96 \times 10^{-4}$ & 283,360 \\ & $9.91 \times 10^{-5}$ & 564,960 \\ & $4.96 \times 10^{-5}$ & 1,131,680 \\[-13pt] & & \end{tabular} & \begin{tabular}{c c} MSE & $N_{p}^{*}$ \\[1pt] \hline $5.04 \times 10^{-4}$ & 78,720 \\ $2.01 \times 10^{-4}$ & 196,800 \\ $1.01 \times 10^{-4}$ & 393,600 \\ $5.03 \times 10^{-5}$ & 787,200 \\[-13pt] & \end{tabular} & \begin{tabular}{c c} MSE & $N_{p}^{*}$ \\[1pt] \hline \makebox[\wd0]{\makebox[0.6\wd0]{\xrfill[3pt]{0.4pt}}} & \makebox[\wd1]{\makebox[0.6\wd1]{\xrfill[3pt]{0.4pt}}} \\ \makebox[\wd0]{\makebox[0.6\wd0]{\xrfill[3pt]{0.4pt}}} & \makebox[\wd1]{\makebox[0.6\wd1]{\xrfill[3pt]{0.4pt}}} \\ \makebox[\wd0]{\makebox[0.6\wd0]{\xrfill[3pt]{0.4pt}}} & \makebox[\wd1]{\makebox[0.6\wd1]{\xrfill[3pt]{0.4pt}}} \\ \makebox[\wd0]{\makebox[0.6\wd0]{\xrfill[3pt]{0.4pt}}} & \makebox[\wd1]{\makebox[0.6\wd1]{\xrfill[3pt]{0.4pt}}} \\[-13pt] & \end{tabular}\\
				\hskip1pt \begin{tabular}{c c c} \hline & & \\[-13pt] \multirow{4}*{\begin{turn}{90}\large Complex\end{turn}} & $2.32 \times 10^{-3}$ & 111,840 \\ & $9.15 \times 10^{-4}$ & 283,328 \\ & $4.58 \times 10^{-4}$ & 566,656 \\ & $2.30 \times 10^{-4}$ & 1,129,647 \end{tabular} & \begin{tabular}{c c} \hline & \\[-13pt] $4.77 \times 10^{-3}$ & 77,970 \\ $1.90 \times 10^{-3}$ & 196,658 \\ $9.47 \times 10^{-4}$ & 393,154 \\ $4.73 \times 10^{-4}$ & 786,308 \end{tabular} & \begin{tabular}{c c} \hline & \\[-13pt] \makebox[\wd0]{\makebox[0.6\wd0]{\xrfill[3pt]{0.4pt}}} & \makebox[\wd1]{\makebox[0.6\wd1]{\xrfill[3pt]{0.4pt}}} \\ \makebox[\wd0]{\makebox[0.6\wd0]{\xrfill[3pt]{0.4pt}}} & \makebox[\wd1]{\makebox[0.6\wd1]{\xrfill[3pt]{0.4pt}}} \\ \makebox[\wd0]{\makebox[0.6\wd0]{\xrfill[3pt]{0.4pt}}} & \makebox[\wd1]{\makebox[0.6\wd1]{\xrfill[3pt]{0.4pt}}} \\ \makebox[\wd0]{\makebox[0.6\wd0]{\xrfill[3pt]{0.4pt}}} & \makebox[\wd1]{\makebox[0.6\wd1]{\xrfill[3pt]{0.4pt}}} \end{tabular} \\
				\hline \multicolumn{3}{l}{\footnotesize \textbf{Parameters (rectangular):} Angle 1: $\delta_{x} \approx 0.009$, $\delta_{y} = 0.008$, $\tau \approx 0.00046$. Angle 2: $\delta_{x} = 0.01$, } \\[-4pt] \multicolumn{3}{l}{\footnotesize $\delta_{y} \approx 0.0104$, $\tau \approx 0.00059$.} 
			\end{tabular}
		\end{center}
		\begin{center}
			\begin{tabular}{@{}l c r@{}}
				\hline \begin{tabular}{c c c} b) & MSE & $N_{p}^{*}$ \\[1pt] \hline \multirow{4}*{\begin{turn}{90}\large Simple\end{turn}} & $5.01 \times 10^{-4}$ & 172,800 \\ & $2.02 \times 10^{-4}$ & 429,300 \\ & $1.01 \times 10^{-4}$ & 858,600 \\ & $5.04 \times 10^{-5}$ & 1,717,200 \\[-13pt] & & \end{tabular} & \begin{tabular}{c c} MSE & $N_{p}^{*}$ \\[1pt] \hline $5.14 \times 10^{-4}$ & 89,271 \\ $2.04 \times 10^{-4}$ & 225,303 \\ $\bf{1.02 \times 10^{-4}}$ & \textbf{449,189} \\ $5.11 \times 10^{-5}$ & 898,378 \\[-13pt] & \end{tabular} & \begin{tabular}{c c} MSE & $N_{p}^{*}$ \\[1pt] \hline $5.16 \times 10^{-4}$ & 160,776 \\ $2.07 \times 10^{-4}$ & 400,664 \\ $1.03 \times 10^{-4}$ & 803,880 \\ $5.16 \times 10^{-5}$ & 1,605,208 \\[-13pt] & \end{tabular} \\ \hskip1pt \begin{tabular}{c c c} \hline & & \\[-13pt] \multirow{4}*{\begin{turn}{90}\large Complex\end{turn}} & $3.14 \times 10^{-3}$ & 174,144 \\ & $1.28 \times 10^{-3}$ & 428,661 \\ & $6.39 \times 10^{-4}$ & 857,323 \\ & $3.19 \times 10^{-4}$ & 1,714,645 \\[-13pt] & & \end{tabular} & \begin{tabular}{c c} \hline & \\[-13pt] $5.82 \times 10^{-3}$ & 88,439 \\ $2.31 \times 10^{-3}$ & 223,320 \\ $\bf{1.14 \times 10^{-3}}$ & \textbf{451,197} \\ $5.73 \times 10^{-4}$ & 897,974 \\[-13pt] & \end{tabular} & \begin{tabular}{c c} \hline & \\[-13pt] $8.01 \times 10^{-3}$ & 156,315 \\ $3.13 \times 10^{-3}$ & 400,615 \\ $1.56 \times 10^{-3}$ & 801,113 \\ $7.81 \times 10^{-4}$ & 1,602,323 \\[-13pt] & \end{tabular}\\
				\hline \multicolumn{3}{l}{\footnotesize \textbf{Parameters (flat-top):} Angle 1: $\delta_{x} \approx 0.0059$, $\delta_{y} = 0.016$, $\tau \approx 0.00013$. Angle 2: $\delta_{x} = 0.01$,} \\[-4pt] \multicolumn{3}{l}{\footnotesize $\delta_{y} \approx 0.0182$, $\tau \approx 0.00043$. Angle 3: $\delta_{x} = 0.01$, $\delta_{y} \approx 0.0101$, $\tau \approx 0.00019$.}
			\end{tabular}
		\end{center}
		\begin{center}
			\begin{tabular}{@{}l c r@{}}
				\hline \begin{tabular}{c c c} c) & MSE & $N_{p}^{*}$ \\[1pt] \hline \multirow{4}*{\begin{turn}{90}\large Simple\end{turn}} & $5.00 \times 10^{-4}$ & 130,560 \\ & $2.00 \times 10^{-4}$ & 326,400 \\ & $1.00 \times 10^{-4}$ & 650,760 \\ & $5.01 \times 10^{-5}$ & 1,303,560 \\[-13pt] & & \end{tabular} & \begin{tabular}{c c} MSE & $N_{p}^{*}$ \\[1pt] \hline $5.27 \times 10^{-4}$ & 65,835 \\ $2.11 \times 10^{-4}$ & 164,065 \\ $1.06 \times 10^{-4}$ & 328,130 \\ $5.28 \times 10^{-5}$ & 656,260 \\[-13pt] & \end{tabular} & \begin{tabular}{c c} MSE & $N_{p}^{*}$ \\[1pt] \hline $5.37 \times 10^{-4}$ & 67,394 \\ $2.14 \times 10^{-4}$ & 169,572 \\ $1.07 \times 10^{-4}$ & 338,057 \\ $5.34 \times 10^{-5}$ & \makebox[\wd1]{677,201} \\[-13pt] & \end{tabular} \\ \hskip1pt \begin{tabular}{c c c} \hline & & \\[-13pt] \multirow{4}*{\begin{turn}{90}\large Complex\end{turn}} & $8.02 \times 10^{-3}$ & 126,765 \\ & $3.13 \times 10^{-3}$ & 324,882 \\ & $1.57 \times 10^{-3}$ & 649,672 \\ & $7.83 \times 10^{-4}$ & 1,299,391 \\[-13pt] & & \end{tabular} & \begin{tabular}{c c} \hline & \\[-13pt] $3.72 \times 10^{-3}$ & 65,044 \\ $1.46 \times 10^{-3}$ & 165,257 \\ $7.38 \times 10^{-4}$ & 327,948 \\ $3.69 \times 10^{-4}$ & 655,860 \\[-13pt] & \end{tabular} & \begin{tabular}{c c} \hline & \\[-13pt] $8.36 \times 10^{-3}$ & 67,003 \\ $3.34 \times 10^{-3}$ & 167,507 \\ $1.65 \times 10^{-3}$ & 339,160 \\ $8.26 \times 10^{-4}$ & \makebox[\wd1]{678,402} \\[-13pt] & \end{tabular}\\ \hline \multicolumn{3}{l}{\footnotesize \textbf{Parameters (pointy-top):} Angle 1: $\delta_{x} \approx 0.0156$, $\delta_{y} = 0.008$, $\tau \approx 0.00032$. Angle 2: $\delta_{x} = 0.02$,} \\[-4pt] \multicolumn{3}{l}{\footnotesize $\delta_{y} = 0.0125$, $\tau \approx 0.0005$. Angle 3: $\delta_{x} = 0.02$, $\delta_{y} \approx 0.012$, $\tau = 0.00047$.}
			\end{tabular}
		\end{center}
	\end{table}
	
	Generally, the MSE (\ref{eq:MSE}) is small and reduces as the number of particles increases. For the simple initial condition (\ref{eq:f_simple}), these values are consistent (or similar) when comparing results between diffusion tensors and lattice configurations. This trend is not observed when assuming a complex initial condition (\ref{eq:f_complex}), but this is due to assuming a new aggregation of particle density for each test case. This is a requirement when changing the diffusion tensor (different number of lattice sites) or lattice configuration (alternative density aggregation algorithm (\ref{eq:rec_complex}), (\ref{eq:flat-top_complex}) or (\ref{eq:pointy-top_complex})), whereas the simple initial condition (\ref{eq:f_simple}) is applied across all test cases without modification. 
	
	We also observe that simulations of the random walk model on a rectangular lattice show consistent overall agreement (similar MSE) with the deterministic model solution for a smaller number of particles in comparison to implementation on the hexagonal lattices. This is attributed to the different functions used for interpolating flux terms in the spatial discretisation of the deterministic model (see section \ref{sec:deterministic}). The bilinear interpolating function (\ref{eq:rec_interpolant}) employed for structured rectangular elements (rectangular lattice) has an additional term in comparison to the linear function (\ref{eq:hex_interpolant}) used for structured triangular elements (hexagonal lattices). Additionally, the implementation process is simpler on a rectangular lattice. Thus, it is more advantageous to implement the deterministic and random walk models on a rectangular lattice if the specified diffusion tensor satisfies the constraint $\det(\mathbf{D}) \geq 3D_{xy}^2$.
	
	If a rectangular lattice is not suitable (diffusion tensor does not satisfy $\det(\mathbf{D}) \geq 3D_{xy}^2$), either of the hexagonal lattices can be used given their suitability for any valid diffusion tensor. Interestingly, implementation of the deterministic and random walk models on a pointy-top hexagonal lattice, as opposed to a flat-top hexagonal lattice, yields consistent overall agreement (similar MSE) between the models for a smaller number of particles. This holds across all three prototype tensors, suggesting that implementation on a pointy-top hexagonal lattice is preferable (in terms of obtaining sufficient model agreement) over a flat-top hexagonal lattice for diffusion tensors with eigenvalues satisfying $\lambda_{x} > \lambda_{y}$. Quantitative results for $\lambda_{y} > \lambda_{x}$ (omitted from this work) show that a flat-top hexagonal lattice is more suitable for that case. Regardless, the quantitative results presented in Table \ref{tab:metrics}, in addition to the visual comparisons presented in Figures \ref{fig:2D_flat-top_simple}--\ref{fig:1D_flat-top_complex}, provide strong evidence to support the equivalence of the deterministic and random walk models presented in this work. MATLAB code available on GitHub (\href{https://github.com/lukefilippini/Filippini2025}{https://github.com/lukefilippini/Filippini2025}) can be used to compare the models for other parameter choices and view more informative animations of the results.
	
	\section{Conclusions}\label{sec:conclusions}
	In summary, we derived a set of equivalent random walk models for the deterministic model (\ref{eq:diffusion_eq})--(\ref{eq:D_tensor}) in a two-dimensional domain with no-flux boundary conditions and a spatially-invariant diffusion tensor. Our approach involved discretising the deterministic model in space to give a homogeneous Markov chain governing the movement of particles between (spatial) lattice sites over each time step. The spatial discretisation was carried out using a vertex-centred element-based finite volume method on rectangular and hexagonal lattices, and a forward Euler discretisation was applied in time to give a stochastic matrix governing the movement of particles in a nearest-neighbour random walk. This time discretisation method gave simple analytical expressions for the transition probabilities that facilitated direct insight into the conditions on model parameters (spatial steps, time step duration and diffusion tensor) required to ensure all probabilities were between zero and one. For each lattice configuration, results (visual and quantitative) demonstrated that simulations of the random walk model matched well with the deterministic model solution.
	
	The rectangular and hexagonal lattices have comparable advantages and disadvantages in terms of their suitability for model implementation. A rectangular lattice allows for simpler implementation and offers similar overall (mean-squared error) agreement between the deterministic model solution and random walk simulations for a smaller number of particles (for the spatial discretisation methods applied in this work). However, while this approach is appropriate for isotropic or orthotropic diffusion, the rectangular configuration is only suitable (valid transition probabilities) for a restricted range of diffusion tensors. Implementation of the models on a hexagonal (flat-top or pointy-top) lattice overcomes this limitation, as the conditions required to ensure each transition probability is between zero and one can be satisfied for any valid diffusion tensor.
	
	Our approach has not been previously applied to anisotropic diffusion and would be of interest to a cross-disciplinary audience, primarily because it yields simple random walk models that readily facilitate analytical insight into the effect of any spatially-invariant diffusion tensor on particle transport. Although, we acknowledge that the equivalent random walk models outlined in this work are limited to the specific anisotropic diffusion model considered. Different transition probabilities would be obtained if this analysis was extended to non-uniform spatial steps, other transport models, lattice-free random walk models or a spatially-varying diffusion tensor. The latter scenario is of particular interest to pursue in the future.
	
	\section{CRediT authorship statement}
	\textbf{Luke P. Filippini:} Conceptualization, Formal analysis, Funding acquisition, Investigation, Methodology, Project administration, Resources, Software, Validation, Visualisation, Writing - original draft, Writing - review and editing. \textbf{Adrianne L. Jenner:} Conceptualization, Funding acquisition, Methodology, Project administration, Resources, Supervision, Writing - review and editing. \textbf{Elliot J. Carr:} Conceptualization, Formal analysis, Funding acquisition, Methodology, Project administration, Resources, Supervision, Writing - review and editing.
	
	\section{Data statement}
	MATLAB code is provided on GitHub (\href{https://github.com/lukefilippini/Filippini2025}{https://github.com/lukefilippini/Filippini2025}) for reproducing the results in this paper and testing alternative parameter choices.
	
	\section{Acknowledgments}
	L. P. F. completed this work as part of his Doctor of Philosophy (PhD) project and acknowledges funding support from an Australian Research Training Program (RTP) scholarship: \href{https://doi.org/10.82133/C42F-K220}{doi.org/10.82133/C42F-K220}. Additionally, he received top-up funding from the Computational Bioimaging Group (CBG) within the School of Mathematical Sciences at Queensland University of Technology (QUT). A. L. J. acknowledges support from the Australian Research Council (ARC) Discovery Project (DP) DP230100025 and the ARC Discovery Early Career Researcher Award (DECRA) DE240100650.
	
	\appendix
	\section{Appendix}
	
	\subsection{Coefficients for interpolating functions}\label{sec:appendix_interp}
	In this appendix, we provide further details in regards to the interpolating functions $g_{m}(\mathbf{x})$ used to approximate the flux terms $\mathbf{q}(\bar{\mathbf{x}}_{\sigma},t)\cdot\hat{\mathbf{n}}_{\sigma}$ in the EbFVM discretisation of the deterministic model (\ref{eq:diffusion_eq})--(\ref{eq:D_tensor}) (see section \ref{sec:deterministic}). Recall that, within each element $E_{m}$, we calculate the flux $\mathbf{q}(\bar{\mathbf{x}}_{\sigma},t)\cdot\hat{\mathbf{n}}_{\sigma}$ under the assumption that the particle density $u(\mathbf{x},t)$ varies linearly or bilinearly in space (within $E_{m}$). This is achieved using the interpolating functions (\ref{eq:rec_interpolant}) and (\ref{eq:hex_interpolant}), given by
	\begin{gather*}
		g_{m}(\mathbf{x}) = \alpha_{m,1}x + \alpha_{m,2}y + \alpha_{m,3}xy + \alpha_{m,4},
	\end{gather*}
	and
	\begin{gather*}
		g_{m}(\mathbf{x}) = \alpha_{m,1}x + \alpha_{m,2}y + \alpha_{m,3},
	\end{gather*}
	for rectangular (bilinear) and triangular (linear) elements, respectively, where the coefficients, which enforce the equality of $g_{m}(\mathbf{x})$ and $u(\mathbf{x},t)$ at the element vertices, are well-known \cite{oliver_et_al_2025,sheikholeslami_2019,voller_2009} and are reproduced in this appendix. Consider an element $E_{m}$ defined, in an anti-clockwise manner, by the arbitrary sites $\tilde{\mathbf{x}}_{m,i} = (\tilde{x}_{m,i},\tilde{y}_{m,i})$ for $i=1,\hdots,N_{m}$, where $N_{m} = 4$ and $N_{m} = 3$ for rectangular and triangular elements, respectively. We obtain a linear system $\mathbf{A}_{m}\boldsymbol{\alpha}_{m} = \mathbf{b}_{m}$ for the coefficients, where
	\begin{gather*}
		\mathbf{A}_{m} = \begin{bmatrix}\tilde{x}_{m,1} & \tilde{y}_{m,1} & \tilde{x}_{m,1}\tilde{y}_{m,1} & 1 \\ \tilde{x}_{m,2} & \tilde{y}_{m,2} & \tilde{x}_{m,2}\tilde{y}_{m,2} & 1 \\ \tilde{x}_{m,3} & \tilde{y}_{m,3} & \tilde{x}_{m,3}\tilde{y}_{m,3} & 1 \\ \tilde{x}_{m,4} & \tilde{y}_{m,4} & \tilde{x}_{m,4}\tilde{y}_{m,4} & 1\end{bmatrix}, \quad \boldsymbol{\alpha}_{m} = \begin{bmatrix}\alpha_{m,1} \\ \alpha_{m,2} \\ \alpha_{m,3} \\ \alpha_{m,4}\end{bmatrix}, \quad \mathbf{b}_{m} = \begin{bmatrix}\tilde{u}_{m,1} \\ \tilde{u}_{m,2} \\ \tilde{u}_{m,3} \\ \tilde{u}_{m,4}\end{bmatrix},
	\end{gather*}
	or
	\begin{gather*}
		\mathbf{A}_{m} = \begin{bmatrix}\tilde{x}_{m,1} & \tilde{y}_{m,1} & 1 \\ \tilde{x}_{m,2} & \tilde{y}_{m,2} & 1 \\ \tilde{x}_{m,3} & \tilde{y}_{m,3} & 1\end{bmatrix}, \quad \boldsymbol{\alpha}_{m} = \begin{bmatrix}\alpha_{m,1} \\ \alpha_{m,2} \\ \alpha_{m,3}\end{bmatrix}, \quad \mathbf{b}_{m} = \begin{bmatrix}\tilde{u}_{m,1} \\ \tilde{u}_{m,2} \\ \tilde{u}_{m,3}\end{bmatrix},
	\end{gather*}
	noting $\tilde{u}_{m,i} \approx u(\tilde{\mathbf{x}}_{m,i},t)$. These linear systems yield the expressions
	\begin{gather*}
		\boldsymbol{\alpha}_{m} = \begin{bmatrix}s_{m,1}\tilde{u}_{m,1} + s_{m,2}\tilde{u}_{m,2} + s_{m,3}\tilde{u}_{m,3} + s_{m,4}\tilde{u}_{m,4} \\ s_{m,5}\tilde{u}_{m,1} + s_{m,6}\tilde{u}_{m,2} + s_{m,7}\tilde{u}_{m,3} + s_{m,8}\tilde{u}_{m,4} \\ s_{m,9}\tilde{u}_{m,1} + s_{m,10}\tilde{u}_{m,2} + s_{m,11}\tilde{u}_{m,3} + s_{m,12}\tilde{u}_{m,4} \\ s_{m,13}\tilde{u}_{m,1} + s_{m,14}\tilde{u}_{m,2} + s_{m,15}\tilde{u}_{m,3} + s_{m,16}\tilde{u}_{m,4} \end{bmatrix},
	\end{gather*}
	or
	\begin{gather*}
		\boldsymbol{\alpha}_{m} = \begin{bmatrix}s_{m,1}\tilde{u}_{m,1} + s_{m,2}\tilde{u}_{m,2} + s_{m,3}\tilde{u}_{m,3} \\ s_{m,4}\tilde{u}_{m,1} + s_{m,5}\tilde{u}_{m,2} + s_{m,6}\tilde{u}_{m,3} \\ s_{m,7}\tilde{u}_{m,1} + s_{m,8}\tilde{u}_{m,2} + s_{m,9}\tilde{u}_{m,3} \end{bmatrix},
	\end{gather*}
	for the coefficients $\alpha_{m,i}$ of (\ref{eq:rec_interpolant}) and (\ref{eq:hex_interpolant}), respectively. The shape coefficients $s_{m,j}$ are given by Cramer's rule,
	\begin{gather*}
		s_{m,j} = \frac{\det(\mathbf{A}_{m,j})}{\det(\mathbf{A}_{m})},
	\end{gather*}
	for $j=1,2,\cdots,N_{m}^2$, where the matrix $\mathbf{A}_{m,j}$ is obtained by replacing column $\lfloor (j-1)/N_{m} \rfloor + 1$ of the coefficient matrix $\mathbf{A}$ with column $\text{mod}(j-1,N_{m}) + 1$ of the $N_{m}\times N_{m}$ identity matrix, noting that $\lfloor x \rfloor$ is the floor function.

	\subsection{Conditions on number of spatial intervals}\label{sec:appendix_alt_constraints}
	In this appendix, we provide alternative representations of the constraints on the spatial steps $\delta_{y}$ and $\delta_{x}$ (see sections \ref{sec:rec_probs}--\ref{sec:pointy-top_probs}) to simplify implementation of the deterministic and random walk models. To elaborate, we modify the existing constraints to give analogous conditions on the number of spatial intervals (between lattice sites) in the horizontal or vertical direction, denoted by $I_{x}$ and $I_{y}$, respectively. We utilise the expressions $\delta_{x}=L_{x}/I_{x}$ and $\delta_{y}=L_{y}/I_{y}$ for the spatial steps, and rearrange the existing constraints to give $I_{y}^{\min} \leq I_{y} \leq I_{y}^{\max}$ or $I_{x}^{\min} \leq I_{x} \leq I_{x}^{\max}$, where
	\begin{gather*}
		I_{y}^{\min} = \frac{I_{x}L_{y}}{L_{x}} \cdot \begin{cases}\sqrt{D_{xx}/(3D_{yy})}, & \text{if rectangular}, \\ |D_{xy}|/D_{yy}, & \text{if flat-top}, \\ \sqrt{D_{xx}/D_{yy}}, & \text{if pointy-top},\end{cases} \quad I_{y}^{\max} = \frac{I_{x}L_{y}}{L_{x}} \cdot \begin{cases}\sqrt{3D_{xx}/D_{yy}}, & \text{if rectangular}, \\ \sqrt{D_{xx}/D_{yy}}, & \text{if flat-top}, \\ D_{xx}/|D_{xy}|, & \text{if pointy-top},\end{cases}
	\end{gather*}
	and
	\begin{gather*}
		I_{x}^{\min} = \frac{I_{y}L_{x}}{L_{y}} \cdot \begin{cases}\sqrt{D_{yy}/(3D_{xx})}, & \text{if rectangular}, \\ \sqrt{D_{yy}/D_{xx}}, & \text{if flat-top}, \\ |D_{xy}|/D_{xx}, & \text{if pointy-top},\end{cases} \quad I_{x}^{\max} = \frac{I_{y}L_{x}}{L_{y}} \cdot \begin{cases}\sqrt{3D_{yy}/D_{xx}}, & \text{if rectangular}, \\ D_{yy}/|D_{xy}|, & \text{if flat-top}, \\ \sqrt{D_{yy}/D_{xx}}, & \text{if pointy-top}.\end{cases}
	\end{gather*}
	The results presented in this work are generated using the largest suitable choice for $I_{y}$ or $I_{x}$, (smallest allowable $\delta_{y}$ or $\delta_{x}$) given by $I_{y} = \lfloor I_{y}^{\max} \rfloor$ or $I_{x} = \lfloor I_{x}^{\max} \rfloor$, respectively, where $\lfloor x \rfloor$ is the floor function. The number of lattice sites in the $x$ and $y$ directions can then be defined as 
	\begin{gather*}
		N_{x} = \begin{cases} I_{x} + 1, & \text{if rectangular}, \\ 2(I_{x} + 1), & \text{if flat-top}, \\ 2I_{x} + 1, & \text{if pointy-top}, \end{cases} \quad N_{y} = \begin{cases}I_{y} + 1, & \text{if rectangular}, \\ 2I_{y} + 1, & \text{if flat-top}, \\2(I_{y} + 1), & \text{if pointy-top}, \end{cases}
	\end{gather*}
	which ensures that $N_{x}$ is even and $N_{y}$ is odd for the flat-top hexagonal lattice and vice versa for the pointy-top hexagonal lattice.
	
	\printbibliography
	
\end{document}